\documentclass[twocolumn,twocolappendix]{aastex63}

\usepackage{amsmath} 
\usepackage[style=iso]{datetime2} 
\usepackage{booktabs} 
\usepackage{ragged2e} 
\usepackage{needspace} 

\usepackage{color} 
\definecolor{green}{rgb}{0.0,0.725,0.024}
\definecolor{red}{rgb}{1.0,0.0,0.0}
\definecolor{RED}{rgb}{1.0,0.0,0.0}

\newcommand{\hersc}{{\it Herschel}}
\newcommand{\planck}{{\it Planck}}
\newcommand{\spitz}{{\it Spitzer}}

\newcommand{\sne}{supernov\ae}

\newcommand{\HI}{H{\sc i}}

\newcommand{\dustbff}{{\texttt{DustBFF}}}
\newcommand{\st}{\textsuperscript{st}}
\newcommand{\nd}{\textsuperscript{nd}}
\newcommand{\rd}{\textsuperscript{rd}}
\renewcommand{\th}{\textsuperscript{th}}

\renewcommand{\baselinestretch}{0.9} 

\received{\DTMDate{2021-06-02}}
\revised{\DTMDate{2021-07-16}}
\accepted{\DTMDate{2021-07-20}}
\needspace{3\baselineskip} \submitjournal{(and accepted for publication in) ApJ}

\makeatletter 
\newcommand\footnoteref[1]{\protected@xdef\@thefnmark{\ref{#1}}\@footnotemark}
\makeatother

\shorttitle{The Quest for the Missing Dust: I}
\shortauthors{Clark et al.}

\begin{document}

\title{The Quest for the Missing Dust:\\ I -- Restoring Large Scale Emission in \textbf{\textit{Herschel}} Maps of Local Group Galaxies}

\correspondingauthor{Christopher J. R. Clark}
\email{cclark@stsci.edu}

\author[0000-0001-7959-4902]{Christopher J. R. Clark}
\affiliation{Space Telescope Science Institute, 3700 San Martin Drive, Baltimore, Maryland 21218-2463, United States of America}

\author[0000-0001-6326-7069]{Julia C. Roman-Duval}
\affiliation{Space Telescope Science Institute, 3700 San Martin Drive, Baltimore, Maryland 21218-2463, United States of America}

\author[0000-0001-5340-6774]{Karl D. Gordon}
\affiliation{Space Telescope Science Institute, 3700 San Martin Drive, Baltimore, Maryland 21218-2463, United States of America}

\author[0000-0001-6118-2985]{Caroline Bot}
\affiliation{Observatoire Astronomique de Strasbourg, Universit\'e de Strasbourg, UMR 7550, 11 rue de l’Universit\'e, F-67000 Strasbourg, France}

\author[0000-0002-3532-6970]{Matthew W. L. Smith}
\affiliation{School of Physics and Astronomy, Cardiff University, Queen's Buildings, The Parade, Cardiff, Wales, CF24 3AA, United Kingdom}




\begin{abstract}

Because the galaxies of the Local Group have such large angular sizes, much of their diffuse, large-angular-scale emission is filtered out by the \hersc\ data reduction process. In this work, we restore this previously missed dust in \hersc\ observations of the Large Magellanic Cloud, Small Magellanic Cloud, M\,31, and M\,33. We do this by combining \hersc\ data (including new reductions for the Magellanic Clouds), in Fourier space, with lower-resolution data from all-sky surveys (\planck, IRAS, and COBE) that did not miss the extended emission. With these new maps, we find that a significant amount of emission was missing from uncorrected \hersc\ data of these galaxies; over 20\% in some bands. Our new photometry also resolves the disagreement between fluxes reported from older HERITAGE Magellanic Cloud \hersc\ reductions, and fluxes reported from other telescopes. More emission is restored in shorter wavelength bands, especially in the galaxies' peripheries, making these regions 20--40\% bluer than before. We also find that the \hersc-PACS instrument response conflicts with the all-sky data, over the 20--90\arcmin\ angular scales to which they are both sensitive, by up to 31\%. By binning our new data based on hydrogen column density, we are able to detect emission from dust at low ISM densities (at $\Sigma_{\rm H} < 1\,{\rm M_{\odot} pc^{-2}}$ in some cases), and are able to detect emission at much lower densities (a factor of 2.2 lower on average, and more than a factor of 7 lower in several cases) than was possible with uncorrected data.

\end{abstract}

\keywords{Dwarf galaxies (416), Far infrared astronomy (529), Interstellar dust (836), Local Group (929), Submillimeter astronomy (1647), Astronomy data reduction (1861).}


\needspace{3\baselineskip} \section{Introduction} \label{Section:Introduction}

The life cycle of interstellar dust in galaxies is highly dynamic. Dust grains are understood to be primarily manufactured through stellar death -- by core-collapse \sne\ \citep{Barlow2010,Matsuura2011E,Gomez2012B} and asymptotic giant branch stars \citep{Hofner2018A}. Grains are then processed in the interstellar environment. In denser regions of the InterStellar Medium (ISM), dust can coagulate \citep{Stepnik2003B}, and gas-phase metals can accrete onto the grains \citep{Kohler2015A,Zhukovska2016A,Jones2017A}, decreasing the gas-to-dust ratio ($G/D$). Conversely, the reduced shielding in low-density environs increases the rate of photo-destruction of dust by high-energy photons from massive young stars \citep{Boulanger1998A,Beirao2006A} and sputtering by \sne\ shocks \citep{Jones2004B,Bocchio2014B,Slavin2015C}, returning metals to the gas phase. 


Dust is found to have properties that differ between environments \citep{Cardelli1989F,Gordon2003B}. The elemental composition of dust, as inferred from observed depletions of elements from the gas phase, is known to vary markedly between regions \citep{Jenkins2009B,Parvathi2012A,Roman-Duval2021B}; and ice spectral features are found in certain areas, indicating the formation of icy mantles \citep{Boogert2015A}. The dependence of dust's emissivity with wavelength -- typically expressed in terms of emissivity spectral index, $\beta$ -- is known to vary within galaxies \citep{MWLSmith2012A,Kirkpatrick2014B,Rigby2018B}, and traces variations in the physical properties of grains \citep{Demyk2017A,Demyk2017B,Ysard2018A}. In some situations, $\beta$ seems to exhibit a `break' at submillimetre (submm) wavelengths, manifesting as submm excess emission. This excess is most commonly found in dwarf galaxies \citep{Galliano2003A,Bot2010A,Remy-Ruyer2013A,Gordon2014B}, and in the periphery of larger late-type galaxies \citep{Paradis2012B,Hunt2015A}; the presence of submm excess is therefore associated with environments of lower density and of lower metallicity. Disentangling the interplay between the relative influence of density and metallicity on dust evolution therefore requires as much data as possible, that samples a wide range in both parameters.


The Local Group is the prime laboratory for understanding internal processes of galaxies, including as those governing dust and the ISM. When observing astrophysical processes in the Local Group, we can enjoy exceptionally high-fidelity data, but also benefit from being able to place our observations in the broader context of the entire galaxies within which those astrophysical processes are observed. Whilst we can of course study the Milky Way with resolution that can't be matched in external galaxies, such studies are complicated by the fact that distances to Milky Way features are often uncertain (especially for the ISM), large swathes of the Galaxy are obscured or confused from our observing position, and ultimately we are only observing one type of environment -- a high-mass high-metallicity moderately-star-forming spiral galaxy. 

\needspace{3\baselineskip} \subsection{Challenges of FIR--Submm Observations in the Local Group} \label{Subsection:Local_Group_Challenges}

Studies of dust emission in Local Group galaxies often suffer from the specific complexities of Far-InfraRed (FIR) and submm observing. Instruments in this regime generally operate by scanning the sky, with the resulting detector timelines compared and combined in the data reduction process to produce maps. However, the change in flux density recorded by the detector as it scans will not only be due to emission from the target source(s), but can also be caused by drift in instrumental temperature, varying atmospheric foreground emission (for ground-based telescopes), $1/f$ noise, and other effects. This will introduce considerable artefacts on larger angular scales in the resulting maps (see \citealp{Roussel2012B}, \citealp{Piazzo2015A}, and references therein). 

A standard way of minimising such artefacts is by observing a wide area of background devoid of emission from the target source. Source emission is then constrained relative to this background by comparing overlapping scans that cover both \citep{Meixner2013A}. However, observing large amounts of sky is not possible for all observing programmes. Plus, background scanning will struggle to save observations from noise that manifests at the same angular scales as the emission from the target object, or where the `empty' background also contains large-scale structure. It is practically impossible to recover emission on scales larger than the size of the map being scanned.

Another technique for removing large scale instrumental artefacts consists of high-pass filtering the detector timelines \citep{Griffin2010D,Roussel2012B}. However, this runs the risk of removing genuine large-scale astronomical emission in the observations \citep{Pascale2011A,Valiante2016A}. Additionally, some degree of ringing is also likely to be introduced around areas of compact bright emission \citep{Chapin2013B,Kirk2018B}; this ringing will manifest on the same scale as the applied filter.

Ultimately, most reduction treatments for suppressing large scale artefacts and noise from FIR--submm scan data will result in astrophysical emission on large angular scales {\it also} being filtered out of the observations. This is particularly problematic for the Local Group, as it contains the most extended galaxies on the sky -- observations of which therefore stand to suffer the most from any removal of FIR--submm emission on larger angular scales. Of particular concern is that the dust giving rise to this filtered-out flux will be the most diffuse dust in these galaxies, found in their outer regions and other low-density environments. This diffuse dust is likely to have distinct properties not found in denser areas, and is important for understanding the evolution of the ISM with environment, especially with regards to $G/D$ \citep{Roman-Duval2017B}. The loss of this flux will therefore systematically skew our understanding of the ISM.

There are, however, certain FIR--submm observations that are unaffected by the large scale noise and filtering issues. All-sky surveys with absolute photometric calibration, such as the COsmic Background Explorer (COBE; \citealp{Boggess1992B}) and \planck\ \citep{Planck2011I}, can can accurately capture emission on all angular scales, down to their resolving limit. However, these observatories have resolutions at least an order of magnitude worse than that achieved by larger telescopes such as the \hersc\ Space Observatory \citep{Pilbratt2010D}; but, of course, larger telescopes are the ones whose data most suffers from suppression of large-scale astrophysical emission\footnote{The \spitz\ \citep{Werner2004B} Multiband Imaging Photometer (MIPS; \citealp{Rieke2004K}) instrument was also remarkably good at preserving large-scale emission, including for the Magellanic Clouds \citep{Meixner2006C}, but instead suffered from non-linearity issues at 160\,\micron\ \citep{Meixner2013A}.}.

This situation – high-resolution data that is missing large-scale flux, and low-resolution data that preserves it – is one that radio astronomers handle frequently.  When performing multi-dish interferometry, it is common practice to use low-resolution single-dish data to restore the large-scale flux to which the high-resolution interferometric data is not sensitive, by combining the two datasets in Fourier space; this process is often referred to as `feathering'. Despite being a long-established technique in radio interferometry \citep{Bajaja1979E}, it has only rarely been applied to single-dish FIR--submm observations (eg, \citealp{Csengeri2016A}, \citealp{Abreu-Vicente2017A}, Smith et al. {\it subm.}).

\needspace{3\baselineskip} \subsection{Paper Overview} \label{Subsection:Paper_Overview}

In this paper, we use the `feathering' Fourier combination approach to produce corrected versions of the \hersc\ maps of the Local Group galaxies M\,31, M\,33, the Large Magellanic Cloud (LMC), and the Small Magellanic Cloud (SMC). 

In Section \ref{Section:Input_Data}, we detail all of the high- and low-resolution input data we employ, including our new \hersc\ reductions. To feathering together two observations requires us to first infer how the emission detected by the low-resolution telescopes {\it would appear} if observed in the bandpass used by the high-resolution telescopes; we describe this process in Section~\ref{Section:SED_Interpolation}. In Section \ref{Section:Feathering}, we cover the specifics of our Fourier combination process. In Section \ref{Section:Foreground_Subtraction}, we describe how we subtract Milky Way foreground emission from the maps we produce; then in Section \ref{Section:Initial_Results}, we present some initial results obtained with our new maps, exploring the properties of the newly-restored dust emission in our target galaxies.

\needspace{3\baselineskip} \section{Sample Galaxies} \label{Section:Sample_Galaxies}

\begin{figure*}
\centering
\includegraphics[width=0.975\textwidth]{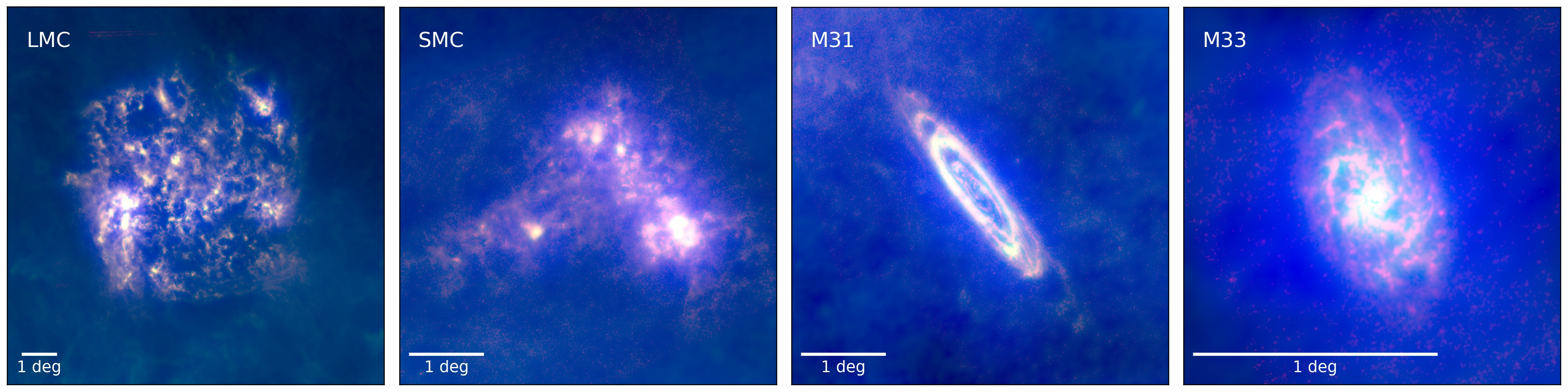}
\caption{FIR--submm three-colour images of the galaxies we study in this work, also illustrating the range of resolution and sensitivity in the data we employ. Blue is mapped to COBE-DIRBE 100\,\micron\ data (FWHM = 0.7\degr), green is mapped to \planck\ 350\,\micron\ data (FWHM = 4.6\arcmin), and red is mapped to \hersc-SPIRE 500\,\micron\ data (FWHM = 35\arcsec).}
\label{Fig:Sample_Multiwavelenth_Grid}
\end{figure*}

\begin{table}
\centering
\caption{Basic properties of the Local Group galaxies considered in this work. Values taken from the {\sc Nasa/ipac} Extragalactic Database; except for axial ratios and position angles, taken from the HyperLEDA database.}
\label{Table:Galaxy_Properties}
\begin{tabular}{lrrrr}
\toprule \toprule
\multicolumn{1}{c}{} &
\multicolumn{1}{c}{M\,31} &
\multicolumn{1}{c}{M\,33} &
\multicolumn{1}{c}{LMC} &
\multicolumn{1}{c}{SMC} \\
\cmidrule(lr){2-5}
$\alpha$ (J2000) & 10.69\degr & 23.46\degr & 80.89\degr & 13.16\degr \\
$\delta$ (J2000) & +41.27\degr &  +30.66\degr & -69.76\degr & -72.80\degr \\
Distance (kpc) & 790 & 840 & 50 & 62 \\
Hubble Type & SAb & SAcd & SBm & Irr \\
$R_{25}$ (arcmin) & 89 & 32 & 323 & 151 \\
$R_{25}$ (kpc) & 20.5 & 7.5 & 5 & 2.5 \\
Pos. Angle (deg) & 35 & 23 & 170 & 45 \\
Axial Ratio & 2.57 & 1.70 & 1.17 & 1.66 \\
\bottomrule
\end{tabular}
\end{table}

For this work, we produced feathered 100--500\,\micron\ maps of the Local Group galaxies M\,31, M\,33, the LMC, and the SMC. We explored the possibility of extending our reprocessing to other extended nearby galaxies, such as M101 and M51. However, as discussed in Section~\ref{Section:Feathering}, feathering together two observations requires there to be a range of angular scales over which they are both sensitive. A consequence of this is that the target source must be well-resolved in both sets of observations, otherwise the low-resolution data will only be providing information about emission on angular scales larger than the size of the target source. Even the most extended galaxies outside the Local Group, such as M101, are at best only marginally resolved in the available low-resolution data, meaning that feathering cannot be relied upon to restore extended emission missed by \hersc. On the other hand, the fact that these more distant galaxies have much smaller angular sizes means that they should not be susceptible to the removal of extended flux. In Sections~\ref{Section:Feathering} and \label{Section:Validation} we examine the scales at which flux does begin to be lost, and do indeed find it to be at large enough scales that galaxies outside the Local Group should not be effected.

The four Local Group galaxies which are suitable for feathering represent a broad range of galaxy types and properties: M\,31 gives us a high-mass disc galaxy passing through the green valley \citep{Mutch2011B}; M\,33 is a lower-mass gas-rich spiral featuring the highest star-formation efficiency in the Local Group \citep{Gardan2007A}; the LMC is a 0.5\,Z$_{\odot}$ galaxy on the spiral / dwarf-irregular border hosting the Local Group's most aggressive site of star formation \citep{Schneider2018A,Ruiz-Lara2020A}; and the SMC is a highly-disturbed 0.25\,Z$_{\odot}$ dwarf galaxy displaying the unusual ISM properties common to low-mass low-metallicity systems \citep{Jenkins2017A,Murray2019D}. There is clear value to producing corrected versions of the high-resolution FIR--submm observations of these key local laboratories. The basic properties of each of these galaxies is provided in Table~\ref{Table:Galaxy_Properties}, whilst FIR--submm colour images of each are shown in Figure~\ref{Fig:Sample_Multiwavelenth_Grid}.

\needspace{3\baselineskip} \section{Input Data} \label{Section:Input_Data}

The obvious low-resolution data to use for restoring large-scale emission to \hersc\ observations are the all-sky surveys produced by \planck, and the InfraRed Astronomical Satellite (IRAS; \citealp{Neugebauer1984}). The wavelength coverage of IRAS and \planck\ fully encompasses our bands of interest for \hersc, and their angular resolution is still high enough compared to \hersc\ that there is good overlap in the angular scales to which both instruments are sensitive. An additional complication arises from the fact that IRAS has {\it also} been found to have flux discrepancies on large angular scales, and lacks independent absolute calibration (see Section~\ref{Subsection:IRAS-IRIS}). We therefore take a two-stage Fourier combination approach. First, we feather together the IRAS data with COBE data. Then, in the second step, the \planck\ and rectified IRAS data is feathered with the \hersc\ data to produce our final maps. 

In this section we describe all of the input data we employ. For the \hersc\ PACS and SPIRE data, this includes a description of how we created our new reductions for the observations of the Magellanic Clouds. For each instrument, we also describe corresponding Point Spread Function (PSF), as these are of particular importance to the feathering process.

\needspace{3\baselineskip} \subsection{Herschel-PACS Data} \label{Subsection:Herschel-PACS}

The Photodetector Array Camera and Spectrometer (PACS; \citealp{Poglitsch2010B}) on board \hersc\ performed photometric imaging in 3 FIR bands, at 70, 100, and 160\,\micron\ (although the design of the filter wheel meant that only two bands could be observed simultaneously; 160\,\micron\ plus one of the others); the typical\footnote{The PACS beam size was dependent upon observing mode; the values given here are roughly average values, to be representative.} FWHM resolution in these bands is 9\arcsec, 10\arcsec,  and 13\arcsec\ respectively. 

For M\,31, these used the PACS data from \hersc\ programme \texttt{GT1\_okrause\_4}, as re-reduced by M.\,W.\,L Smith ({\it Priv. Comm.}). The detector timeline reduction was performed using the \hersc\ Interactive Processing Environment (\texttt{HIPE}; \citealp{Ott2010B}) v12\footnote{\label{Footnote:HIPE_v12}\texttt{HIPE} v12 uses the same photometric calibration products as \texttt{HIPE} v15.0.1,  the most recent version as of the time of writing.}. Map-making was then performed using the \texttt{Scanamorphos} pipeline \citep{Roussel2012B,Roussel2013A} v24.0, a pipeline designed to be particularly effective at preserving larger-scale emission, and minimise negative bowl features.

For M\,33, we retrieved the latest reductions from the \hersc\ Science Archive (HSA\footnote{\url{http://www.cosmos.esa.int/web/herschel/science-archive}}). Specifically, we used the Standard Product Generation (SPG; the standardised automated reductions provided by the HSA) maps, for which the detector timeline reduction was performed by the HSA using \texttt{HIPE} v14, and for which the map-making was carried out using the \texttt{JScanam} pipeline (a modified version of \texttt{Scanamorphos}, now included with \texttt{HIPE}, designed to be robust when run in an automated way; \citealp{Gracia-Carpio2017A}). At 70 and 160\,\micron, we used the PACS data from \hersc\ programme \texttt{OT2\_mboquien\_4}; at 100\,\micron, where data from that programme was not available, we instead used the shallower (although wider-area) data from programme \texttt{KPOT\_ckrame01\_1}.

\needspace{3\baselineskip} \subsubsection{PACS Re-Reduction for the Magellanic Clouds} \label{Subsubsection:PACS_Re-Reduction}

\begin{figure*}
\centering
\includegraphics[width=0.975\textwidth]{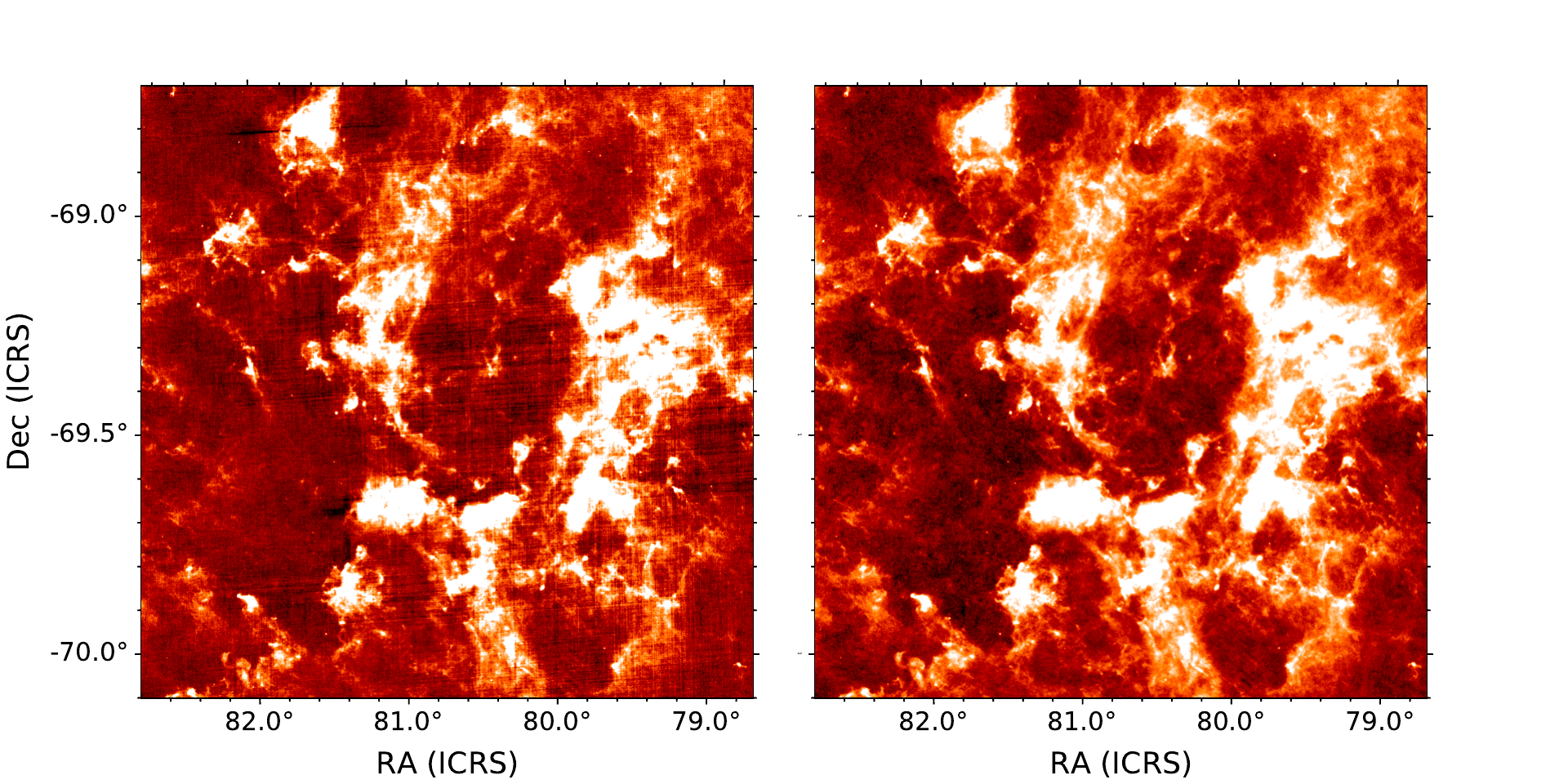}
\caption{Comparison of the old HERITAGE reduction of the PACS data ({\it left}), to our new reduction ({\it right}), for a portion of the LMC at 100\,\micron. Note the significant artefacts in the old reductions, such as the cross-hatching, and the linear negative features around bright regions. Also note the improved sensitivity in the new reduction revealing faint features not visible in the old map. (NB: The new map shown here does {\it not} incorporate the restored large-scale emission, as covered in Section~\ref{Section:Feathering}.)}
\label{Fig:PACS_Reduction_Old_New}
\end{figure*}

For the Magellanic Clouds, we performed our own re-reductions. The Magellanic Clouds were observed by \hersc\ as part of the {\it Herschel} Inventory of The Agents of Galaxy Evolution (HERITAGE; \citealp{Meixner2013A}) key programme (\texttt{KPOT1\_mmeixner1\_1}). The HERITAGE PACS data observed both the LMC and SMC at 100 and 160\,\micron. HERITAGE employed an unusual `basket-weave' observing strategy, where the full width of each Magellanic Cloud was observed with long continuous scans legs in alternating directions; then 6 months later, a matching set of orthogonal cross-scans was observed. This approach was taken for several reasons: it ensured that all scans sampled the `empty' sky beyond the target galaxies; it maximised the number of cross-scans that each scan overlapped, to allow instrumental variation to be disentangled more effectively; and it increased the fraction of the observing time that was spent integrating along scan legs, as opposed to time spent in turnaround or slewing. 

However, this observing strategy had the unintended consequence of leaving the PACS data severely affected by $1/f$ noise, arising from the extremely long scan legs (8\degr\ for the LMC, and 6\degr\ the SMC), which are the longest of any \hersc\ observations. In the PACS detector timelines, instrumental baseline drift entirely dominates over astrophysical signal. Whilst the redundancy from the multiple orthogonal cross-scans means this drift can be well accounted for, doing so requires holding all of the scans in memory, which was not plausible at the time of the original HERITAGE data release. Instead, they adopted a number of other strategies to remove the baseline drift from the scans; see Section~3 of \citet{Meixner2013A} for details. The baseline drift removal strategies employed by \citet{Meixner2013A} led to significant artefacts around bright sources (illustrated in the left panel of Figure~\ref{Fig:PACS_Reduction_Old_New}), and relied upon tying the surface brightness at the end of each scan leg to a linear interpolation from IRAS and COBE data, leading to conspicuous discontinuities.

We therefore needed to perform our own new reductions of the HERITAGE observations of the Magellanic Clouds. We obtained the detector timelines for each individual observation from the HSA, where they had already been reduced using SPG pipeline in \texttt{HIPE} v14, with the most recent calibration products. To create combined maps from these processed timelines, we used the \texttt{UNIMAP} pipeline \citep{Piazzo2015A}, v7.1.0. \texttt{UNIMAP} is a generalised least-squares map-maker, specifically designed to handle data where the detector timelines suffer from $1/f$ noise. Whilst most other PACS map-makers (such as \texttt{JScanam}) require specific matched pairs of scan and cross-scan observations in order to function, \texttt{UNIMAP} can operate with arbitrary sets of overlapping scans -- a necessary feature given the HERITAGE observing strategy. Recent releases of \texttt{HIPE} provide the option to call \texttt{UNIMAP} for the map-making stage of PACS data reduction, and this was how we used \texttt{UNIMAP} to produce maps for the LMC and SMC, with all processing options set to the default settings. This makes our LMC and SMC reductions substantially similar to those produced by the \hersc\ Science Centre as part of its PACS `Highly Processed Data Products' data release \citep{Calzoletti2017A}, which also used \texttt{UNIMAP}. However, we use a slightly newer version of \texttt{HIPE} than \citet{Calzoletti2017A}; also, their reductions for the SMC break the data into separate maps for the bar and bridge of the SMC, whereas we produce one contiguous map. A comparison of the old HERITAGE reduction and our new reduction for a portion of the SMC is shown in Figure~\ref{Fig:PACS_Reduction_Old_New}. As an example, our PACS 100\,\micron\ reduction for the LMC is shown in the right panel of Figure~\ref{Fig:LMC_Data_Example}.

\needspace{3\baselineskip} \subsubsection{Herschel-PACS PSF} \label{Subsubsection:Herschel-PACS_PSF}

For the PACS PSF, we used the azimuthally-averaged PSFs produced by \citet{Aniano2011A}\footnote{\label{Footnote:Aniano}\url{https://www.astro.princeton.edu/~draine/Kernels.html}}. For the LMC and SMC, the fact that the observations conducted during multiple epochs, spaced at six month intervals means that the effective PSF will be averaged over multiple orientations, meaning that the standard PACS PSF is unsuitable, as it has a great deal of azimuthal variation \citep{Bocchio2016C} that will not be reflected in our maps. For consistency, we therefore also use the \citet{Aniano2011A} PSFs when working with the data for M\,31 and M\,33.

\needspace{3\baselineskip} \subsection{Herschel-SPIRE Data} \label{Subsection:Herschel-SPIRE}

The Spectral and Photometric Imaging REceiver (SPIRE; \citealp{Griffin2010D}) on board \hersc\ provided simultaneous photometric imaging in 3 submm bands, centred at 250, 350, and 500\,\micron; these bands achieved resolutions of 18\arcsec, 25\arcsec, and 36\arcsec\ FWHM, respectively. 

For M\,31, we used the SPIRE data from \hersc\ programme \texttt{GT1\_jfritz\_1}, as re-reduced by M.\,W.\,L Smith ({\it Priv. Comm.}). The detector timeline reduction was performed using the Bright Galaxy Adaptive Element ({\sc BriGAdE}; \citealp{MWLSmith2012A,MWLSmith2013A}) pipeline, run with \texttt{HIPE} v12\footnoteref{Footnote:HIPE_v12}. In the case of M\,31, this data has fewer thermal drift artefacts than the SPG reductions from the HSA.

For M\,33, we took the latest SPG reductions from the HSA, which use the up-to-date \texttt{HIPE} v14 photometric calibrations. These maps use the data from \hersc\ programme \texttt{KPOT\_ckrame01\_1}.

In all instances, the data were reduced with the relative gains of the SPIRE bolometers optimised for detecting extended emission, using the values for the beam area values provided in \texttt{HIPE} v15; specifically, 469.7, 831.7, and 1793.5 arcsec$^{2}$ at 250, 350, and 500\,\micron\ respectively. Additionally, all maps were produced using the SPIRE de-striper, to mitigate large scale artefacts arising from instrumental drift.

\needspace{3\baselineskip} \subsubsection{SPIRE Re-Reduction for the Magellanic Clouds}  \label{Subsubsection:SPIRE_Re-Reduction}

For the LMC and SMC, we again used the data from HERITAGE (\hersc\ programme \texttt{KPOT1\_mmeixner1\_1}). We had to perform our own reductions, as the HSA cannot automatically generate reductions for observations with unusual scan strategies, like HERITAGE. The original HERITAGE reductions relied upon setting the surface brightness at the end of each scan leg to 0 -- making the final maps insensitive to emission at their edges, such as might arise from cirrus, extended Magellanic Cloud dust, etc.

We obtained the detector timelines for all of the HERITAGE observations from the HSA, where they had already been reduced through the SPG pipeline using the \texttt{HIPE} v14 calibrations. We merged all of the scan legs for all of the observations for each galaxy, and then carried out map-making using the SPIRE destriper. 

Running the destriper on the SPIRE observations of the LMC and SMC was a complex proposition. Firstly, the destriper can struggle to achieve good results when a map contains extremely bright regions -- which the Magellanic Clouds certainly do (30 Doradus, LHA 120-N 55A, etc). The destriper provides the option to specify regions that may prove problematic, so they can be masked from the destriping algorithm; we  therefore manually identified and masked 8 such regions  in the LMC, and 6  in the SMC (listed in Appendix~\ref{AppendixSection:SPIRE_Masked_Regions}). Secondly, the destriper is highly memory-intensive. The scan timelines are compared simultaneously, through an iterative-map-making process with which the destriper isolates instrumental drift; as such, a change in the value of any one pixel can potentially change the value of every other pixel. The memory requirements for carrying out this process for the HERITAGE observations of the Magellanic Clouds were exceptional. In particular, for the LMC, destriping all three SPIRE bands required 1 petabyte-hour of memory, with instantaneous memory usage peaking at almost 10 terabytes for the 250\,\micron\ data. This is in all likelihood the most computationally-intensive \hersc\ data reduction that will ever be carried out (whilst larger SPIRE maps do exist, they can be produced by reducing individual tiles of scans and cross-scans, and then mosaicking them together; the HERITAGE observing strategy precludes this).

\needspace{3\baselineskip} \subsubsection{Herschel-SPIRE PSF} \label{Subsubsection:Herschel-SPIRE_PSF}

In keeping with PACS, and for the same reasons given in Section~\ref{Subsubsection:Herschel-PACS_PSF}, also we used the azimuthally-averaged PSFs produced by \citet{Aniano2011A}\footnoteref{Footnote:Aniano} for SPIRE.

\begin{figure*}
\centering
\includegraphics[width=0.975\textwidth]{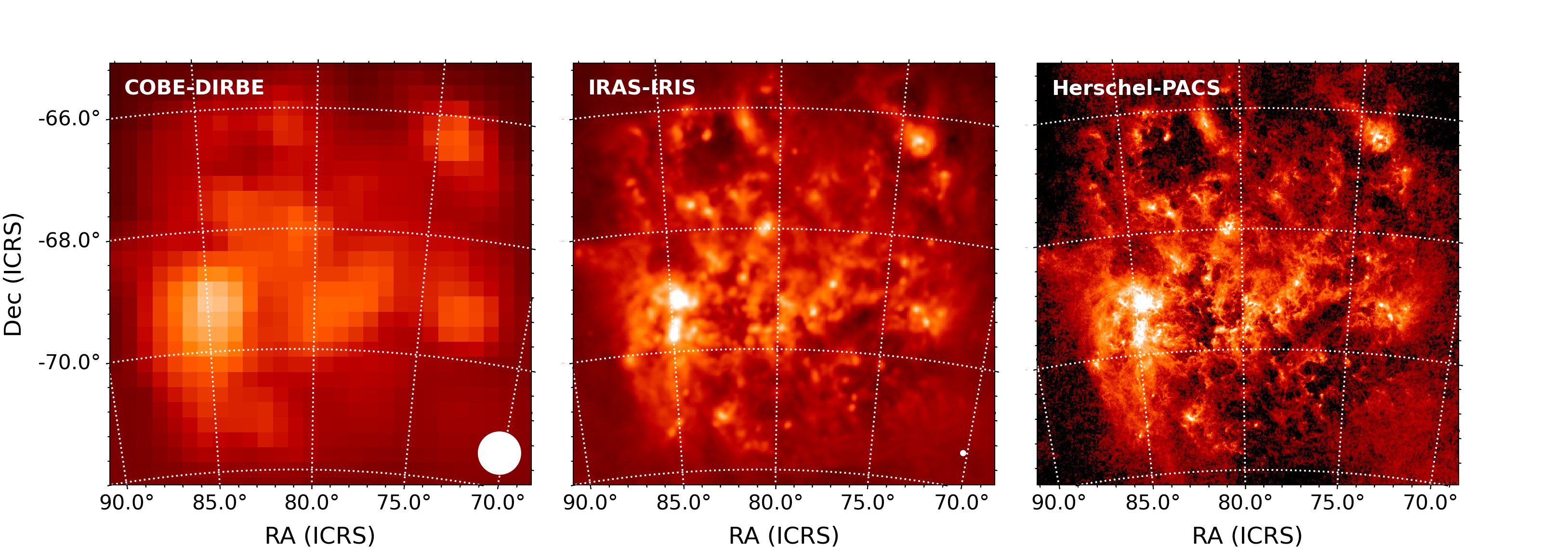}
\caption{The 100\,\micron\ observations of the LMC, from the three instruments that observed at that wavelength; DIRBE ({\it left}), IRIS ({\it centre}), and PACS ({\it right}). These IRIS and PACS images have not had large-scale emission restored by feathering. The beam FWHM for each instrument is shown by the white circle in each panel; the PACS beam is not visible at this scale. All three images have been matched to approximately the same colour scale; the dark `holes' in the PACS image, and the inconsistent surface brightness levels around the galaxy's outskirts, illustrate the difficulty PACS suffered in retrieving larger-scale diffuse emission.}
\label{Fig:LMC_Data_Example}
\end{figure*}

\needspace{3\baselineskip} \subsection{Planck Data} \label{Subsection:Planck}

The \planck\ satellite \citep{Planck2011I} surveyed the entire sky in 9 submm--microwave, bands from 350\,\micron\ to 10\,mm. Its mission to map the cosmic microwave background required its instruments to have highly accurate absolute calibration, with sensitivity to emission on angular scales spanning the whole range from its primary beam up to the all-sky dipole \citep{Planck2015VIII,Planck2015V}. This makes it perfectly suited to providing the absolutely-calibrated low-resolution data we need at wavelengths $\geq$\,350\,\micron.

We retrieved the 2018 \planck\ data release all-sky intensity maps from the \planck\ Legacy Archive\footnote{\url{https://pla.esac.esa.int/maps}}. Specifically we obtained the 350 and 550\,\micron\ maps, both from \planck's high frequency instrument \citep{Planck2018III}. These maps have had the zodiacal light contribution removed, are already provided in our desired surface brightness units of MJy\,sr$^{-1}$, and are calibrated assuming a constant $\nu S_{\nu}$ reference spectrum. The telescope achieves FWHM resolution of 4.6\arcmin\ and 4.8\arcmin\ at 350 and 550\,\micron\ respectively.

For each of our galaxies, we produced a square cutout map centred on the target, with a width equal to 20 times its $R_{25}$ (the isophotal semi-major axis at which the optical surface brightness falls beneath 25\,${\rm mag\,arcsec^{2}}$; see Table~\ref{Table:Galaxy_Properties}). We made these cutouts by reprojecting from the HEALPix projection \citep{Gorski2005A} of the all-sky maps to a standard east--north gnomic tan projection, using the \texttt{Python} package \texttt{reproject}.

\needspace{3\baselineskip} \subsubsection{Planck PSF} \label{Subsubsection:Planck_PSF}

Different parts of the sky were scanned a varying number of times by \planck, and each with different scan orientations, so the \planck\ PSF varies as a function of sky position. The \planck\ Legacy Archive is able to provide the average \planck\ PSF for a given area of sky. We obtained the average PSF for each of our sample galaxies, with the average taken over the $10 D_{25}$ square cutout area. In reality, the effective PSF will be slightly different over different parts of our sample galaxies, but these differences will be extremely small -- the variation in the solid angle of the \planck\ beam across the sky has a standard deviation of \textless 0.8\% in the bands we use \citep{Planck2015VII}. Performing feathering with a positionally-varying PSF would add exceptional complication to our work for very little gain. The average PSFs we obtained for each galaxy well match the average all-sky PSFs, but differ in terms of the structure of the wings.

\needspace{3\baselineskip} \subsection{IRAS-IRIS Data} \label{Subsection:IRAS-IRIS}

The InfraRed Astronomical Satellite (IRAS; \citealp{Neugebauer1984}) mapped 96\% of the sky in 4 IR bands, from 12--100\,\micron. The IRAS observations at 100\,\micron\ provide the necessary shorter-wavelength counterpart to the \planck\ data, with a well-matched angular resolution of 5\arcmin. Together, IRAS and \planck\ cover the entire wavelength range of the \hersc\ data we are reprocessing.

There exist two sets of all-sky IRAS maps; the \textsc{Iras} Sky Survey Atlas (ISSA; \citealp{Wheelock1994A}), and the Improved Reprocessing of the \textsc{Iras} Survey (IRIS; \citealp{Miville-Deschenes2005A}). For our purposes, both sets of maps have pros and cons. The newer IRIS maps have been de-striped, de-glitched, zodiacal light subtracted, and had their absolute flux calibration pegged to that of COBE. However, Bot et al.\,({\it in prep.}) have found that the IRIS maps nonetheless show considerable (up to $\sim$20\%) deviation from the COBE data over large angular scales (\textgreater\ a few degrees) in certain regions, such as the Magellanic Clouds -- precisely where we need to be able to depend on large-scale fidelity in the IRAS data. Whilst the ISSA maps do not suffer from this specific systematic, they do however lack the improved absolute calibration as the IRIS data, and have a highly non-linear response function that varies with both the angular scale and surface brightness of the emission being observed. Indeed, the ISSA explanatory supplement \citep{Wheelock1994A} suggests that users should assume photometric uncertainties of up to 30\% and 60\% at 60\,\micron\ and 100\,\micron\ respectively.

For this reason, we opt to use IRIS for our source of IRAS maps. But this requires us to first find a way to correct the extended surface brightness deviations in the IRIS data around the Magellanic Clouds. Fortunately, this problem can be corrected in exactly the same manner as the poor preservation of extended emission in the \hersc\ data.

First, we correct the IRIS data by feathering it with data from an even-lower-resolution telescope, one that has accurate absolute calibration and preservation of large-scale-emission. Then, we use this corrected IRIS data (in concert with the \planck\ data) to provide the low-resolution counterpart to correct the \hersc\ and \spitz\ data in a second round of feathering, to produce our final high-resolution maps. The obvious choice for the even-lower-resolution data with which to correct the IRIS maps is COBE, which we describe below, in Section~\ref{Subsection:COBE-DIRBE}.

We retrieved the IRAS-IRIS data we used from the InfraRed Science Archive (IRSA\footnote{\url{https://irsa.ipac.caltech.edu/data/IRIS/images/}}). For this work, we only required the IRAS 100\,\micron\ maps. For each our galaxies, we produced a square cutout map centred on the target, with a width equal to 20 times its $R_{25}$. We made these cutouts by mosaicking together all of the individual IRIS maps that covered the region of interest – by first projecting these individual maps to a shared pixel grid (using the \texttt{Python} package \texttt{reproject}), then taking the mean value in any pixel where multiple maps overlapped. This is the same process used in the mosaicking scripts released by the IRIS team\footnote{\url{https://www.cita.utoronto.ca/~mamd/IRIS/IrisDownload.html}}. As an example, the IRIS 100\,\micron\ map of LMC is shown in the centre panel of Figure~\ref{Fig:LMC_Data_Example}.

\needspace{3\baselineskip} \subsubsection{IRAS-IRIS PSF} \label{Subsubsection:IRAS-IRIS_PSF}

Unfortunately, the IRIS PSF is not especially well characterised. Because IRIS maps are constructed from scanning data, the instrumental PSF is worse in the cross-scan direction \citep{Wheelock1994A}. To mitigate this, the IRIS reductions were made such that all data was smoothed to this worst, cross-scan direction, rendering the effective PSF circular. \citet{Miville-Deschenes2005A} verified that the power spectrum of the resulting maps was compatible with the PSF being Gaussian, but performed no further characterisation.

We attempted to determine the true 100\,\micron\ IRIS PSF empirically, by stacking on the positions of bright point sources from the IRIS Point Source Catalog \citep{Beichman1988B}. However, the noise in the stack was high enough that we were unable to trace the PSF past 20\arcmin, beyond which the noise dominated. We tried limiting the stack to only sources with higher Signal-to-Noise Ratios (SNR), or to sources out of the Galactic plane, but in these cases the reduction in the number of sources being stacked meant that the noise in the stack remained similarly poor. However, we were able to verify that the radial profile of the IRIS PSF is well-modelled by a Gaussian with a FWHM of 4.4\arcmin\ out to the maximum radius we could trace. This value compares well to the 4.3\arcmin\ FWHM stated in \citet{Miville-Deschenes2005A}; we therefore opt to follow that official value, and use a Gaussian with FWHM of 4.3\,\arcmin\ as our 100\,\micron\ IRIS PSF.

\needspace{3\baselineskip} \subsection{COBE-DIRBE Data} \label{Subsection:COBE-DIRBE}

The Diffuse InfraRed Background Experiment (DIRBE; \citealp{Silverberg1993B}) instrument on COBE was a single-pixel absolute photometer that observed the entire sky in 10 bands from 1.25--240\,\micron, at an instrumentally-limited resolution of 0.7\degr. DIRBE required accurate absolute calibration for COBE's mission to map both the cosmic infrared background and cosmic microwave background. As a result, its accuracy has been well characterised  \citep{Fixsen1997C}, and DIRBE serves as the primary photometric reference standard for extended emission across most of its wavelength range \citep{Miville-Deschenes2005A,Gordon2006A}.

We retrieved the all-sky COBE-DIRBE maps at 100, 140, and 240\,\micron\ from the Legacy Archive for Microwave Background Data Analysis (LAMBDA\footnote{\label{Footnote:DIRBE_Products}\url{https://lambda.gsfc.nasa.gov/product/cobe/dirbe_products.cfm}}). Specifically, we used the Zodical-Light-Subtracted Mission Average Maps. These all-sky are presented in the pre-HEALPix quadrilateralised spherical cube projection \citep{Torres1989A}. We used this all-sky data to create square cutouts, of width 30\degr, centred on each of our sample galaxies. To reproject the maps from the quadrilateralised spherical cube projection to the standard gnomonic TAN projection for our cutouts, we used the algorithm supplied by the COBE team\footnoteref{Footnote:DIRBE_Products}, albeit adapted from the original \texttt{FORTRAN} code into \texttt{Python} (Oliver Lomax, {\it priv. comm.}).  As an example, the DIRBE 100\,\micron\ map of LMC is shown in the left panel of Figure~\ref{Fig:LMC_Data_Example}.

We note that there is an alternate set of DIRBE maps hosted at IRSA, stored in a HEALPix projection. However, the surface brightness in these maps, compared to those hosted at LAMBDA, do not agree. When re-gridded to the same projection to allow direct comparison, they can disagree by up to $\pm$15\%, particularly surrounding bright sources. It seems that this is due to a slight difference in resolution between the two maps. In communication with personnel at IRSA, it appears that these maps were produced and delivered by the \planck\ team as part of their data releases (potentially just for quick-look purposes). As the maps hosted at LAMBDA are the original and official maps delivered by the COBE team, we opt to use those. People seeking DIRBE data online should be aware that the different archives provide differing versions of the data.

\needspace{3\baselineskip} \subsubsection{COBE-DIRBE PSF} \label{Subsubsection:COBE-DIRBE_PSF}

The resolution of DIRBE is instrument-limited, not diffraction-limited, and is dictated by how the single pixel of DIRBE scanned a given sky position numerous times, along numerous direction. However, the DIRBE data products hosted on LAMBDA only include the instrument's {\it scanning} PSF – the spread function response from a single pass by the detector, in a particular scanning direction. We, however, need the effective PSF resulting from all the scans in all the directions. We therefore produced maps of the effective DIRBE PSFs for ourselves, by making an azimuthally-averaged version of each band's one-dimensional scanning PSF. The azimuthal average should well replicate the effect of the \textgreater\,100 scans  \citep{Hauser1998C} DIRBE conducted at every point on the sky. Our resulting PSFs have the smoothed top-hat shape expected of the DIRBE beam \citep{Kashlinsky1996D,Barreiro2004A}, with the correct width of 0.7\degr.

\needspace{3\baselineskip} \section{SED-Driven Interpolation} \label{Section:SED_Interpolation}

In order to correctly feather together two observations, those observations must be observing the sky with the same wavelength sensitivity. For instance, we cannot simply feather SPIRE 500\,\micron\ and \planck\ 550\,\micron\ observations together; although their spectral responses have significant overlap (having bandwidths of $\sim \lambda/3$), they will nonetheless sample different parts of any source SED, meaning observed brightness will vary as a function of source spectrum. This is even true for bands with the same effective wavelengths, such as the DIRBE, IRAS, and PACS 100\,\micron\ bands – each has a different response function, so the relative brightnesses reported by each will differ by source spectrum.

Therefore, we need to model the underlying source spectrum. Specifically, we take the approach of modelling the SED in each pixel of a given set of low-resolution maps, and then use this to predict what emission {\it would} be seen by the high-resolution filters. These predicted maps have the resolution (and preserved large-scale emission) of the low-resolution maps, translated into the high-resolution bands. This method also allows us to create these predicted maps for high-resolution bands that have no direct equivalent amongst the low-resolution data; for instance, by modelling the 100, 350, and 550\,\micron\ IRAS and \planck, we can predict the low-resolution emission that would be observed in the SPIRE 250\,\micron\ filter. In effect, we are performing a `physically motivated interpolation' to infer the emission in the high-resolution bands.

The FIR--submm emission in our observations will have an SED that is dominated by dust emission. This dust SED can be approximated by a Modified BlackBody (MBB), whereby the dust mass opacity varies with wavelength according to the emissivity law:

\begin{equation}
\kappa(\lambda) = \frac{\kappa(\lambda_{\it ref})}{\lambda_{\it ref}^{-\beta}}\lambda^{-\beta}
\label{Equation:SED_MBB}
\end{equation}

\noindent where $\kappa$ is the dust mass opacity (ie, the grain absorption cross section per unit mass) at wavelength $\lambda$, and $\Sigma_{d}$ is the dust column density.

The surface brightness, $S(\lambda)$, of dust emission at a given wavelength, $\lambda$, can be expressed by:

\begin{equation}
S(\lambda) = \kappa(\lambda)\ B(\lambda, T_{d})\ \Sigma_{d}
\label{Equation:SED_Surface_Brightness}
\end{equation} 

where $\kappa(\lambda_{\it ref})$ is the dust mass opacity at some reference wavelength $(\lambda_{\it ref})$, $B$ is the Planck function evaluated at wavelength $\lambda$ and dust temperature $T_{d}$, and $\beta$ is the dust emissivity spectral index. \citet{Gordon2014B} showed that the FIR--submm dust SED of the Magellanic clouds is particularly well fit by a Broken-Emissivity Modified BlackBody (BEMBB). In this model, the value of $\beta$ changes at some break wavelength $\lambda_{\it break}$, allowing us to fit SEDs that exhibit `submm excess' emission above what would be predicted from a MBB alone \citep{Galliano2003A,Bot2010A}, by changing to a shallower $\beta$ at longer wavelengths. The BEMBB emissivity law therefore takes the form:

\begin{equation}
\kappa(\lambda) = \frac{\kappa(\lambda_{\it ref})}{\lambda_{\it ref}^{-\beta}} E(\lambda)
\label{Equation:SED_BEMBB}
\end{equation}

\noindent where

\begin{equation}
E(\lambda) =
\begin{cases}
\lambda^{-\beta_{1}} & \text{for}\ \lambda < \lambda_{break}\\
\lambda^{\beta_{2}-\beta{1}}\,\lambda^{-\beta_{1}} & \text{for}\ \lambda \geq \lambda_{break}
\end{cases}       
\label{Equation:SED_BEMBB_Break}
\end{equation}

\noindent for which $\beta_{1}$ is the value of $\beta$ at wavelengths $< \lambda_{\it break}$, and $\beta_{2}$ is the value of $\beta$ at wavelengths $\geq \lambda_{\it break}$. 

Adopting a BEMBB model allows our SED fitting to be flexible. In the portions of the Magellanic Clouds and M\,33 known to exhibit submm excess \citep{Planck2011XVII,Gordon2014B,Relano2018A}, it will provide a good fit where a MBB model would have had larger (and systematic) residuals, Meanwhile, in areas where the SED would already have been well fit by a MBB model (or when fitting data at wavelengths too short to be sensitive to break; see Section~\ref{Subsection:DustBFF_DIRBE}), a BEMBB model can simply have $\beta_{1} = \beta_{2}$. Because all of our data is at wavelengths \textgreater\,100\,\micron, we do not need to be concerned about the power-law emission that results from stochastic heating small grains at $\lesssim$\,70\,\micron\ \citep{Boulanger1988,Desert1990}. The BEMBB model should therefore provide reliable interpolation to predict the surface brightness expected in the filters of interest.

Note that although we are modelling the SED with physical parameters (such as the dust temperature,  mass surface density, emissivity spectral index, etc), we are not, at this point, at all concerned with the actual physical implications of any of the models – we are only performing the SED fitting to achieve our physically motivated interpolation. We are therefore unconcerned by the fact that each pixel will contain a combination of emission from the target galaxy and the foreground Milky Way cirrus, or by the fact that some pixels might be more properly modelled by some other model (two temperature, etc) – as long as our SED modelling provides good fits for reliable interpolation, it has achieved its purpose. We only consider the various model parameters insomuch as they provide a diagnostic for the success of the fitting. 

For $\kappa(\lambda_{\it ref})$, we use the \citet{Roman-Duval2017B} value of $1.24\,{\rm m^{2}\,kg^{-1}}$ at $\lambda_{\it ref} = 160\,\micron$. For our purposes, the specific value of $\kappa(\lambda_{\it ref})$ is essentially arbitrary; however, using the \citet{Roman-Duval2017B} value allows ease of comparison to their IRAS and \planck\ \dustbff\ analysis of the Magellanic Clouds, which provides a useful additional check.

\needspace{3\baselineskip} \subsection{SED fitting with DustBFF} \label{Subsection:DustBFF}

We perform our SED fitting using the Dust Brute Force Fitter (\dustbff; \citealp{Gordon2014B}). \dustbff\ is a grid-based SED fitting code. This is the most efficient solution for fitting the SED of large numbers of pixels in a set of maps. 

The BEMBB model implementation in \dustbff\ does not explicitly parameterise $\beta_{2}$, instead using the 500\,\micron\ excess, $e_{500}$, which describes the relative excess in flux at 500\,\micron, above what would be expected from a MBB model with no break (where  negative values indicate a 500\,\micron\ deficit), given by:

\begin{equation}
e_{500} = \left( \frac{\lambda_{\it break}}{\rm 500\,\mu m} \right)^{\beta_{2}-\beta_{1}} - 1
\label{Equation:DustBFF_e500}
\end{equation}


With our BEMBB model, the free parameters are therefore $\Sigma_{d}$, $T_{d}$, $\beta_{1}$, $\lambda_{\it break}$, and $e_{500}$. The parameter grid we use is described in Table~\ref{Table:DustBFF_Grid}. When fitting DIRBE data alone (see Section \ref{Subsection:DustBFF_DIRBE}), we set $e_{500} = 0$ and fix $\lambda_{\it break}$, effectively adopting a MBB model, as DIRBE data does not provide sufficient long wavelength coverage to constrain any SED break. 

\begin{table}
\centering
\caption{SED model grid parameter ranges and step sizes, used for pixel-by-pixel SED fitting using \dustbff, with a BEMBB model. For the logarithmically-spaced parameters $\Sigma_{d}$ and $T_{c}$, we also give the percentage difference between grid steps.}
\label{Table:DustBFF_Grid}
\begin{tabular}{lrrr}
\toprule \toprule
\multicolumn{1}{c}{Parameter} &
\multicolumn{1}{c}{Minimum} &
\multicolumn{1}{c}{Maximum} &
\multicolumn{1}{c}{Step} \\
\cmidrule(lr){1-4}
$\Sigma_{d}$ (${\rm M_{\odot}\,pc^{-2}}$) & $10^{-3.5}$ & $10^{1}$ & 0.025\,dex (9.6\%) \\
$T_{d}$ (K) & 10 & 50 & 0.04\,dex (5.9\%) \\
$\beta_{1}$ & 0 & 3 & 0.1 \\
$^{\rm a }$ $\lambda_{\it break}$ (\micron) & 150 & 350 & 25 \\
$^{\rm a }$ $e_{500}$ & -0.5 & 2.0 & 0.075 \\
\bottomrule
\end{tabular}
\footnotesize
\justify
$^{\rm a }$ When fitting only DIRBE data, we set $e_{500} = 0$ and fix $\lambda_{\it break}$, effectively adopting a MBB model. \\
\end{table}

The grid spacing for $\Sigma_{d}$ and $T_{c}$ is base-10 logarithmic; this is to capture the wide dynamic range of possible dust mass surface densities, and the fact that the FIR--submm bands we are modelling are more sensitive to smaller changes in cooler dust temperatures than hotter ones. Our parameter grid effectively imposes a flat prior across the range of values modelled for each parameter (or flat in logarithmic space for $\Sigma_{d}$ and $T_{c}$), with zero likelihoods outside these ranges. The full grid contains over 24 million models.
\dustbff\ computes the probability of fitting the data for each model in the grid. The mathematical formalism under which \dustbff\ operates is provided in full in Section~4 of \citet{Gordon2014B}.

The \dustbff\ formalism incorporates a full covariance matrix, $\mathcal{C}$, given by:

\begin{equation}
\mathcal{C} = \mathcal{C_{\it calib}} + \mathcal{C_{\it instr}}
\label{Equation:DustBFF_Covariance}
\end{equation}

\noindent where $\mathcal{C_{\it calib}}$ is a matrix incorporating the covariances between bands due to calibration uncertainty, and $\mathcal{C_{\it instr}}$ is a matrix used to capture how the instrumental noise will effect the uncertainty for each model.

It is vital to take account of the correlated uncertainties, as they can lead to systematic errors in model results \citep{Veneziani2013B}. For instance, if correlated error in the calibration of the long-wavelength points of a dust SED leads to those fluxes being recorded as greater than they are, this would be erroneously interpreted as favouring a flatter emissivity slope $\beta$; accounting for the correlated uncertainties via $\mathcal{C_{\it calib}}$ protects against this. We provide $\mathcal{U}_{\it uncorr}$ and $\mathcal{U}_{\it corr}$ for each instrument in their respective subsections below. This effect can be accounted for by explicitly parameterising the correlated uncertainties \citep{Galliano2011B,CJRClark2019B}, or by treated them as hyperparameters in a hierarchical Bayesian model \citep{Kelly2012B}, or by including them as off-diagonal terms in the covariance matrix \citep{MWLSmith2012A,Gordon2014B}, as we do here.

We also incorporate the effect of the instrumental noise in the observations, via the matrix $\mathcal{C_{\it instr}}$ (which was not included in the original \citealp{Gordon2014B} presentation of \dustbff). This allows our fitting to take account of the fact that models which predict a higher surface brightness will be less effected by uncertainty, simply because the noise will be lower relative to the emission. All of the observations we will be fitting using \dustbff\ have effectively flat noise over the map areas being modelled. We therefore calculate the diagonal values of $\mathcal{C_{\it instr}}$ by taking the median of each observation's uncertainty map; the off-diagonal elements are zero. We provide  $\mathcal{C_{\it instr}}$ for each instrument in their respective subsections below, with a different  $\mathcal{C_{\it instr}}$ calculated for each galaxy's observations.

In previous uses of \dustbff, $\mathcal{C}$ also incorporated the correlated uncertainty between bands arising from the uncertainty on the background subtraction \citep{Gordon2014B,Roman-Duval2017B,Chastenet2017A}. However, as we are concerned with the entirety of the flux recorded in each pixel, we are not performing background subtraction, so background subtraction covariance is not present. 

\begin{figure*}
\centering
\includegraphics[width=0.975\textwidth]{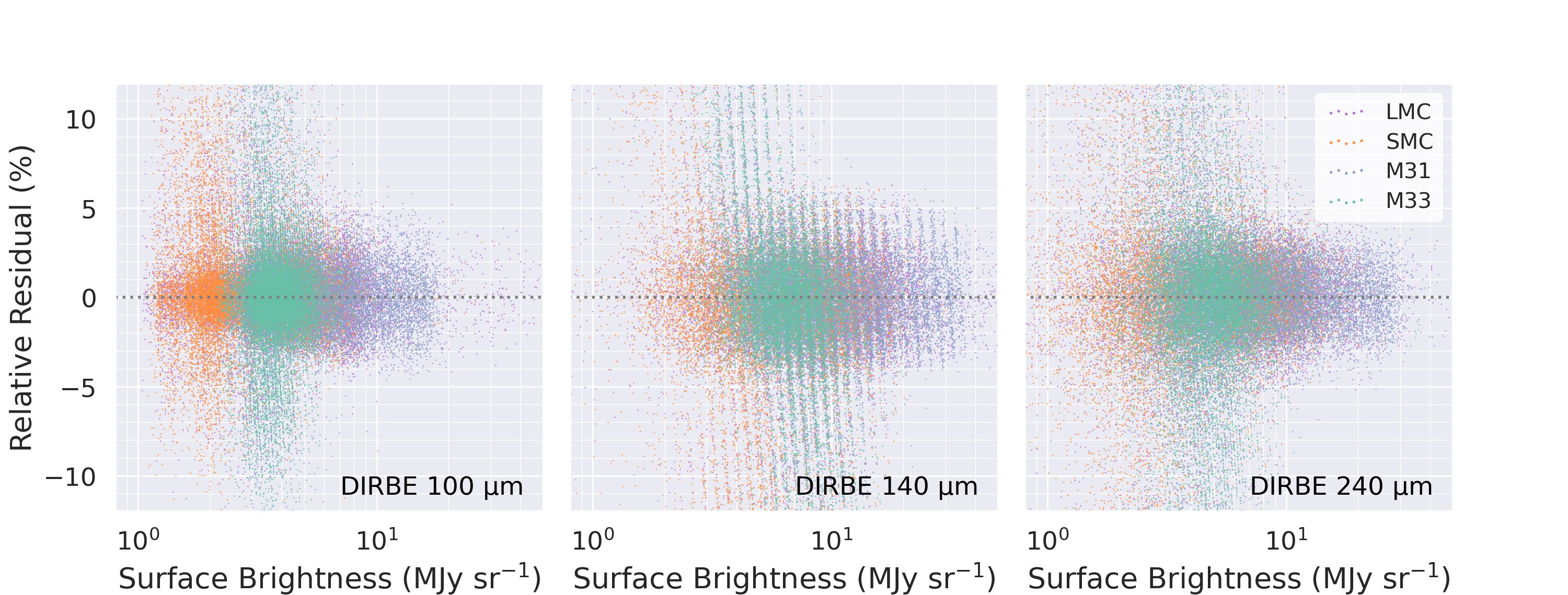}
\caption{Relative residuals (ie, $\frac{S_{\it mod} - S_{\it obs}}{S_{\it obs}}$) of \dustbff\ SED fits to DIRBE pixels for our target galaxies, for the MAP models. The plotted parameter space contains \textgreater\,98\% of pixels. The striations, most apparent at 140\,\micron, are due to the discrete nature of the model grid.}
\label{Fig:DIRBE_SED_Resid_vs_Bright}
\end{figure*}

The full posterior probability distributions returned by \dustbff\ are extremely large, containing the probabilities of all 24 million models, calculated for every single pixel (of which there are are over one million, in the case of our \planck\ and IRIS maps of the LMC). This is an impractical amount of data to handle, especially given that the vast majority of the models will have essentially nil probability for any given pixel. Therefore, for each pixel, we draw 500 random models from the posterior (with replacement), where the probably of any given model being drawn is proportional to its posterior probability. These 500 realisations of the posterior are what we use to constrain parameter medians, uncertainties, covariances, etc, in each pixel, and a propagated throughout our analysis. For our fiducial model in each pixel, we use the Maximum A-Posteriori likelihood (MAP) model; our fiducial output maps are comprised of the MAP model values for each pixel.

\needspace{3\baselineskip} \subsection{SED Fitting with DIRBE to Infer Large-Scale Emission in IRIS 100\,\micron\ Band} \label{Subsection:DustBFF_DIRBE}

As described in Section~\ref{Section:Input_Data}, we require DIRBE data to correct large-scale artefacts in the IRIS maps, and feathering together these two datasets first requires fitting the DIRBE SED, pixel-by-pixel, to infer how the emission observed by DIRBE would appear in the IRIS 100\,\micron\ band. To do this, we model the pixel-by-pixel SEDs of the DIRBE 100, 140, and 240\,\micron\ (ie, DIRBE bands 8, 9, and 10) observations of our target galaxies, using \dustbff.

All of the data are already at the same 0.7\degr\ instrumentally-limited DIRBE resolution, so no convolution was required prior to fitting. 

We fit the SED of each pixel using \dustbff. We kept $e_{500} = 0$ and fixed $\lambda_{\it break}$ at an arbitrary value, because DIRBE does not provide sufficiently long-wavelength coverage to constrain these parameters. Moreover, because we are only performing this fitting to predict the IRIS flux at 100\,\micron, constraining the model at the longest wavelengths is less vital (unlike for the IRIS--\planck\ SED fitting in Section~\ref{Subsection:DustBFF_IRIS-Planck}). This resulted in a grid of 79\,200 models. 

The uncertainty on these bands' calibration is dominated by the 10\% uncertainty on the absolute gain, which is shared between the bands; the remaining 1--3\% uncertainty for each band represents the photometric repeatability error arising from fluctuations in the relative gains, and is uncorrelated between bands. Therefore, each element of the DIRBE $\mathcal{U}_{\it corr}$ matrix has a value of 0.1; whilst the diagonal elements of the $\mathcal{U}_{\it uncorr}$ matrix have values of 0.03, 0.01, and 0.02 (with matrix rows/columns representing the 100, 140, and 240\,\micron\ values respectively).

%
%
%
%

We calculated the target galaxies' covariance matrices for instrumental noise, $\mathcal{C_{\it instr}}$, by taking the average absolute of each band's uncertainty map as the diagonal elements. The diagonal elements for each galaxy and band are given in Table~\ref{Table:DIRBE_C_Instr}; the off-diagonals are all zero. As can be seen, each band has similar noise across all maps.

\begin{table}
\centering
\caption{Diagonal elements of $\mathcal{C_{\it instr}}$, the instrumental noise covariance matrix for DIRBE SED fitting (off-diagonals are all zero). All values are in map units of MJy\,sr$^{-1}$.}
\label{Table:DIRBE_C_Instr}
\begin{tabular}{lrrr}
\toprule \toprule
\multicolumn{1}{c}{Galaxy} &
\multicolumn{1}{c}{100\,\micron} &
\multicolumn{1}{c}{140\,\micron} &
\multicolumn{1}{c}{240\,\micron} \\
\cmidrule(lr){1-4}
LMC & 0.0529 & 1.90 & 1.08 \\
SMC & 0.0452 & 1.94 & 1.11 \\
M\,31 & 0.0589 & 2.10 & 1.18 \\
M\,33 & 0.0529 & 2.45 & 1.38 \\
\bottomrule
\end{tabular}
\end{table}  

We validated the SED fits by checking the residual between the model and the observed surface brightness. In Figure~\ref{Fig:DIRBE_SED_Resid_vs_Bright}, we plot the relative residuals, defined as $\Delta_{\it rel} = (S_{\it mod} - S_{\it obs}) / S_{\it obs}$, for the MAP model of every pixel. For all galaxies and bands, the median $\Delta_{\it rel}$ was \textless\,0.7\%, with \textgreater\,61\% of pixels having $|\Delta_{\it rel}| < 5\%$. This indicates that \dustbff\ suffered no systematic problems fitting the data.

With the \dustbff\ SED fitting completed, we calculated how each model would appear if it were observed in the IRIS 100\,\micron\ band. This was done by convolving every model SED, in each pixel, through the IRAS 100\,\micron\ filter, colour-corrected to match the IRIS reference spectrum. The result was 500 images, representing the IRIS 100\,\micron\ fluxes predicted from the realisations of the posterior of the SED fit for every pixel, along with the fiducial image, containing the MAP model prediction for every pixel -- all at DIRBE resolution. An illustration of this for a particular pixel can be seen in Figure~\ref{Fig:Flux_Prediction_Example}.

Our predicted IRIS 100\,\micron\ band maps at the DIRBE resolution scale, with uncertainties constrained by our posterior sampling, are now suitable to be feathered with the the unfeathered IRIS maps. This process is described in Section~\ref{Subsection:Feathering_DIRBE_IRIS}.

\needspace{3\baselineskip} \subsection{SED Fitting with IRIS--Planck to Infer Large-Scale Emission in Herschel Bands} \label{Subsection:DustBFF_IRIS-Planck}

\begin{figure}
\centering
\includegraphics[width=0.475\textwidth]{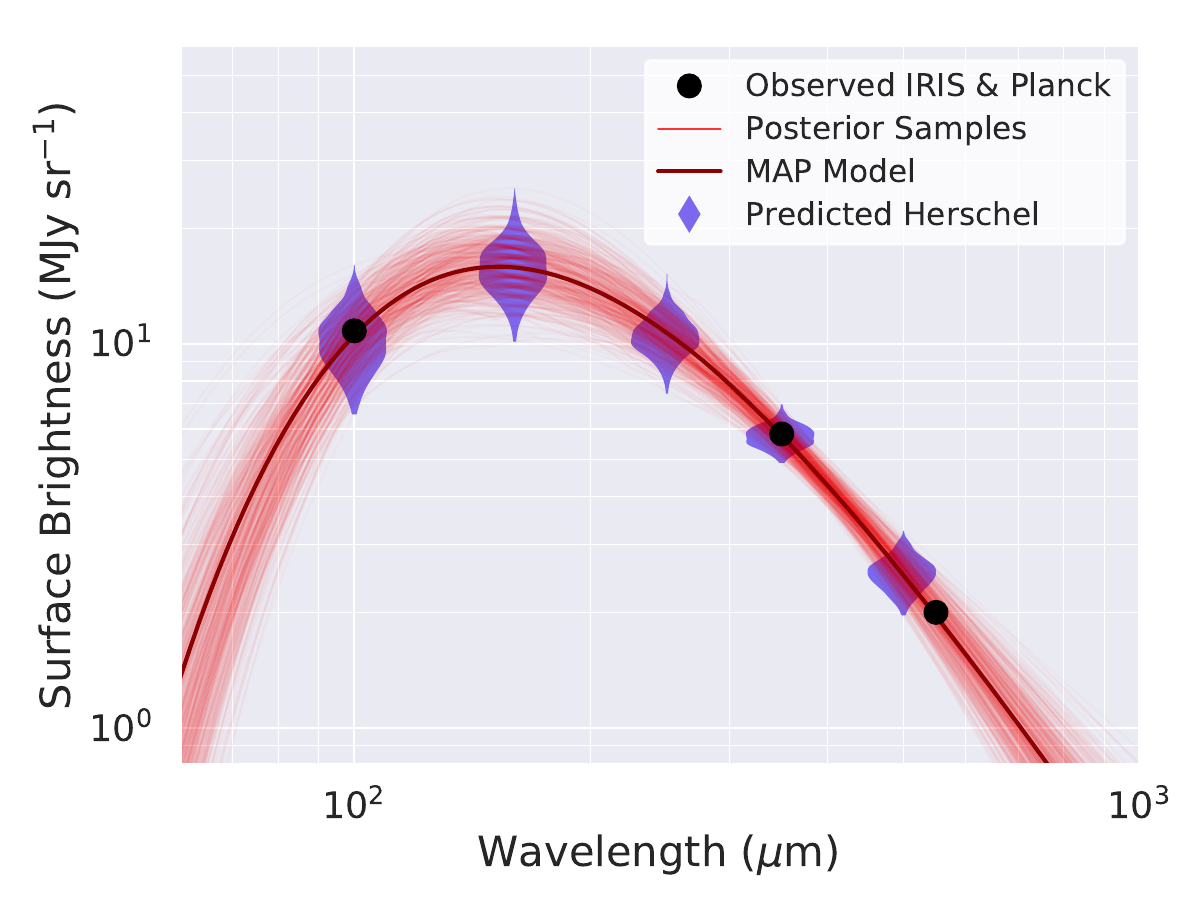}
\caption{Example of how our \dustbff\ SED fitting is used to predict fluxes, for the case of using IRIS--\planck\ data to predict \hersc\ fluxes for the pixel located at $\alpha = 23.637^{\circ}, \delta = 30.510^{\circ}$, in the outskirts of M\,33. The black circles show the observed surface brightness in the pixel, from our IRIS and \planck\ data. The thin red lines show 500 posterior samples from the \dustbff\ fit to the observations. The thick dark red line shows the MAP model. The blue violin plots show the distributions of predicted fluxes in each of the \hersc\ bands.}
\label{Fig:Flux_Prediction_Example}
\end{figure}

We fit the SED of IRIS and \planck\ data in order to predict how the emission observed by these facilities would appear in the \hersc\ bands; these predicted maps are can then be used to feather together the high- and low-resolution data. Note that in this section we use {\it the feathered IRIS data, that has already been combined with the DIRBE data}, as described later in Section~\ref{Subsection:DustBFF_DIRBE}\footnote{We present the SED fitting of the feathered IRIS data here, before we describe the feathering of the IRIS data, in order to keep together our descriptions of all the flux-prediction SED-fitting.}.

\begin{figure*}
\centering
\includegraphics[width=0.975\textwidth]{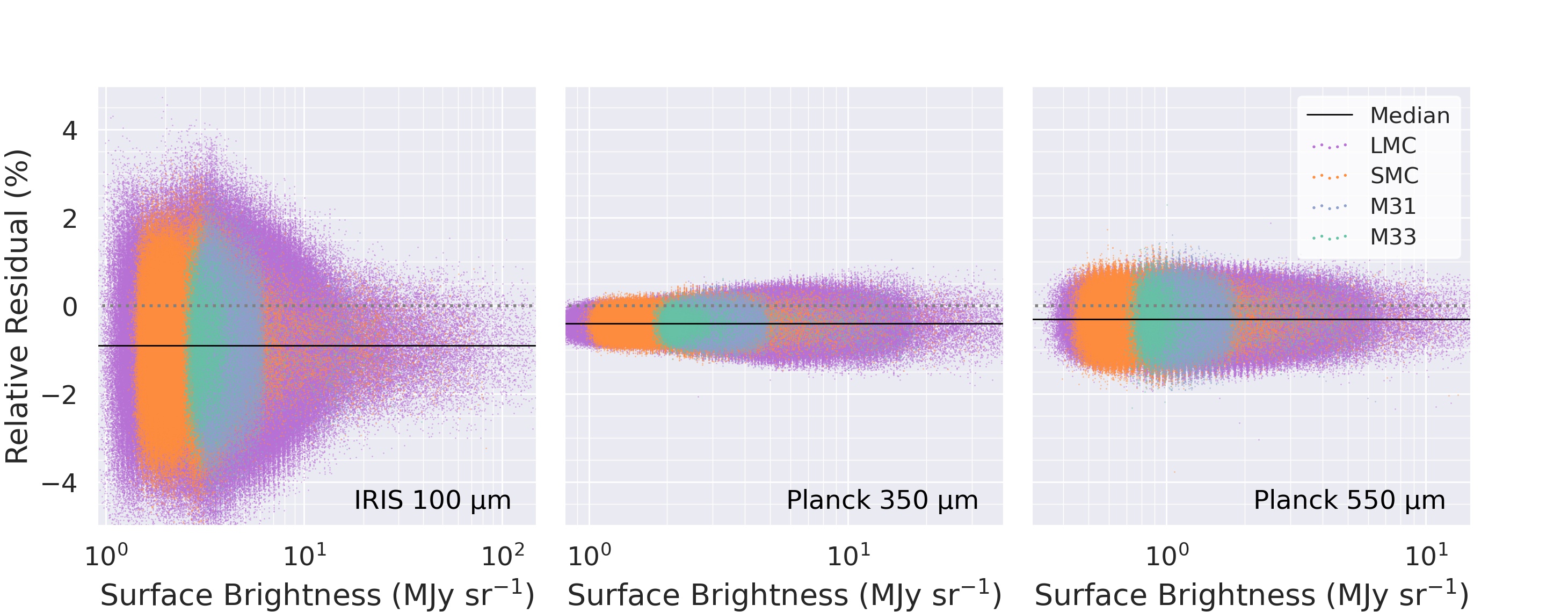}
\caption{Relative residuals (ie, $\frac{S_{\it mod} - S_{\it obs}}{S_{\it obs}}$) of \dustbff\ SED fits to IRIS--\planck\ pixels for our target galaxies, for the MAP models. Over 99\% of pixels fall into the parameter space plotted. The median relative residual for each band is also plotted; for each band, the difference between the galaixes' medians was less than 1 part in 500 in all cases.}
\label{Fig:IRIS-Planck_SED_Resid_vs_Bright}
\end{figure*}

\begin{table}
\centering
\caption{Diagonal elements of $\mathcal{C_{\it instr}}$ instrumental noise covariance matrix for IRIS--\planck\ SED fitting (off-diagonals are all zero). All values are in map units of MJy\,sr$^{-1}$.}
\label{Table:IRIS-Planck_C_Instr}
\begin{tabular}{lrrr}
\toprule \toprule
\multicolumn{1}{c}{Galaxy} &
\multicolumn{1}{c}{100\,\micron} &
\multicolumn{1}{c}{350\,\micron} &
\multicolumn{1}{c}{550\,\micron} \\
\cmidrule(lr){1-4}
LMC & 2.108 & 0.0154 & 0.0166 \\
SMC & 1.276 & 0.0167 & 0.0180 \\
M\,31 & 1.906 & 0.0265 & 0.0287 \\
M\,33 & 1.615 & 0.0308 & 0.0329 \\
\bottomrule
\end{tabular}
\end{table}  

Each of the IRIS and \planck\ bands are at slightly different resolutions, so prior to fitting we convolved each to the 4.8\arcmin\ poorest resolution of the \planck\ 550\,\micron\ band. We did this using conversion kernels creating with the \texttt{Python} package \texttt{photutils}, using the instrumental PSFs for each band.

We modelled the pixel-by-pixel SEDs of the IRIS 100\,\micron\, and \planck\ 350 and 500\,\micron, observations of our target galaxies, using \dustbff. We used the full grid of 24\,235\,200 models. The correlated uncertainty matrix used  for the IRIS--\planck\ SED fitting was:

\begin{equation}
\mathcal{U}_{\it corr} = 
\begin{bmatrix}
0 & 0 & 0 \\
0 & 0.05 & 0.05 \\
0 & 0.05 & 0.05
\end{bmatrix}
\label{Equation:DustBFF_IRIS-Planck_Matrix_Corr}
\end{equation}

\noindent where the rows/columns represent the 100, 350, and 550\,\micron\ values respectively.  The uncorrelated uncertainty matrix, $\mathcal{U}_{\it uncorr}$, had diagonal elements of 0.13, 0.014, and 0.011 (with off-diagonals of zero).

The 5\% correlated error between the \planck\ bands come from uncertainty on the emission models of Uranus and Neptune, the calibrator sources for \planck's high frequency instrument \citep{Bertincourt2016A}. As the IRIS 100\,\micron\ band is calibrated independently from the \planck\ bands, it has no correlated uncertainty here. The  13\% uncorrelated uncertainty on the IRIS 100\,\micron\ calibration is taken from \citet{Miville-Deschenes2005A}. This combines the 10\% uncertainty on the absolute calibration of DIRBE, with which IRIS is calibrated; and an additional 3.7\% from uncertainty in the colour corrections used by \citet{Miville-Deschenes2005A} to relate IRIS to DIRBE. The 1.4\% and 1.1\% uncorrelated uncertainties on the \planck\ 350 and 500\,\micron\ bands are the statistical repeatability noise \citep{Bertincourt2016A}. 

The diagonal values of the instrumental noise covariance matrices, $\mathcal{C_{\it instr}}$, are given in Table~\ref{Table:IRIS-Planck_C_Instr}, having been calculated from the average of the absolute values in the uncertainty map for each galaxy and band. For the 100\,\micron\ IRIS data, the uncertainty maps are part of the output from the DIRBE--IRIS feathering (detailed in Section~\ref{Subsection:Feathering_DIRBE_IRIS}).

As with the DIRBE SED fitting, we checked the SED fits by inspecting the relative residuals between the MAP model and observed surface brightnesses, for every pixel. This is plotted in Figure~\ref{Fig:IRIS-Planck_SED_Resid_vs_Bright}. There are definite systematic offsets apparent in the relative residuals. Fortunately, however, they remain very small in absolute terms. For IRIS 100\,\micron, the band with the worst offset, the median $\Delta_{\it rel}$ was only -0.91\%, with a standard deviation of 0.99\%, and \textgreater\,95\% of pixels having $|\Delta_{\it rel}| < 2.1\%$. For the \planck\ bands, median $\Delta_{\it rel}$ values are -0.40\% and -0.29\% at 350 and 500\,\micron\ respectively, with \textgreater\,95\% having $|\Delta_{\it rel}| < 0.91\%$ in both cases. Given that the systematics are so small, being at the sub-percent level in all cases, with symmetrical distributions, we believe the SED fits to be satisfactory. It is odd that the offsets are negative in {\it all} bands, but we suspect this may arise from the logarithmic spacing of the dust mass surface density (and/or temperature) increments in the parameter grid. 

Having completed the IRIS--\planck\ SED fitting\footnote{The IRIS--\planck\ SED fitting took about a week apiece for the Magellanic Clouds, and about a day each for M\,31 and M\,33, on a 3.2\,GHz $\times$ 32 core computer.}, we used the resulting posterior to predict how the modelled emission would appear in the \hersc\ 100--500\,\micron\ bands, convolving each model through the \hersc\ filters, colour-corrected to match the \hersc\ reference spectrum. And example of this for the model SEDs of a particular pixel are shown in Figure~\ref{Fig:Flux_Prediction_Example}. The output was 500 images, representing the \hersc\ fluxes in each band predicted from the realisations of the posterior of the SED fit for every IRIS--\planck\ pixel, along with that of the fiducial Maximum A-Posteriori (MAP) image, containing the MAP prediction for every pixel, all at DIRBE resolution.

We originally attempted this SED fitting process also using the \planck\ 850\,\micron\ band as well, to provide additional constraints for the models. However, we found that this resulted in considerably larger residuals, positive and negative, dominated by the structure of cosmic microwave background radiation, which is more prominent at 850\,\micron. For this reason, we opted to not use the \planck\ 850\,\micron\ data.

\needspace{3\baselineskip} \needspace{3\baselineskip} \section{Feathering} \label{Section:Feathering}

The process of feathering is widely employed to allow low-resolution single-dish data to restore missing large-angular-scale emission to interferometric observations. Large-angular-scale emission is ofter missing from interferometric data due to the fact that no arrangement of dishes can sample the shortest spacings, therefore making interferometers insensitive to the lowest frequency parts of the {\it u-v} plane (ie, the lowest frequency parts of the Fourier domain). However, this shortcoming of interferometry can be overcome by replacing the poorly-sampled parts of the {\it u-v} plane with data from single-dish observations, which don't suffer from the short-spacing problem \citep{Weiss2001A,Stanimirovic2002E}.

\begin{figure}
\centering
\includegraphics[width=0.475\textwidth]{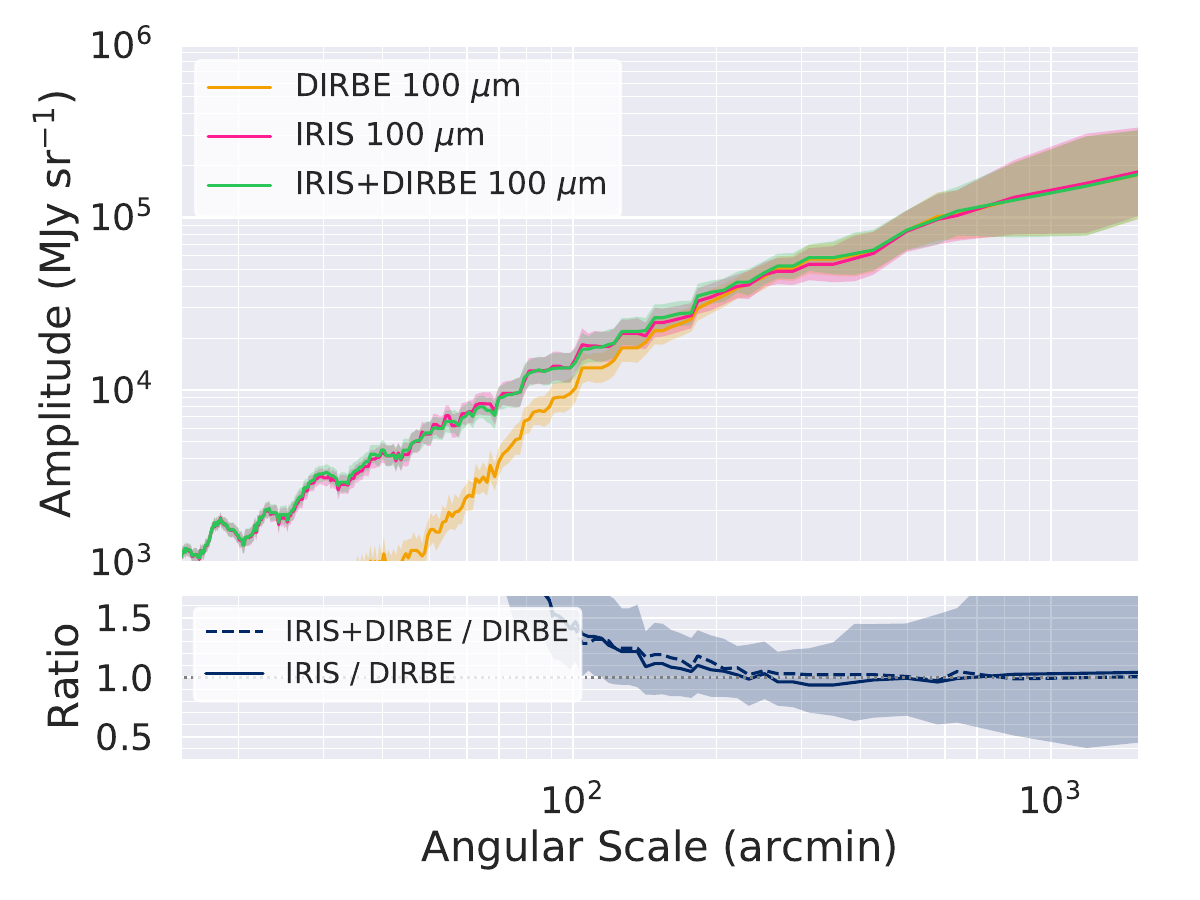}
\caption{The power spectra of the DIRBE data, IRIS data, and IRIS+DIRBE feathered data, for the SMC. The shaded region around each line shows the uncertainty on the average binned spectrum amplitudes at each scale. The panel beneath the main plot shows the ratios of the IRIS amplitudes to the DIRBE amplitudes, and of the feathered IRIS+DIRBE amplitudes to the DIRBE amplitudes; the horizontal dotted line indicates a ratio of unity.}
\label{Fig:IRIS-DIRBE_Power_Spec_SMC}
\end{figure}

The feathering technique of combining high- and low-resolution data in Fourier space is especially well-suited to single-dish FIR--submm observations. In multi-dish interferometry, the {\it u-v} plane is sparsely sampled due to the fact that there are only a finite number of dishes (and therefore baselines). This complicates the process of successfully combining the data with single-dish observations to provide complete sampling of all relevant spatial frequencies. With both of our datasets being single-dish, however, the situation is more straightforward. 

The low-resolution data provides full coverage of the {\it u-v} plane at spatial frequencies below the instrument's large PSF. The high-resolution data provides coverage of the {\it u-v} plane at spatial frequencies between the instrument's small PSF, up to some cutoff dictated by the observing and reduction strategy, above which emission is filtered (see Section~\ref{Subsection:Local_Group_Challenges}). As long as the resolution of the low-resolution data is better than the largest scale to which the high-resolution data is sensitive -- ie, as long as the spatial scales sampled by the two instruments have some overlap -- then the observations can be combined via feathering.

In practice, there are several steps to the feathering process:
\begin{enumerate}
\item Place the two observations on the same pixel grid.
\item Transform both observations to Fourier space.
\item Deconvolve the low-resolution data with the low-resolution beam.
\item Re-convolve the low-resolution data with the high-resolution beam.
\item If possible, correct for any gain differences between the brightness scales of the observations.
\item Replace the low-frequency components of the high-resolution data with the low-frequency components of the low-resolution data, with some weighting function imposing a smooth transition between the two.
\item Transform the combined data back out of Fourier space.
\end{enumerate}
\noindent For a more formal description of the mathematics behind the feathering process, see \citet{Vogel1984B}, \citet{Stanimirovic2002E}, and \citet{Weiss2001A}.

As outlined in Section~\ref{Subsection:Paper_Overview},  producing our final data entails two stages of feathering. First, we combine DIRBE and IRIS data; secondly the resulting feathered IRIS maps, together with \planck\ data, are combined with \hersc\ data to produce our final maps. Because of differences between these various data sets, the specific implementations of the feathering process for each of the two stages vary somewhat; descriptions of both are provided in the following subsections. In Appendix~\ref{AppendixFig:M101_Feathering_Sim}, we perform in/out simulation tests to verify that the feathering methodology we use is effective.

\begin{figure}
\centering
\includegraphics[width=0.475\textwidth]{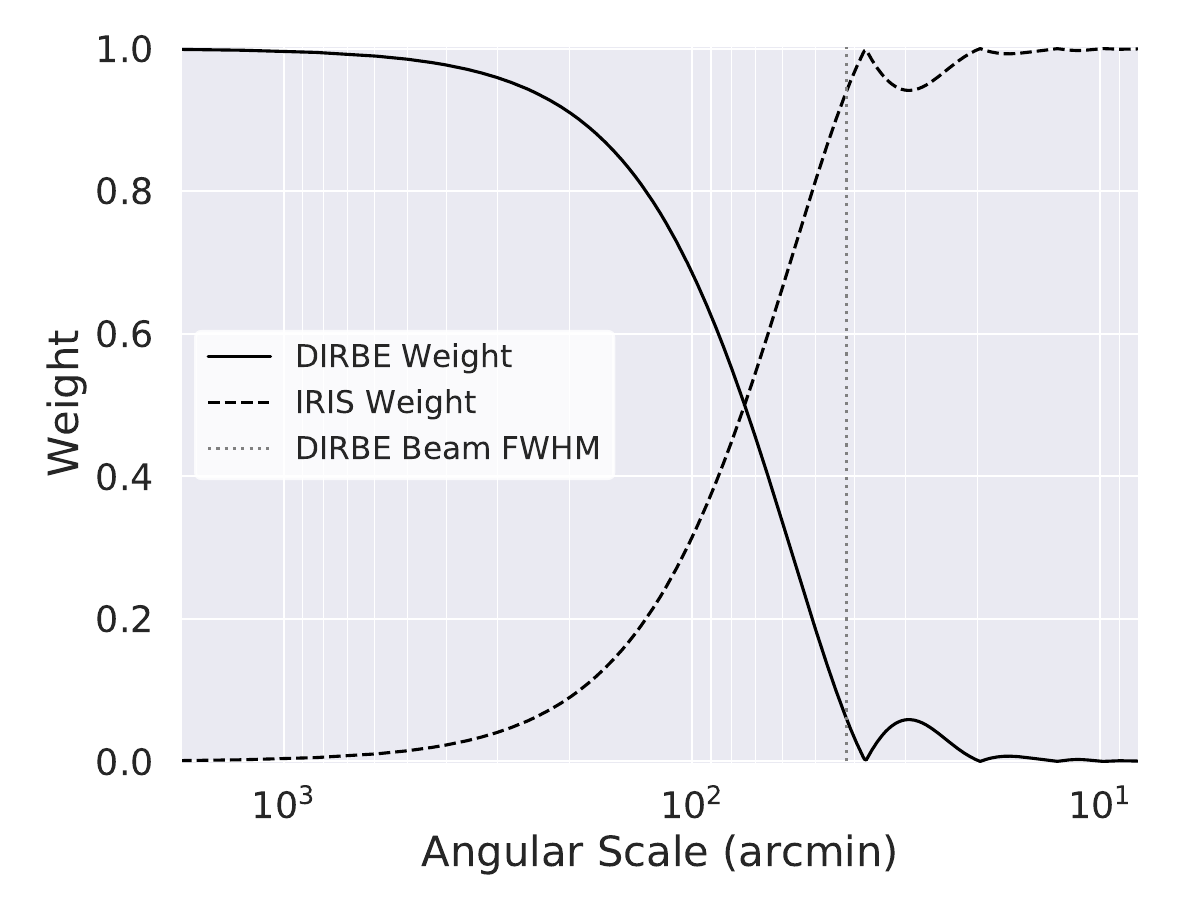}
\caption{The weighting applied at different angular scales (ie, in Fourier space) when feathering together IRIS and DIRBE data. The 0.7\degr\ scale of the DIRBE beam FWHM is shown for reference. As the DIRBE beam is not Gaussian, the crossover point in Fourier space does not occur exactly at the angular frequency corresponding to 0.7\degr\ -- rather,  because the DIRBE beam is flatter than a Gaussian, with relatively more weight at larger radii, the crossover is also at larger radii. The beam non-Gaussianity is also what causes the slight ringing structure at smaller scales}
\label{Fig:IRIS-DIRBE_Weighting}
\end{figure}

The code we used to perform the feathering process is available online at: \url{https://doi.org/10.5281/zenodo.4776266}. That code includes a main function to conduct the feathering itself (\texttt{FourierCombine}), along with sub-functions to handle the cross-calibration and construction of a tapering function (\texttt{FourierCalibrate} and \texttt{FourierTaper}).

\needspace{3\baselineskip} \subsection{Feathering Together IRIS and DIRBE} \label{Subsection:Feathering_DIRBE_IRIS}

\begin{figure*}
\centering
\includegraphics[width=0.975\textwidth]{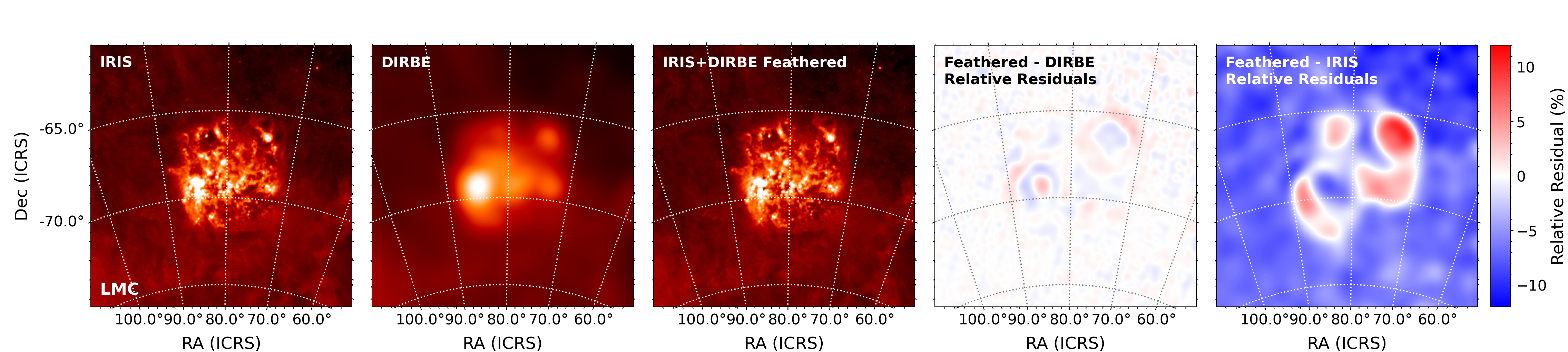}
\includegraphics[width=0.975\textwidth]{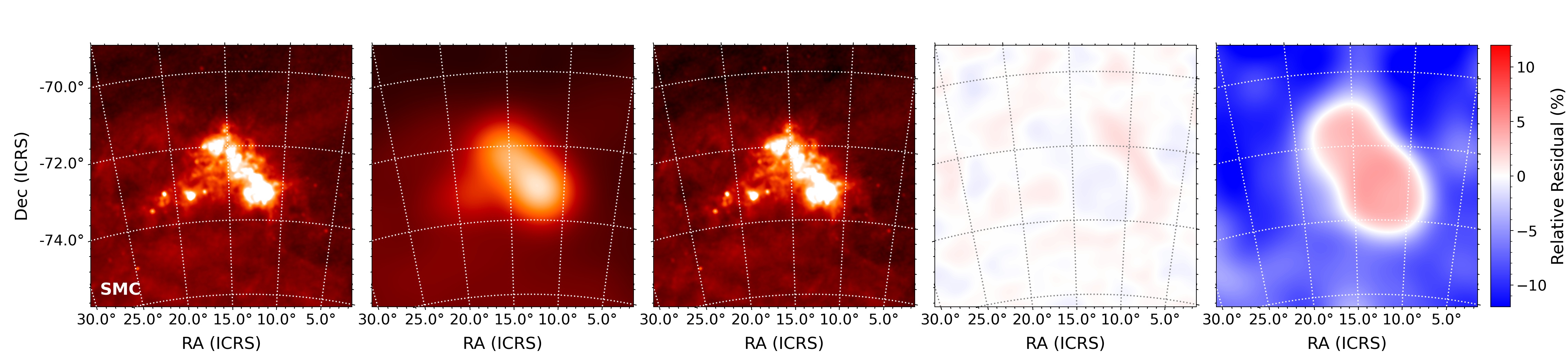}
\includegraphics[width=0.975\textwidth]{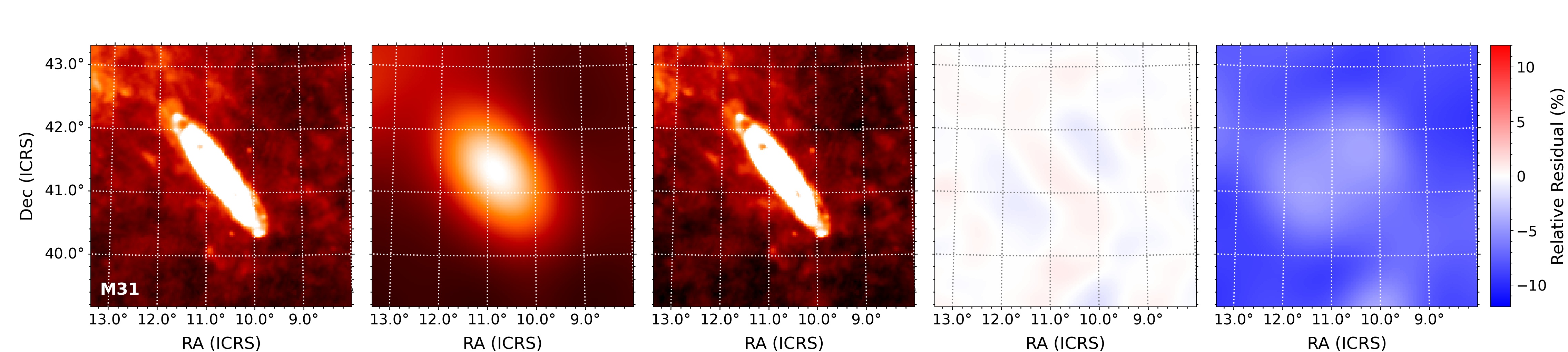}
\includegraphics[width=0.975\textwidth]{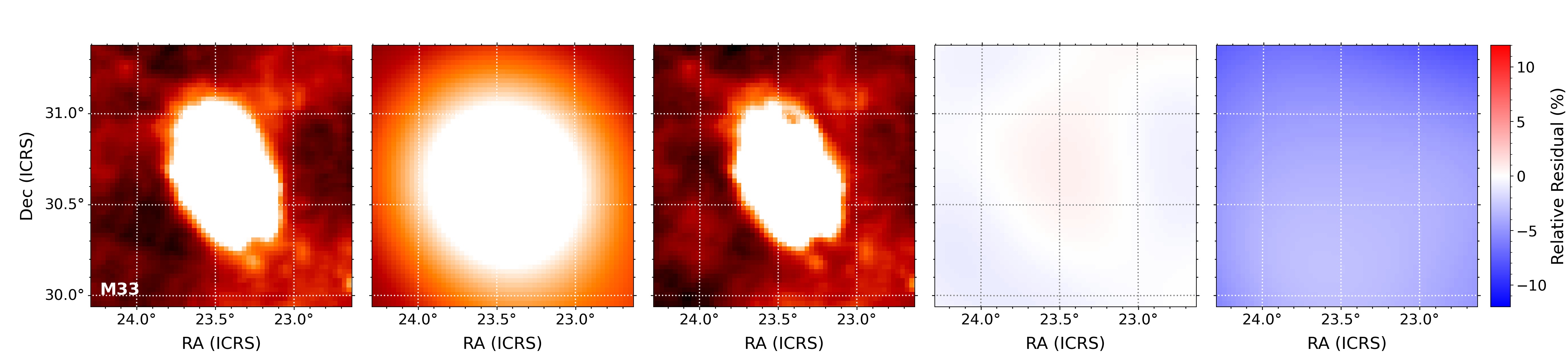}
\begin{minipage}{\linewidth}
\caption{The results of feathering together IRIS and DIRBE data at 100\,\micron\ for each of our galaxies, along with diagnostic plots. In all cases, our fiducial data are shown. {\it 1\st\ Panel:} The un-corrected IRIS data. {\it 2\nd\ Panel:} The DIRBE data. {\it 3\rd\ Panel:} The combined IRIS+DIRBE data after feathering (all three surface brightness maps for each galaxy use the same colour scale). {\it 4\th\ Panel:} The relative residuals between the feathered IRIS+DIRBE data and the input DIRBE data, compared at the DIRBE resolution; positive residuals mean the feathered data has greater surface brightness than the DIRBE data, negative mean less. Ideally, this should be zero everywhere. {\it 5\th\ Panel:} The relative residuals between the IRIS+DIRBE data and the unfeathered IRIS data; positive residuals indicate where the IRISD surface brightness was increased by feathering, negative where it was decreased.}
\label{Fig:IRIS-DIRBE_Feathering_Output}
\end{minipage}
\end{figure*}

The only band for which we feathered together DIRBE and IRIS data was at 100\,\micron. The DIRBE data we used are the maps we produced in Section~\ref{Subsection:DustBFF_DIRBE}, predicting the emission that DIRBE would have observed in the IRIS 100\,\micron\ band. For our fiducial feathered maps, we combine the IRIS cutout for each galaxy with the image produced from the MAP outputs for the pixel-by-pixel DIRBE SED-fitting (Section~\ref{Subsection:DustBFF_DIRBE}).

For each target galaxy, we reprojected this DIRBE data to the IRIS pixel grid. We also reprojected the DIRBE and IRIS PSF maps to have the same dimensions and pixel widths as the IRIS observation. As part of the reprojection, we also had to apodise the DIRBE maps, to remove pixel-edge artefacts from the re-gridded maps; we detail this in Appendix~\ref{AppendixSection:Apodisation}.

With all of the data on the same pixel grid, we transform the observations and beams to Fourier space. Next, we deconvolve the DIRBE data with the DIRBE beam, then convolve it with the IRIS beam. This accounts for the brightness differences between the data sets that were caused {\it only} by the differences in their resolution; ie, where a given source would have had a lower peak brightness in the low-resolution data. The DIRBE data now has the same beam amplitude and effective resolution as the IRIS data\footnote{The deconvolution of the low-resolution data introduces almost infinite noise to the low-resolution data at small spatial scales; however, because we will only be extracting large spatial scale information from this data when feathering, this causes no problems.}, meaning the two can be directly compared and combined.

The spatial scales at which systematic response errors are observed in the IRIS data ($\sim$\,1\degr; Bot et al., {\it in prep.}) are only somewhat larger than the spatial scales above which DIRBE is sensitive, as dictated by its 0.7\degr\ beam. The power spectra of both datasets, for the example of the SMC, are shown in Figure~\ref{Fig:IRIS-DIRBE_Power_Spec_SMC}. The amplitude of the IRIS power spectrum is not considerably different from that of DIRBE at the largest scales, because the errors in IRIS are localised only around certain parts of the LMC and SMC; the IRIS data is otherwise well-behaved. This can also be seen in the 5\th\ column of Figure~\ref{Fig:IRIS-DIRBE_Feathering_Output}.

For feathering together the DIRBE and IRIS data in Fourier space, we chose to follow the weighting strategy used in the Common Astronomy Software Applications (CASA; \citealp{McMullin2007A}) task \texttt{feather}\footnote{\url{https://casa.nrao.edu/casadocs/casa-5.4.1/image-combination/feather}}. In this approach, the Fourier transform of the low-resolution beam is used as the weighting function. Therefore, at the largest scales (ie, the zeroth Fourier mode), the DIRBE data is assigned a weight of 1. Going to smaller spatial scales, the weighting of the DIRBE data decreases, whilst the weighting of the IRIS data increases; the sum of both weights remains 1 everywhere. At scales smaller than the DIRBE beam FWHM, the weights cross over so that the IRIS data dominates, and becomes increasingly heavily weighted. This is illustrated in Figure~\ref{Fig:IRIS-DIRBE_Weighting}. 

The CASA algorithm minimises ringing by making the weighting transition as smooth as possible, tapering according to the low-resolution beam, using its FWHM as the natural crossover scale.  A disadvantage of this approach is that the high-resolution data will nonetheless be assigned some (albeit minimal) weighting at the larger scales, where it has poor sensitivity (and vice-a-versa for the low-resolution data). However, If one were to instead attempt to `squeeze' the transition region into a smaller range of scales, where the two datasets' sensitivity overlaps, there will be considerable ringing in the combined image if the transition window is too narrow. This is what would happen if we attempted to combine the DIRBE and IRIS data with such a transition window; the small-scale edge of such a window would be dictated by the 0.7\degr\ DIRBE beam, whilst the large-scale edge of the window would be dictated by the $\sim$\,1\degr\ scale of the IRIS response errors around the LMC and SMC. 

We therefore opt instead to take advantage of the ringing minimisation provided by the CASA algorithm. We test this beam tapering method, by feathering simulated high- and low-resolution data, in Appendix~\ref{AppendixSection:Feathering_In-Out_Tests}.

Comparison of the DIRBE and IRIS power spectra (Figure~\ref{Fig:IRIS-DIRBE_Power_Spec_SMC})) suggests there are no significant deviations between the brightness scales. IRIS was constructed to be cross-calibration to the DIRBE brightness scales, so this is not surprising (indeed, it seems that the IRIS response errors around the LMC and SMC are localised failures of this cross-calibration).  Because the largest scales of the feathered maps are dictated entirely by the DIRBE data, the zero-level of the output maps will match that of DIRBE, by construction.

We carried out the feathering process 500 times, once for each sample of the DIRBE SED-fitting posterior interpolations produced in Section~\ref{Subsection:DustBFF_DIRBE}. We also feathered the fiducial DIRBE data with the unfeathered IRIS data to produce our fiducial IRIS+DIRBE map. For each of these 500 iterations, we permutated the surface brightness values in each IRIS pixel according to the IRIS uncertainty maps, assuming a random Gaussian noise distribution. The 500 resulting feathered maps therefore provide a bootstrapped Monte Carlo estimate of the uncertainty on the feathered output in each pixel. 

\needspace{3\baselineskip} \subsubsection{Validation of Feathering IRIS with DIRBE} \label{Subsubsection:Validating_Feathering_DIRBE_IRIS}

The results of the IRIS+DIRBE feathering are shown in Figure~\ref{Fig:IRIS-DIRBE_Feathering_Output}. The first three columns compare the IRIS, DIRBE, and feathered IRIS+DIRBE data for each galaxy. The last two columns show our two main validation tests for the feathering process. 

\begin{figure*}
\centering
\includegraphics[width=0.475\textwidth]{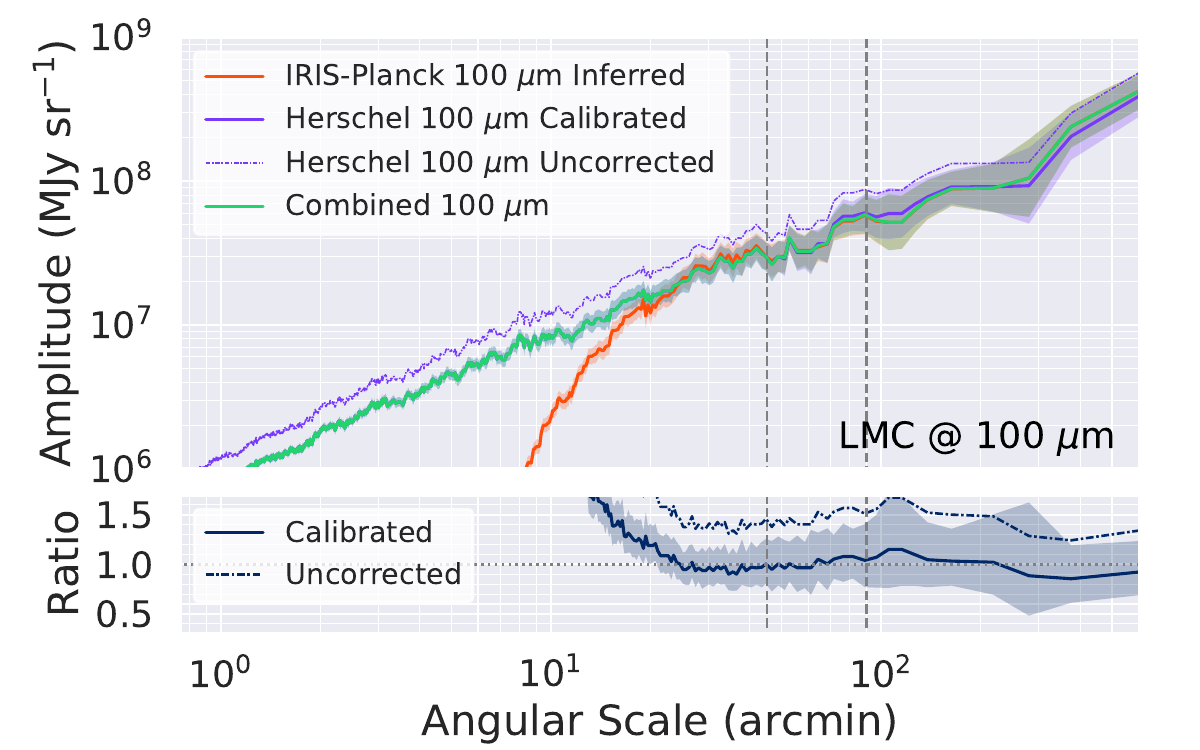}
\includegraphics[width=0.475\textwidth]{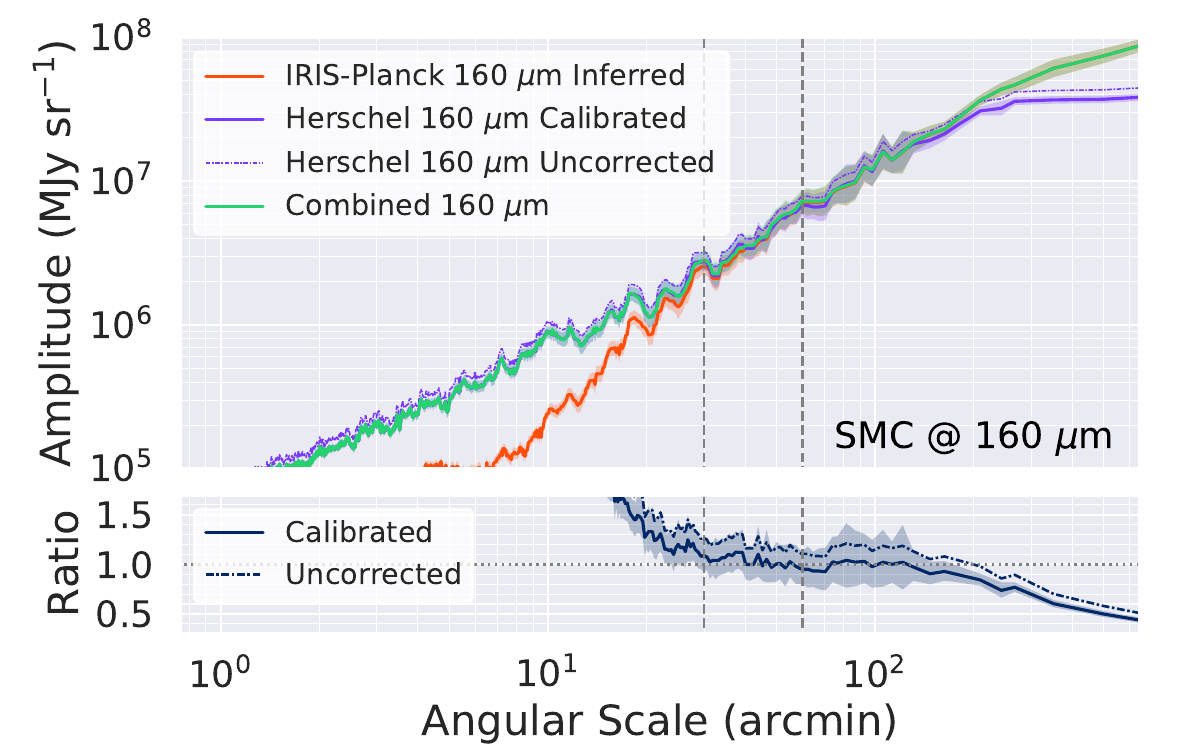}
\includegraphics[width=0.475\textwidth]{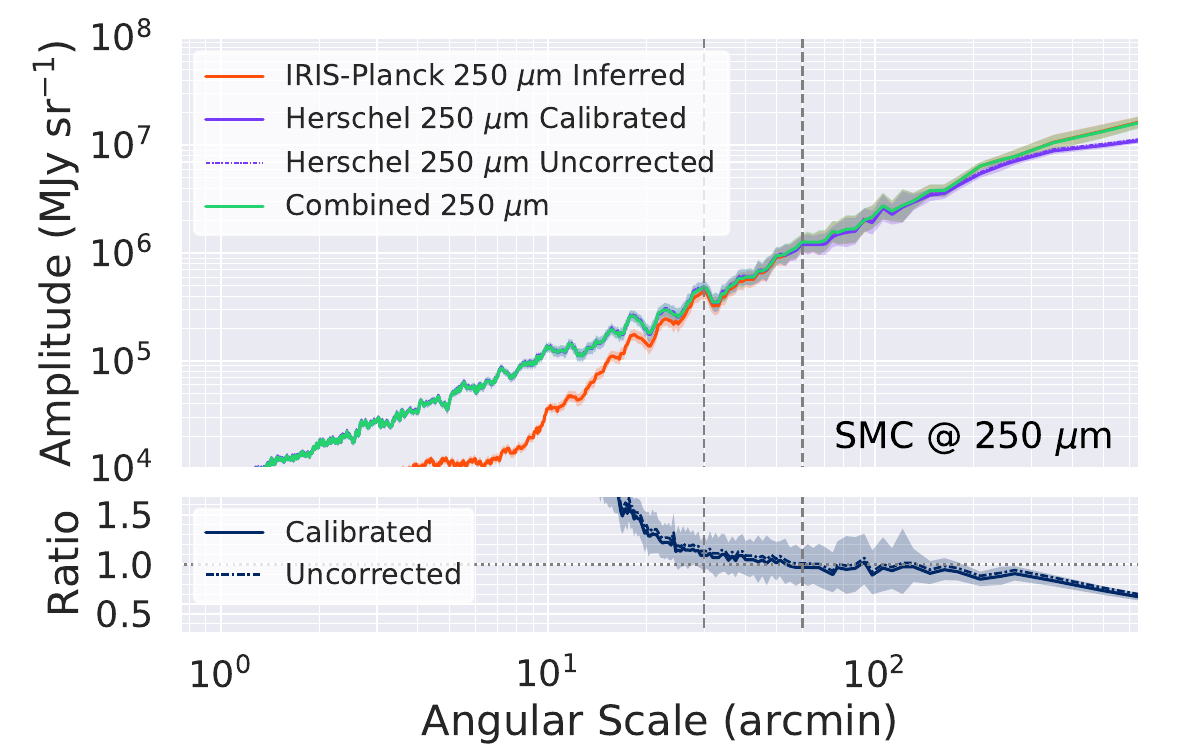}
\includegraphics[width=0.475\textwidth]{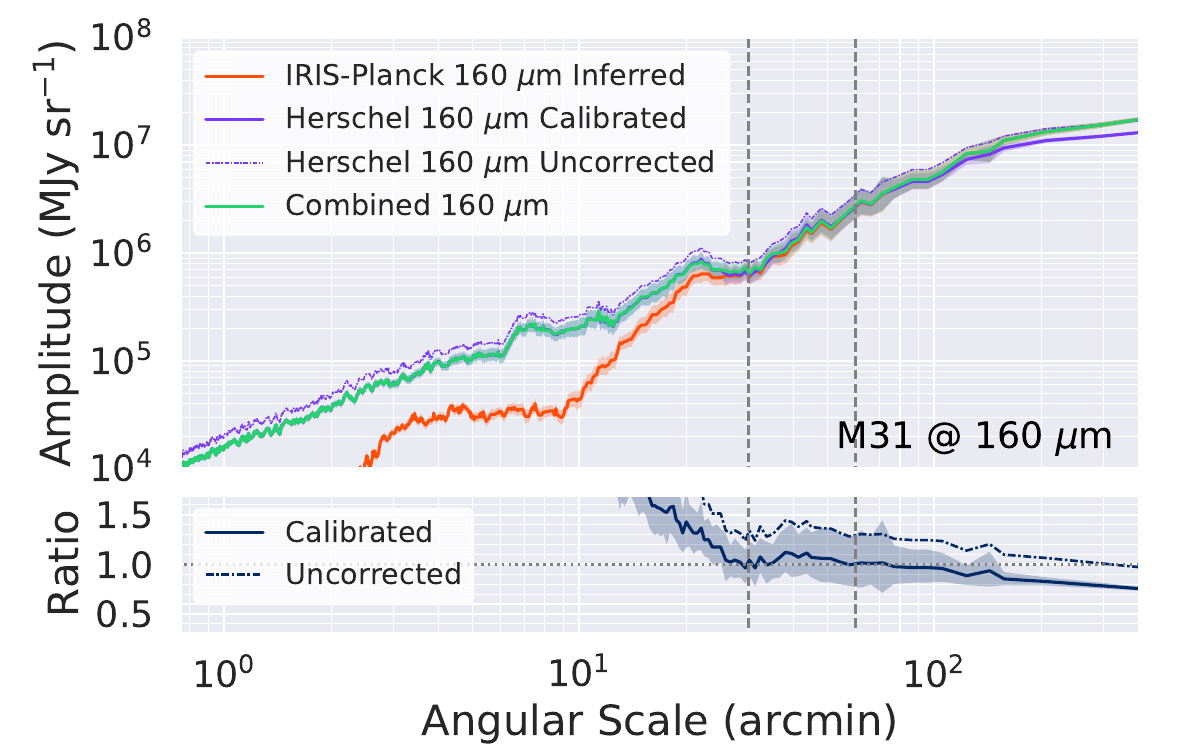}
\includegraphics[width=0.475\textwidth]{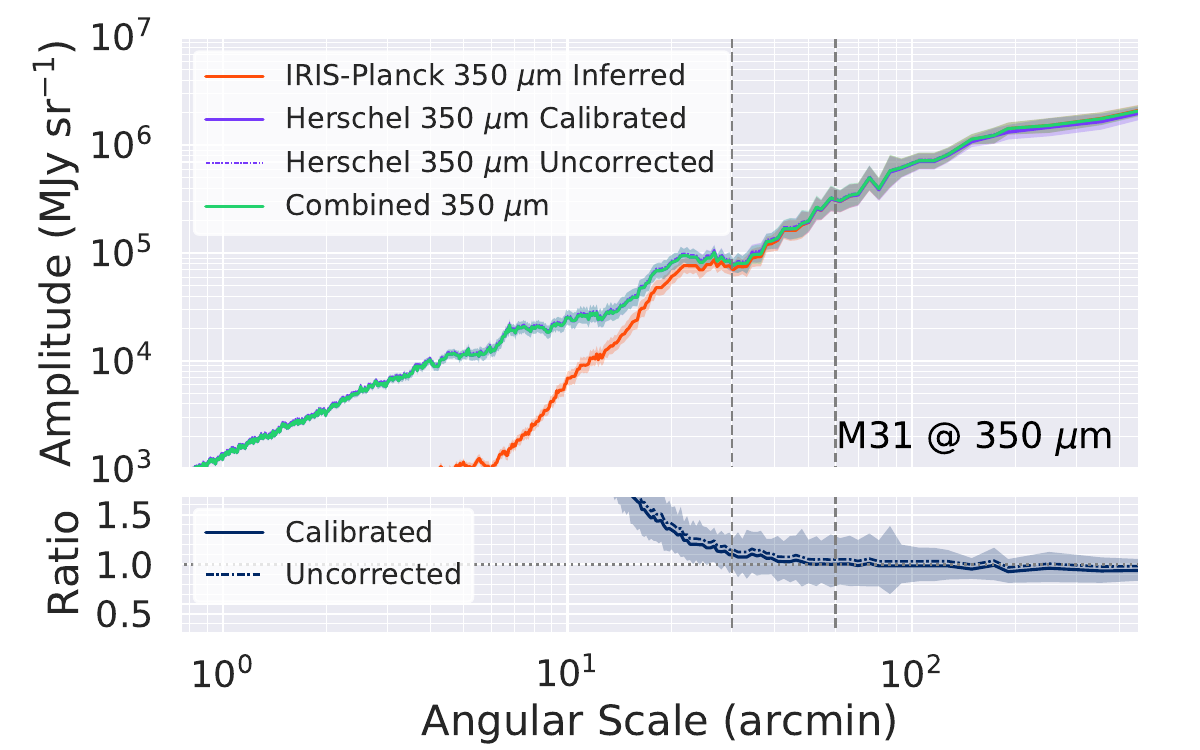}
\includegraphics[width=0.475\textwidth]{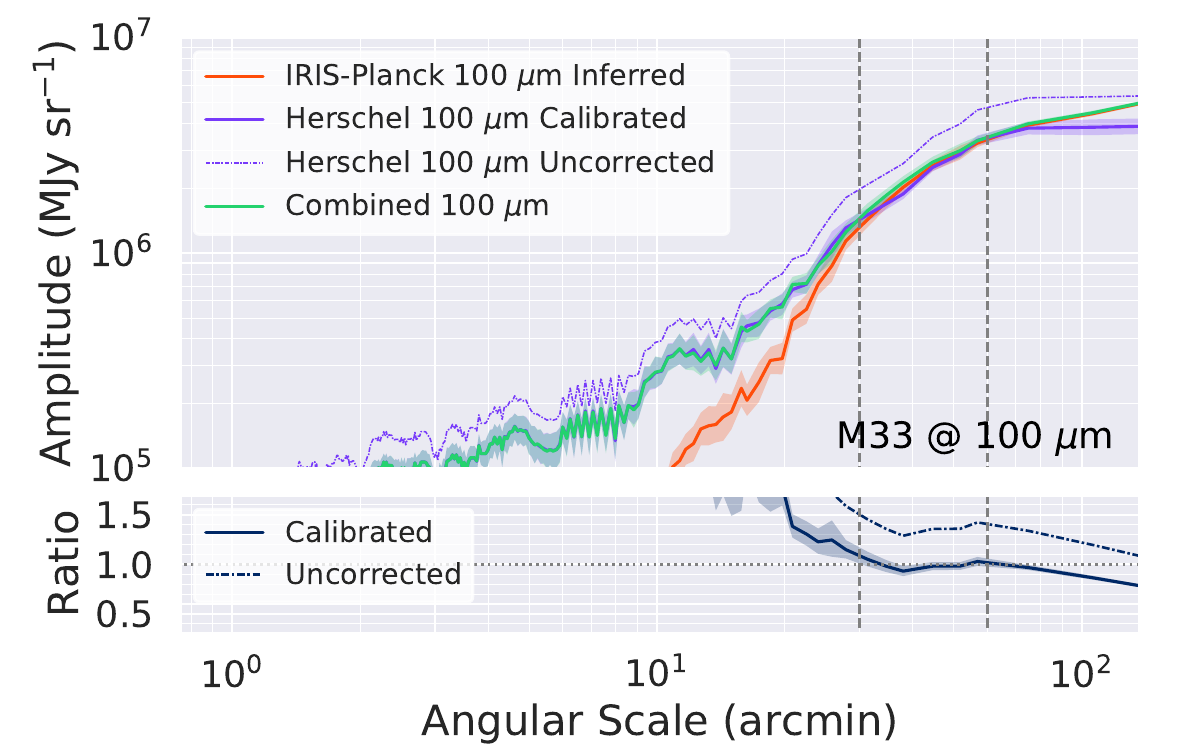}
\caption{Example power spectra. The shaded region around each line shows the uncertainty on the average spectrum amplitude at each scale. Vertical dashed lines demark the 30\arcmin--60\arcmin\ tapering and calibration window (45\arcmin--90\arcmin\ for the LMC). In most bands the power in the \hersc\ data falls significantly beneath that in the IRIS--\planck\ data are large angular scales. Both the cross-calibrated and uncorrected \hersc\ power spectra are shown, to illustrate the offset between the two. The panel beneath each plot shows the ratio of the high-resolution \hersc\ amplitudes to the low-resolution IRIS--\planck\ amplitudes; the horizontal dotted line indicates a ratio of unity.}
\label{Fig:IRIS_Planck_Herschel_Power_Spec}
\end{figure*}

\begin{figure}
\centering
\includegraphics[width=0.475\textwidth]{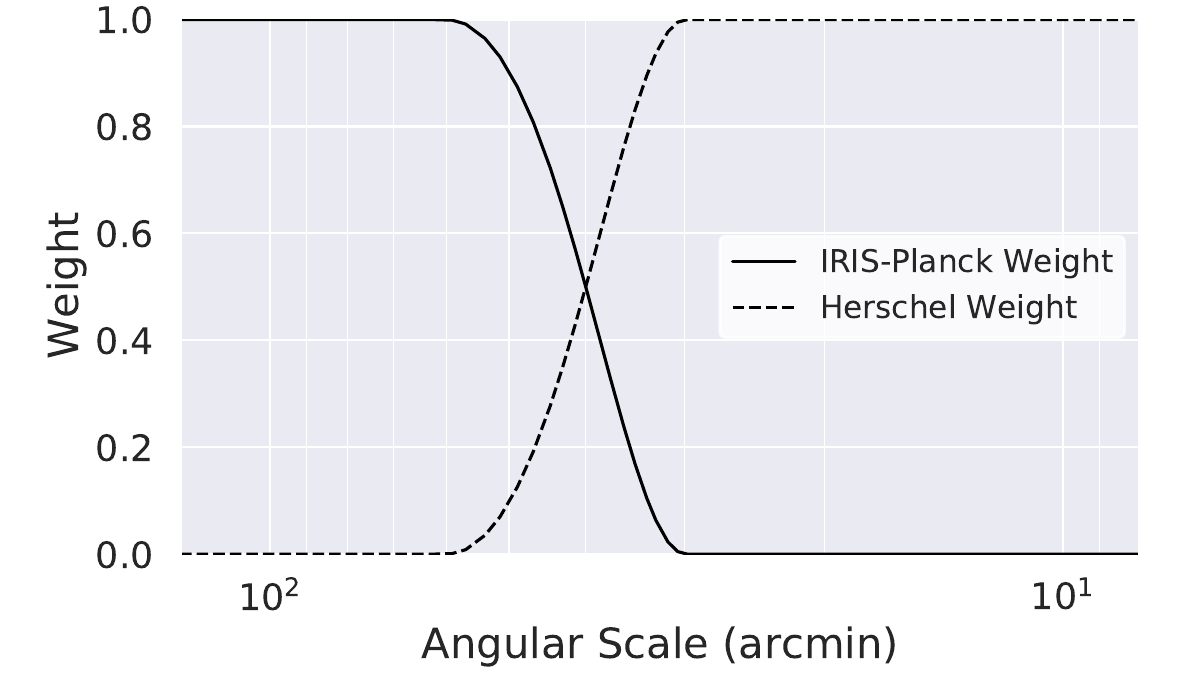}
\caption{The weighting applied at different angular scales (ie, in Fourier space) when feathering together \hersc\ and IRIS--\planck\ data. Below 30\arcmin, the \hersc\ data has a weight of 1, and the IRIS--\planck\ data has a weight of 0; above 60\arcmin, these are reversed. Within the 30\arcmin--60\arcmin\ tapering window, the weights smoothly swap, following a half cosine bell function.}
\label{Fig:Planck-IRIS-Herschel_Weighting}
\end{figure}

The 4\th\ column of Figure~\ref{Fig:IRIS-DIRBE_Feathering_Output} shows the relative residuals between the DIRBE map and the feathered map; for this comparison, we convolved the feathered map to the DIRBE resolution. Ideally this residual should be 0\% everywhere, as the feathered maps should conform to the DIRBE data at scales $\geq$ the DIRBE beam. There are slight ringing artefacts visible around the LMC and SMC, but at worst they are $\pm$\,3\% (for the LMC). This is much lower than either instruments' calibration uncertainty.

The very small residuals between the feathered and IRIS data also compare very favourably to the 5\th\ column of Figure~\ref{Fig:IRIS-DIRBE_Feathering_Output}. This column shows the relative residuals between the feathered data and the unfeathered IRIS data; this illustrates the magnitude of the correction applied to the IRIS data by the feathering process. As before, this comparison is plotted at the resolution of DIRBE. The feathered data has significantly more surface brightness over the LMC and SMC, versus in the unfeathered IRIS data, correcting the response errors. The change in surface brightness is up to 10\% for the LMC, and up to 5\% for the SMC. M\,31 and M\,33 are not bright and/or extended enough to have suffered from the IRIS response issue, so do not display these changes in the feathered maps. The data for all four galaxies also shows a reduction in surface brightness over the background surrounding all galaxies, of $\sim$\,8\% in all cases; this is due to an apparent difference in the zero-level of the DIRBE and IRIS data. 

The peak of 10\% extra flux in bright regions, super-imposed over an $\sim$\,8\% reduction in background level, corresponds to an up-to 18\% correction to the background-subtracted surface-brightness for the Magellanic Clouds. This correction is substantial, and clearly necessary to obtain good science analysis with the data; it is also in line with the $\approx$\,20\% IRIS response errors measured in other work (Bot et al., {\it in prep.}).

\needspace{3\baselineskip} \subsection{Feathering Together Herschel and IRIS--Planck} \label{Subsection:Feathering_IRIS-Planck_Herschel}

Here, we feather together our \hersc\ reductions with our inferred maps of large-scale emission in \hersc\ bands (produced using IRIS and \planck\ data in Section~\ref{Subsection:DustBFF_IRIS-Planck}). We carried out this process for the 100, 160, 250, 350, and 500\,\micron\ data. 

We first generated power spectra for of the IRIS--\planck\ and \hersc\ data for each of our galaxies and bands, some examples of which are shown in Figure~\ref{Fig:IRIS_Planck_Herschel_Power_Spec}); these plots also show the ratio between the amplitudes of the two power spectra. At the largest angular scales, the \hersc\ power spectrum tends to fall beneath the IRIS-\planck\ power spectrum, by as much as 50\% (eg, see SMC 500\.\micron\ panel of Figure~\ref{Fig:IRIS_Planck_Herschel_Power_Spec}). 

We note that in some instances no power on large scales appears to be missing, such as at 500\,\micron\ for M\,31 or M\,33; this is potentially due to the smaller angular size of these galaxies, care taken during the reduction process, and the reduced filtering suffered needed at 500\,\micron\ thanks to its larger beam. However, it should be remembered that even if the power spectra for \hersc\ shows no deficit in power compared to IRIS--\planck, that does not preclude the possibility of the emission being distributed differently. For instance, as described in Section~\ref{Subsubsection:Validating_Feathering_IRIS-Planck_Herschel}, the LMC, SMC, and M\,31 data all show gradients in the foreground cirrus emission in the IRIS--\planck\ data that is considerably attenuated in the \hersc\ data; the presence (or not) of these gradients significantly effect the emission measured in the outskirts of our target galaxies after background subtraction (see Section~\ref{Section:Foreground_Subtraction}). We therefore apply the full feathering process to all bands, even when no difference appears in the power spectra, to deal with these sorts of problems, and in order to treat all of our data consistently throughout. 

Inspecting the power spectra in Figure~\ref{Fig:IRIS_Planck_Herschel_Power_Spec} indicates that there is a reasonably wide range of spatial scales over which both sets of data are sensitive to emission. We therefore used a feathering technique that could take advantage of this. Within this window of overlapping spatial scales, a smooth tapering function was used to mediate the transition in weights, from using the low- to high-resolution data. At spatial scales larger than the taper window, only the low-resolution data was used; and at scales smaller, only the high-resolution data. The overlap in spatial sensitivity between both sets of data also made it possible to cross-calibrate them, to ensure that the low- and high-resolution data were both on the same brightness scale before being combined. 

To find an appropriate tapering window, we inspected the power spectra, to identify a specific spatial scale interval for which the ratio between the high- and low-resolution amplitudes remained relatively constant. This constant ratio indicates that the two sets of data are maintaining a constant spatial sensitivity, except for a difference in flux calibration (if the ratio is 1, then the two datasets already have the same flux calibration). It was necessary to balance the benefits of a larger tapering window (less ringing in the feathered outputs; more values with which to perform brightness scale cross-calibration), versus the benefits of a narrower tapering window (less risk of extending sampling to spatial scales where either of the datasets starts to lose sensitivity). 

After iterating through a range of options and inspecting the outputs, particularly the residual plots (see Section~\ref{Subsubsection:Validating_Feathering_IRIS-Planck_Herschel} and Figure~\ref{Fig:Herschel+Planck+IRAS+COBE_Feathering_Output_250}), we found that a tapering window of 30\arcmin--60\arcmin\ worked best (except for the LMC; see below); this range is marked in Figure~\ref{Fig:IRIS_Planck_Herschel_Power_Spec}. Specifically, within this window of overlapping sensitivity, we used a half cosine bell function (ie, half of a Hann window function; \citealp{Harris1978A}) to smoothly mediate the transition in weights applied to the datasets. This transition is illustrated in Figure~\ref{Fig:Planck-IRIS-Herschel_Weighting}. 

Even within the tapering window, the ratio between the high- and low-resolution amplitudes does not remain perfectly constant. To check if there were systematic variations in the ratio within each tapering window, we fit straight lines to the amplitudes inside them; ideally the gradient of such a line should be zero. We performed a Monte-Carlo bootstrap resampling to evaluate the gradient uncertainty. For 17 of the 20 power spectra (ie, for each band for each galaxy), the gradient of the ratio within the tapering window was compatible with being 0, to within the uncertainty. For the remaining 3, the deviations were 1.03$\sigma$, 1.20$\sigma$, and 2.15$\sigma$. Given that 1 out of 20 {\it should} deviate by \textgreater\,2$\sigma$ on average for Gaussian statistics, we are satisfied that this indicates the amplitude ratios within the tapering windows are well-behaved.

Based upon a wide suite of feathering simulations, \citet{Kurono2009A} find that the high- and low-resolution observations should have spatial scale overlap spanning at least a factor of 1.7, in order to achieve outputs with minimal errors (above this overlap factor, they find minimal, asymptotic improvements). Given that our 30\arcmin--60\arcmin\ tapering window spans a factor of 2, we comfortably satisfy this criterion.

\begin{table*}
\centering
\caption{Cross-calibration correction factors applied to \hersc\ data for each band and galaxy, after comparison with power spectra of the corresponding IRIS--\planck\ data (within the tapering window), in order to place both on the same brightness scale prior to feathering together.}
\label{Table:Brightness_Corrections}
\begin{tabular}{lrrrr}
\toprule \toprule
\multicolumn{1}{c}{} &
\multicolumn{1}{c}{LMC} &
\multicolumn{1}{c}{SMC} &
\multicolumn{1}{c}{M\,31} &
\multicolumn{1}{c}{M\,33} \\
\cmidrule(lr){2-5}
100\,\micron & 0.688 $\pm$ 0.006 & 0.705 $\pm$ 0.007 & 0.718 $\pm$ 0.014 & 0.723 $\pm$ 0.015 \\
160\,\micron & 0.797 $\pm$ 0.004 & 0.860 $\pm$ 0.006 & 0.777 $\pm$ 0.008 & 0.717 $\pm$ 0.010 \\
250\,\micron & 0.895 $\pm$ 0.002 & 0.962 $\pm$ 0.009 & 0.868 $\pm$ 0.011 & 0.850 $\pm$ 0.014 \\
350\,\micron & 0.943 $\pm$ 0.002 & 0.980 $\pm$ 0.006 & 0.955 $\pm$ 0.009 & 0.903  $\pm$ 0.024 \\
500\,\micron & 0.967 $\pm$ 0.003 & 0.995 $\pm$ 0.006 & 0.996 $\pm$ 0.011 & 0.928  $\pm$ 0.016 \\
\bottomrule
\end{tabular}
\end{table*}

When feathering the PACS data for the LMC, we found that shifting the window function to larger angular scales was necessary, due to an apparent non-linearity in the response function of our IRIS--\planck\ data around the extremely high-surface-brightness star-forming complex of 30 Doradus. This response non-linearity caused the IRIS--\planck\ data to over-estimate the surface brightness at the centre of 30 Dor, and under-estimate it in the surrounding area (similar response issues around 30 Dor were discussed in \citealp{Meixner2013A}). These errors appear at scales of up to $\sim$40\arcmin. As we describe in Appendix~\ref{AppendixSection:IRIS_Around_30-Dor}, it appears that the PACS data is correctly recovering the emission around 30 Dor at these scales. By putting our tapering window at a larger scale, we ensure the emission from 30 Dor is correctly reproduced in the final feathered maps. We experimented with a range of tapering windows, and found that 45\arcmin--90\arcmin\ stops emission being lost around 30 Dor, whilst still allowing the IRIS--\planck\ data to correct emission at larger scales. This maintains a tapering window spanning a factor of 2 in angular scale, so will satisfy the \citet{Kurono2009A} factor \textgreater1.7 overlap criterion just as well as for the other galaxies.

To cross-calibrate the two datasets to the same brightness scale (ie, ensure the gain for both agrees), we divided the low-resolution amplitudes by the high-resolution amplitudes within the overlap window, then took the median of these ratios. We multiplied the high-resolution amplitudes (over all scales) by this ratio to fix them to the brightness scale of the low-resolution data (which has absolute calibration ultimately pegged by DIRBE and \planck). We estimated uncertainties on the cross-calibration factors by performing 100 bootstrap resamplings (with replacement) of all the amplitudes in the overlap window, and re-computing the correction factor each time; the standard deviation of these values was taken as the uncertainty. 

The cross-calibration factors for each band and galaxy are listed in Table~\ref{Table:Brightness_Corrections}. At 500\,\micron, relatively little correction is necessary, with all factors being \textgreater\,0.928. However, going to shorter wavelengths, especially into the PACS bands, larger corrections are necessary; the average correction at 100\,\micron\ is 0.706. The fact that PACS measures emission as being 20--30\% brighter than IRIS (and \planck) is all the more striking given that we have shown that PACS can miss a considerable amount of emission at large scales. This means that PACS must be overestimating the brightness it {\it does} detect by even {\it more} than 20--30\%. When comparing PACS and IRIS--\planck\ at a common resolution {\it without} applying the cross-calibration corrections, we found that regions of compact sources were indeed much brighter in PACS, over 40\% so in some cases. 

PACS is flux calibrated by reference to a set of 5 standard stars \citep{Balog2014B}; the fact this calibration is based off point sources may explain why the calibration appears to struggle at the highly extended scales we are working with. Unlike SPIRE, PACS makes no corrections to the relative gains of the bolometers to account for the different illumination by the beam when making maps of extended emission. This may contribute to the need for a large gain correction. We also note that the two largest correction factors are for the LMC and SMC 100\,\micron\ maps; it is therefore possible that the issue is partly due to the \texttt{UNIMAP} pipeline, perhaps in how it handles the unusual scan strategy for the LMC and SMC observations.

It is conceivable that this calibration discrepancy is not present at scales smaller than the 6\arcmin\ low-resolution beams, beneath which we cannot compare the \hersc\ data to the IRIS--\planck\ data. It is therefore possible that point source and compact sources {\it are} accurately represented in the unfeathered data. However, this would require a very abrupt shift from well-calibrated to poorly-calibrated as angular scale increased. If this is the case, then our cross-calibrated maps would make point sources too faint, whilst accurately representing extended emission (compared to the original maps making diffuse emission too bright, but accurately representing point sources). 

Previous authors have compared \hersc\ data to all-sky survey data, such as from IRAS and \planck, to check for differences in gain. For instance, \citet{Molinari2016B} compared \hersc\ and IRIS data for observations of the Galactic plane (which therefore contain highly extended emission), and didn't find evidence for gain discrepancies \textgreater\,10\%. However, they made this comparison pixel-by-pixel, not in Fourier space, which could give undue weight to compact features. In contrast, \citet{Abreu-Vicente2017A} {\it do} make such a comparison in Fourier space, also for Galactic plane fields -- including data from \citealp{Molinari2016B} -- and find average cross-calibration factors of 0.84, 0.88, 0.91, and 0.97 necessary at 160, 250, 350, and 500\,\micron\, respectively (measured between 7\arcmin--100\arcmin). These factors all lie within the range of values we find for each of these bands.

Regardless, the feathering process requires us to cross-calibrate the two datasets being combined to the same brightness scale, otherwise we find that the differences lead to very pronounced artefacts -- and over the scales we {\it are} able to probe, the \hersc\ response (especially for PACS) is too bright compared to IRIS--\planck\ (and therefore DIRBE also). Given that the science in this paper, and the subsequent papers in this series (Clark et al., {\it in prep.}), is concerned primarily with the diffuse ISM, our priority is that this is correctly calibrated -- which is what our cross-calibration ensures.

We also note that the cross-calibration factors determined are quite robust against the specific choice of tapering window. As can be seen from inspecting Figure~\ref{Fig:IRIS_Planck_Herschel_Power_Spec} (especially the plots for PACS bands), the offsets between the uncorrected \hersc\ amplitudes and the IRIS-\planck\ amplitudes stay remarkably constant, over all scales large enough for IRIS-\planck\ to be sensitive.

\begin{figure*}
\centering
\includegraphics[width=0.975\textwidth]{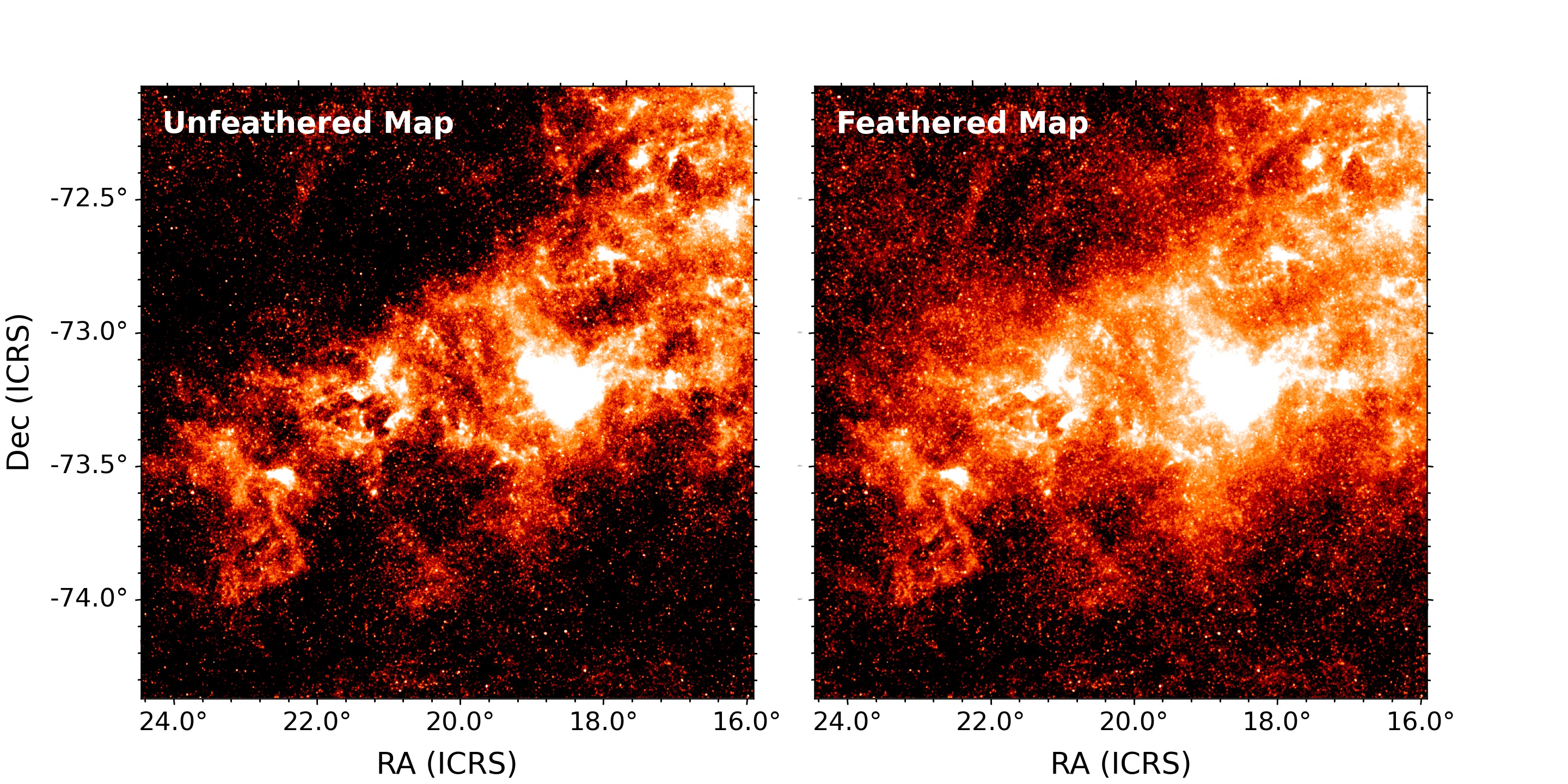}
\caption{The southern portion of the SMC, centered on the SMC wing, as seen at 250\,\micron\ in both the unfeathered map ({\it left}) and our feathered map ({\it right}). Note that these images are on the same colour scale, and have both had foreground Milky Way emission subtracted, as per Section~\ref{Section:Foreground_Subtraction}.}
\label{Fig:SMC_Wing_Before_After}
\end{figure*}

Before feathering, we reprojected the low-resolution data to pixel grid of the \hersc\ data in each band, apodising it to remove pixel-edge artefacts (as per Appendix~\ref{AppendixSection:Apodisation}). For each band, both sets of data, and their corresponding beams, were Fourier transformed. The low-resolution data was deconvolved with its beam, then re-convolved with the band's corresponding \hersc\ beam. Having done this, we combined the two sets of data in Fourier space, using our tapering function to govern the transition within the tapering window. 

As with IRIS+DIRBE, we performed the feathering 500 times for each galaxy and band; once for each sample of the SED-fitting posterior in each pixel. For each Monte Carlo iteration, we drew surface brightness values for each \hersc\ pixel according to the uncertainty map in each band, assuming Gaussian uncertainty. Recall that the uncertainties on the IRIS--\planck\ SED-fitting factored in the uncertainties on the IRIS+DIRBE feathering, which in turn factored in the uncertainties on the DIRBE SED fitting (and the IRIS photometric uncertainties). So the 500 bootstrap samples produced for each feathered map encompass the propagated uncertainty from all previous stages of our process.

The abrupt edge of the \hersc\ maps result in considerable edge effects at the borders of the feathered maps. We therefore constructed a mask to describe the portion of each feathered map far enough from the edge to not be affected. Specifically, we found that excluding a border of three times the low-resolution beam (ie, 14.5\arcmin) provided sufficient buffer. We exclude this region from our various diagnostic plots assessing the feathered data.

\needspace{3\baselineskip} \subsubsection{Validation of Feathering Herschel with IRIS--Planck} \label{Subsubsection:Validating_Feathering_IRIS-Planck_Herschel}

\begin{figure*}
\centering
\includegraphics[width=0.975\textwidth]{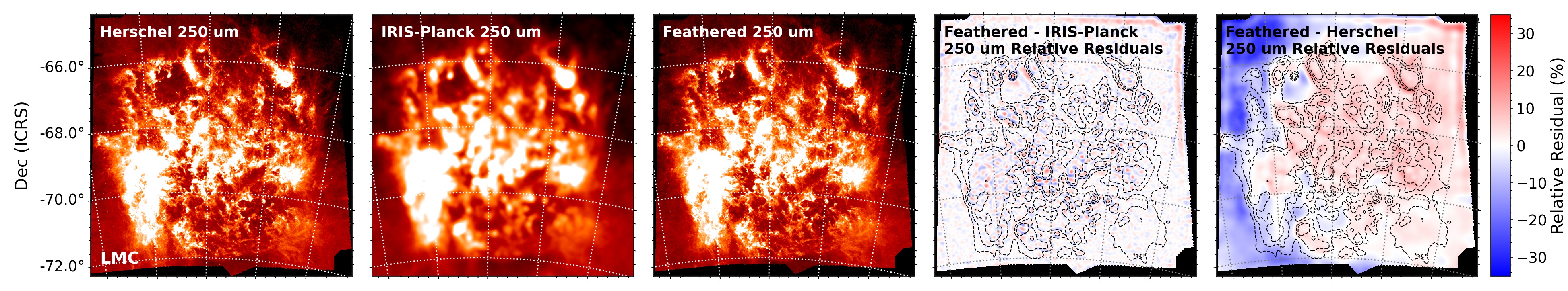}
\includegraphics[width=0.975\textwidth]{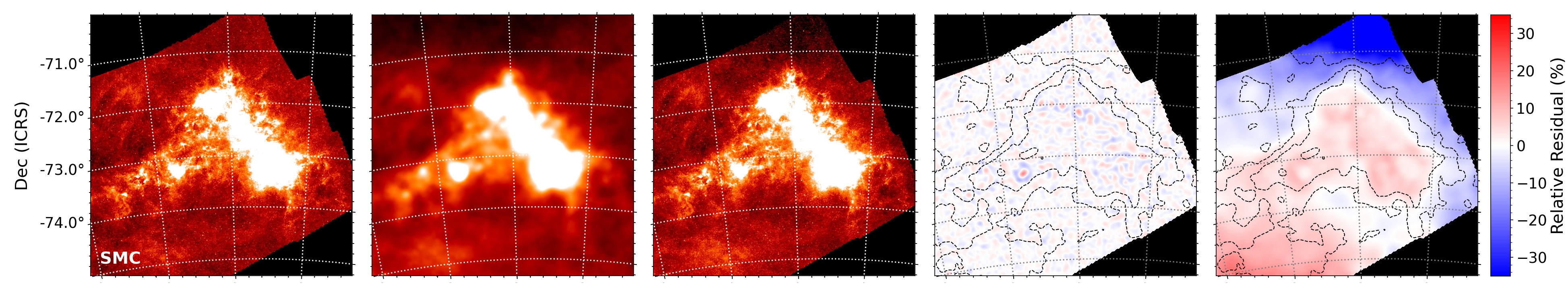}
\includegraphics[width=0.975\textwidth]{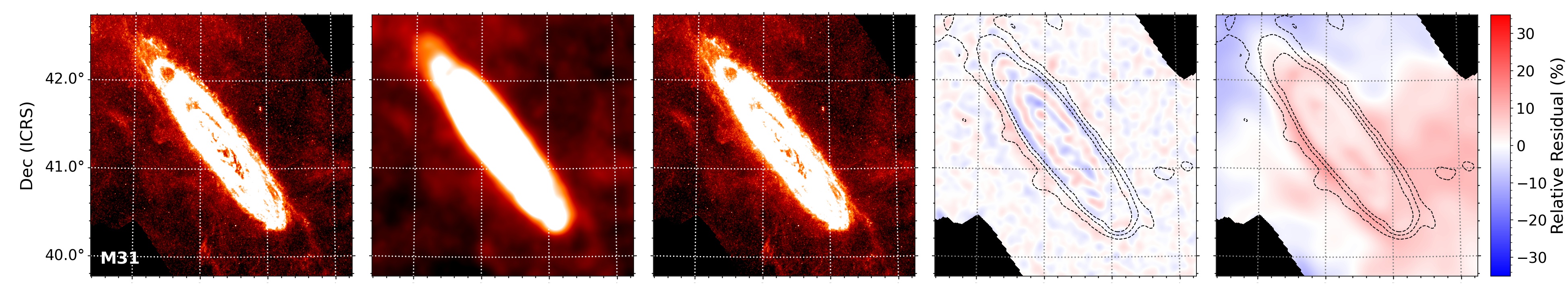}
\includegraphics[width=0.975\textwidth]{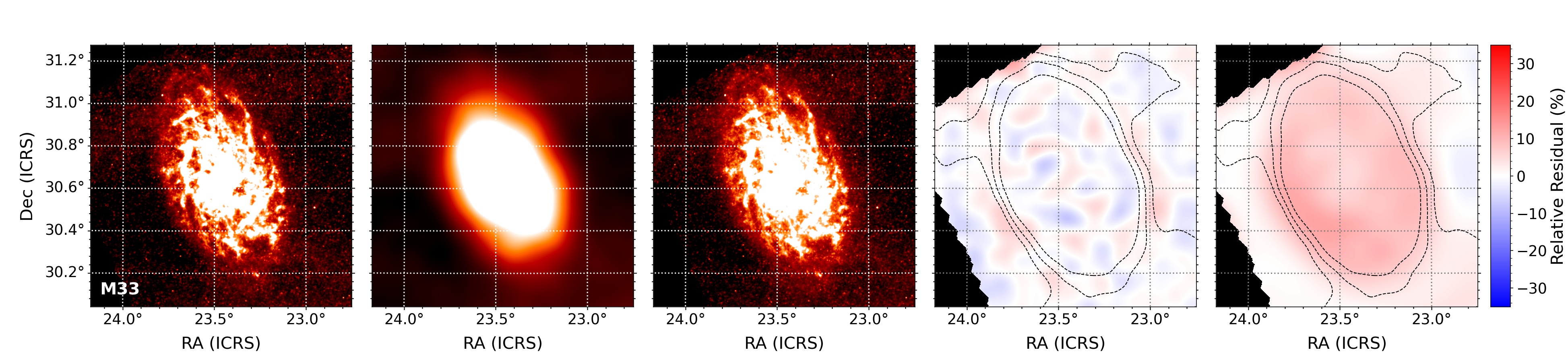}
\caption{The results of feathering together \hersc\ with IRIS--\planck\ data at 250\,\micron\ for each of our galaxies, along with diagnostic plots. {\it 1\st\ Panel:} The un-corrected \hersc\ data. {\it 2\nd\ Panel:} The IRIS--\planck\ data. {\it 3\rd\ Panel:} The combined \hersc\ + IRIS--\planck\ data after feathering (all three surface brightness maps for each galaxy use the same colour scale). {\it 4\th\ Panel:} The relative residuals between the feathered data and the input IRIS--\planck\ data, compared at the IRIS--\planck\ resolution; positive residuals mean the feathered data has greater surface brightness than the DIRBE data, negative mean less. Ideally, this should be zero everywhere. {\it 5\th\ Panel:} The relative residuals between the IRIS--\planck\ data and the unfeathered \hersc\ data (to which the cross-calibration factors from Table~\ref{Table:Brightness_Corrections} have been applied); positive residuals indicate where the IRIS--\planck\ surface brightness was increased by feathering, negative where it was decreased; note that the zero-level for this is somewhat arbitrary\textsuperscript{\hyperlink{FootnoteLink:Rel_Resid_Zero_Footnote}{\getrefnumber{Footnote:Rel_Resid_Zero_Footnote}}}. The contours in the 4\th\ and 5\th\ panels show SNR of 2, 5, and 10, from the IRIS--\planck\ map, using simple pixel noise in calculated in background regions.}
\label{Fig:Herschel+Planck+IRAS+COBE_Feathering_Output_250}
\end{figure*}

The results of feathering together the \hersc\ and IRIS--\planck\ data are shown in Figure~\ref{Fig:Herschel+Planck+IRAS+COBE_Feathering_Output_250}, for the example case of the 250\,\micron\ data for each; corresponding figures for 100, 160, 350, and 500\,\micron\ are shown in Appendinx~\ref{AppendixSection:Feathering_Outputs}. A zoomed-in view of the SMC wing, contrasting the unfeathered and feathered maps, is shown in Figure~\ref{Fig:SMC_Wing_Before_After}.

The 4\th\ column of these figures show the relative residuals between the IRIS--\planck\ map and the feathered map (with the feathered map convolved to the low-resolution beam for comparison); Ideally this residual should be 0\% everywhere. The 5\th\ columns show the relative residuals between the feathered data and the unfeathered \hersc\ data, illustrating the degree to which  the unfeathered was corrected by feathering; note that the zero level of these plots are somewhat arbitrary\footnote{\label{Footnote:Rel_Resid_Zero_Footnote}\hypertarget{FootnoteLink:Rel_Resid_Zero_Footnote}{The zero levels for the plots of relative residuals, between the IRIS--\planck\ data and the unfeathered \hersc\ data, are relative and somewhat arbitrary. This is because the \hersc\ data has no absolute calibration. We have attempted to place the unfeathered \hersc\ maps on a zero level that matches that of the feathered data, by comparing the peaks of their pixel value distributions. But because the surface brightness distributions of the two sets of data are, different due to their different angular sensitivities, this is only approximate.}}. If the correction in the 5\th\ column is larger than the residual in the 4\th\ column, then the feathering process has done `more good than harm'.

In Figures~\ref{AppendixFig:Herschel+Planck+IRAS+COBE_Feathering_Output_100_160} and \ref{AppendixFig:Herschel+Planck+IRAS+COBE_Feathering_Output_350_500}, in Appendix~\ref{AppendixSection:Feathering_Outputs}, we can see that the average residual in the 4\th\ column is zero for every galaxy, as we would wish. There is some ringing apparent surrounding bright features, such a star-forming complex in the SMC wing, and the rings of M\,31; these residuals peak in the $\pm$5--8\% range (the background regions where there is no ringing have residuals $\leq$1\%). In contrast, the flux corrections in the 5\th\ column show that surface brightness was increased by 10--13\%  over most of the outskirts of the target galaxies. The distribution of emission restored by feathering is exactly as we would expect -- regions of bright compact emission show little-to-no correction, whereas in the diffuse outer regions there has been significant addition. We can also see that feathering has corrected the flat backgrounds that \hersc\ reduction tends to impose, restoring the gradients due to foreground cirrus; this is especially clear to the north-west and south-east of the SMC.

These diagnostic plots show some noteworthy differences in the feathering results in different bands. The ringing in the residuals in the 4\th\ column are less pronounced in the other SPIRE bands, at 350 and 500\,\micron, (Figure~\ref{AppendixFig:Herschel+Planck+IRAS+COBE_Feathering_Output_350_500}), falling to $\pm$3.5\% at 500\,\micron. The flux corrections in the 5\th\ column also become smaller towards longer wavelengths. By 500\,\micron, little-to-no-correction is being made to M\,31 and M\,33, with only slight changes to the foreground gradient apparent. This is borne out by the power spectra, which show no significant power missing in the \hersc\ data, compared to the IRIS--\planck\ data at 500\,\micron\ for either galaxy (eg, see lower-left panel of Figure~\ref{Fig:IRIS_Planck_Herschel_Power_Spec}).

The feathering diagnostic plots for the PACS data at 100 and 160\,\micron\ (Figure~\ref{AppendixFig:Herschel+Planck+IRAS+COBE_Feathering_Output_100_160}) show even more dramatic corrections to the surface brightness, as can be clearly seen in the 5\th\ columns. The restored surface brightness represents a \textgreater 30\% increase over large areas around all of the galaxies\footnote{Note that this is separate from the cross-calibration corrections, as listed in Table~\ref{Table:Brightness_Corrections}, applied prior to feathering.}, for both PACS bands, once again in line with expectations from the power spectra in these bands (see Figure~\ref{Fig:IRIS_Planck_Herschel_Power_Spec}). Unfortunately, the noise in the residuals between the feathered data and IRIS--\planck\ data in the 4\th\ columns can be considerable. The ringing around M\,31 reaches 17\% level for 160\,\micron, and  20\% for 100\,\micron. Nonetheless, this is still much smaller than the correction applied by feathering, so the new maps are still more correct than the unfeathered maps, even with this residual noise. The only exception to this is the region of sky to the south-east of M\,33, where there is a large $\approx$40\% positive residual in the 4\th\ column, whilst the correction in the 5\th\ column is only reaches $\approx$30\%. In general, large residuals in the 4\th\ column are limited to the regions of empty background; within the galaxies (as outlined by the contours), the mean absolute residuals of the 4\th\ column are $\approx$4.5\% at 100\,\micron, and $\approx$4.0\% at 160\,\micron. This compares favourably to the 7\% photometric calibration uncertainty of the PACS instrument. 


As noted in Section~\ref{Subsubsection:Validating_Feathering_DIRBE_IRIS} and Appendix~\ref{AppendixSection:IRIS_Around_30-Dor}, our IRIS--\planck\ maps do not correctly recover the emission around 30 Dor in the LMC. So the large residuals around 30 Dor in the 4\th\ column of Figure~\ref{AppendixFig:Herschel+Planck+IRAS+COBE_Feathering_Output_100_160} for PACS is expected (and indeed desirable), indicating where there is disagreement between the feathered data and the IRIS--\planck\ data, due to the PACS data correcting IRIS--\planck\ surface brightness artefacts at smaller scales.

 \begin{figure*}
\centering
\includegraphics[width=0.975\textwidth]{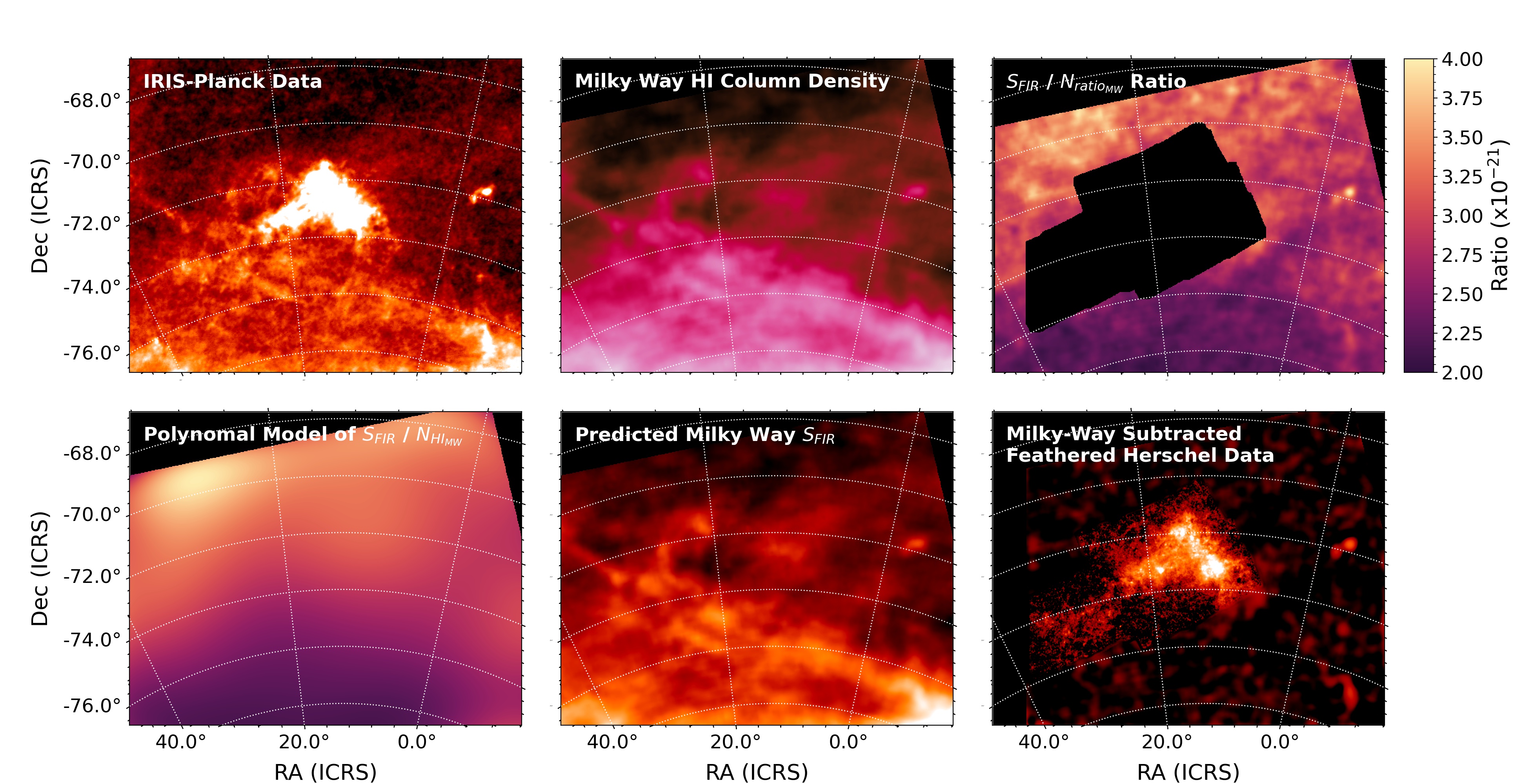}
\caption{An example of our Milky Way foreground subtraction, for the case of the SMC at 500\,\micron. {\it Upper left:} Map of 500\,\micron\ surface brightness, $S_{\it FIR}$ from our IRIS--\planck\ data. {\it Upper centre:} Column density of Galactic \HI\, $N_{\rm HI_{\it MW}}$, from HI4PI 21\,cm data in the Milky Way velocity range. {\it Upper right:} The ratio of FIR surface brightness to \HI\ column density, $S_{\it FIR} / N_{\rm HI_{\it MW}}$. Footprint of \hersc\ observations is masked so that only the Milky Way values are traced. {\it Lower left:} Polynomial model of $S_{\it FIR} / N_{\rm HI_{\it MW}}$, allowing us to estimate the value of the ratio in the masked SMC region. {\it Lower centre:} Predicted foreground FIR surface brightness, via multiplying polynomial $S_{\it FIR} / N_{\rm HI_{\it MW}}$ model with the Galactic HI column density map. {\it Lower right:} Feathered \hersc\ data with predicted Galactic foreground FIR emission subtracted. The region with high-resolution \hersc\ coverage is visible in the centre, whilst regions with only low-resolution data available can be seen surrounding it.}
\label{Fig:Foreground_Subtraction_Example}
\end{figure*}

Each of our feathered maps has an associated uncertainty map, which captures the uncertainties propagated through the entire process. In most bands, the increase in the uncertainty maps is negligible (especially in the PACS bands, where the instrumental uncertainty levels were already quite high). However, in the SPIRE bands the levels in the uncertainty maps can increase by a factor of a few, mainly because the instrumental noise levels in the original maps were very low to start with. Even with this increased uncertainty, the cirrus emission in faint sky regions away from our target galaxies is still consistently detected with SNR\,\textgreater\,10 in the SPIRE bands.
 
 \needspace{3\baselineskip} \section{Foreground Subtraction} \label{Section:Foreground_Subtraction}
 
 The feathered maps produced in Section~\ref{Section:Feathering} include not only the restored large-angular-scale emission from our target galaxies, but also highly extended emission from the foreground of Milky Way cirrus emission. We must subtract this foreground emission from the map before we can investigate the dust in our targets. 
 
 The Milky Way cirrus is highly structured, with much of this structure at angular sizes similar to our target galaxies. As a result, simple foreground subtraction with an annulus or similar will not do a good job of removing the foreground cirrus structure directly in front our targets. We therefore follow \citep{Roman-Duval2017B}, and many previous works, in using all-sky maps of Galactic \HI\ to trace the cirrus structures around, and in front of, our targets, and then subtracting them from our maps by comparison to their dust emission.
 
To do this, we used the all-sky \HI\ data of the HI4PI survey \citep{Bekhti2016B,Kalberla2020C}, which provides 21\,cm coverage of the entire Milky Way velocity range, at 0.7\degr\ resolution. We obtained the HI4PI cubes that provide coverage towards our target galaxies. We inspected the velocity profiles of each, and found that the Milky Way \HI\ emission was contained within the $-75 < v < 50\,{\rm km\,s^{-1}}$ velocity range. The exception to this is M\,31, where the side rotating towards us has $\sim$0 net velocity, making its \HI\ emission confused with that of the Milky Way; we therefore handle foreground subtraction for M\,31 differently, as described in Section~\ref{Subsection:M31_Foreground_Subtraction}. An example of the Milky Way \HI\ emission, for the case of the SMC, is shown in the upper-left panel of Figure~\ref{Fig:Foreground_Subtraction_Example}.

For the other galaxies, we compared the foreground \HI\ column density to our predicted maps of the large-scale emission in each \hersc\ band, as inferred from our IRIS--\planck\ SED fitting in Section~\ref{Subsection:DustBFF_IRIS-Planck}. These maps cover a much larger area than the footprint of the \hersc\ observations, spanning an area with a width of 3 times the ${\it D25}$. In each \hersc\ band, we first smoothed the map of predicted dust emission (with its resolution of 4.8\arcmin) to the 0.7\degr\ resolution of the \HI\ data. We then produced a map of the ratio of the band's surface brightness to the \HI\ column density, $S_{\it FIR} / N_{\rm HI_{\it MW}}$, for the region outside of the \hersc\ observation footprint; an example of this for the case of $S_{\it 500 \mu m} / N_{\rm HI_{\it MW}}$ around the SMC, is shown in the upper-right panel of Figure~\ref{Fig:Foreground_Subtraction_Example}.

The maps of $S_{\it FIR} / N_{\rm HI_{\it MW}}$ show considerable variation, up to 75\% (in line with \citealp{Bianchi2017A}) -- no doubt driven by changes in the Galactic dust-to-gas ratio, the dust temperature, etc. We could potentially use these ratio maps to simply calculate a constant ratio of FIR to \HI\ column density for each band and field, and thereby impute the cirrus FIR surface brightness in front of our targets using the Galactic \HI\ column for that area; however the large variation and structure in $S_{\it FIR} / N_{\rm HI_{\it MW}}$ means this would suffer large errors. Instead, we fit a 7\th-order 2-dimensional polynomial to the $S_{\it FIR} / N_{\rm HI_{\it MW}}$ ratio map, to impute the variation of the ratio over the region observed by \hersc, which we mask out from the polynomial fitting; this is illustrated in the lower-left panel of Figure~\ref{Fig:Foreground_Subtraction_Example}. We found that a 7\th-order polynomial has sufficient freedom to capture the bulk of the structure in the $S_{\it FIR} / N_{\rm HI_{\it MW}}$ map without producing spurious features. We then use this polynomial model of the $S_{\it FIR} / N_{\rm HI_{\it MW}}$ ratio for a given band, along with the map of the Milky Way \HI\ column over the field, to calculate the surface brightness of the foreground cirrus in that band, and subtract it from the feathered \hersc\ map; an example of this is shown in the lower-central and lower-right panels of Figure~\ref{Fig:Foreground_Subtraction_Example}.

To assess the uncertainty in a subtracted map, we found the median value of the subtracted map outside the \hersc\ footprint (ie, over the area where the polynomial was fitted). If the subtraction performed perfectly, this residual map should have a value of zero everywhere. For M\,31, because of the less precise foreground subtraction we had to employ, the residuals are larger, with root-mean-square (RMS) residuals varying between bands from 8.3--9.3\% (being larger in the PACS bands). For the other galaxies, the RMS residuals across all five bands average 6.8\% for the LMC, 4.1\% for the SMC, and 1.9\% for M\,33 -- comparable for the photometric calibration uncertainties. The RMS for each map is always larger than the average residual for that map.

Because of the 0.7\degr\ resolution of HI4PI, our method is unable to subtract foreground structures smaller than 0.7\degr. The {\it total} amount of flux associated with any such smaller structures will be subtracted, however the result will be a compact structure in the final map, surrounded by a negative bowl. An example of this can be seen in the western edge of the final panel of Figure~\ref{Fig:Foreground_Subtraction_Example}. Fortunately, the cirrus is sufficiently close to us that the vast majority of the foreground features in our maps are large enough to be subtracted well. Inspection of the \HI\ data shows that there are no bright compact features in front of, or in the immediate proximity of, our target galaxies. Moreover, the worst negative bowl present in any of the foreground-subtracted maps is the one associated with the bright compact feature several degrees east of the SMC, visible in Figure~\ref{Fig:Foreground_Subtraction_Example}, which at its worst point displays an 18\% residual; the next worst is 13\%, and all the rest are \textless\,10\%. The RMS error on the HI4PI column densities is 6\% outside of the Galactic plane, increasing to 10--15\% in places \citep{Bekhti2016B}. As such, with only a few small exceptions, the magnitiude of the negative bowls is comparable to, or less than, the inherent noise in the HI maps.

This foreground subtraction will have also removed the emission of the Cosmic Infrared Background (CIB) from our maps. Technically the CIB will have manifested as a constant offset, in addition to the $S_{\it FIR}$ emission per $N_{\rm HI_{\it MW}}$. In practice, our $S_{\it FIR} / N_{\rm HI_{\it MW}}$ will have captured this contribution as part of the $S_{\it FIR}$ term. The fact that the residuals in the background of the subtracted maps are so close to zero indicates that the CIB was effectively removed.
 
 \needspace{3\baselineskip} \subsection{Foreground Subtraction for M\,31}  \label{Subsection:M31_Foreground_Subtraction}
 
Because the velocity of the \HI\ emission from the Milky Way and M\,31 overlap, we could not use the method described above for foreground subtraction. Fortunately, M\,31 sits behind a region of Galactic cirrus that exhibits a fairly smooth gradient in surface brightness, so the more detailed subtraction method is not vital for success.
Instead, we first masked the \hersc\ observation footprint from the predicted map of emission in each \hersc\ band derived from the IRIS-\planck\ SED-fitting, the same as above. However we then performed the polynomial modeling directly on the FIR maps, providing an estimate of the foreground emission in front of M\,31 in each band, which we then subtracted from the feathered \hersc\ data. We calculated the uncertainty on this subtraction in the same manner as above.
 
 \needspace{3\baselineskip} \subsection{Foreground Subtraction for the Unfeathered Maps} \label{Subsection:Naive_Foreground_Subtraction}
 
In order to compare our new, feathered \hersc\ maps to the unfeathered original maps, we also needed to foreground subtract the original maps. In the the case of the LMC, and of the PACS data for M\,33, there is very little sky surrounding the target galaxies. Without sufficient area of empty sky, it was not possible to perform the same sort of subtraction as we did in Section~\ref{Section:Foreground_Subtraction}, using Galactic \HI\ data. 
 
 \begin{table*}
\caption{The integrated fluxes we measure for our sample galaxies in each band with our feathered maps, compared to those measured from the unfeathered maps. Note that the new fluxes incorporate the cross-calibration corrections from Table~\ref{Table:Brightness_Corrections}, whereas the unfeathered fluxes do not; this correction imposes a 14--31\% reduction in flux in the PACS bands, and a 0--16\% reduction in the SPIRE bands.}
\label{Table:Fluxes}
\centering
\begin{tabular}{lrlrrlrrlrrlr}
\toprule \toprule
\multicolumn{1}{c}{Band} &
\multicolumn{3}{c}{LMC (kJy)} &
\multicolumn{3}{c}{SMC (kJy)} &
\multicolumn{3}{c}{M\,31 (kJy)} &
\multicolumn{3}{c}{M\,33 (kJy)} \\
\multicolumn{1}{c}{(\micron)} &
\multicolumn{2}{c}{Feathered} &
\multicolumn{1}{c}{Unfeathered} &
\multicolumn{2}{c}{Feathered} &
\multicolumn{1}{c}{Unfeathered} &
\multicolumn{2}{c}{Feathered} &
\multicolumn{1}{c}{Unfeathered} &
\multicolumn{2}{c}{Feathered} &
\multicolumn{1}{c}{Unfeathered} \\
\cmidrule(lr){2-4} \cmidrule(lr){5-7} \cmidrule(lr){8-10} \cmidrule(lr){11-13}
100 & 188 & $\pm$ 19 & 196 & 16.7 & $\pm$ 1.4 & 15.7 & 3.37 & $\pm$ 40 & 3.31 & 1.13 & $\pm$ 0.08 & 1.62 \\
160 & 232 & $\pm$ 22 & 196 & 20.1 & $\pm$ 1.6 & 18.5 & 7.29 & $\pm$ 0.83 & 7.00 & 1.77 & $\pm$ 0.13 & 1.80 \\
250 & 140 & $\pm$ 12 & 142 & 12.9 & $\pm$ 0.8 & 10.7 & 6.18 & $\pm$ 0.65 & 5.46 & 1.29 & $\pm$ 0.08 & 1.31 \\
350 & 71.2 & $\pm$ 6 & 71 & 7.11 & $\pm$ 0.48 & 6.23 & 3.43 & $\pm$ 0.35 & 3.00 & 0.71 & $\pm$ 0.04 & 0.71 \\
500 & 28.5 & $\pm$ 3 & 29 & 3.24 & $\pm$ 0.23 & 3.01 & 1.40 & $\pm$ 0.14 & 1.28 & 0.30 & $\pm$ 0.02 & 0.31 \\
\bottomrule
\end{tabular}
\end{table*}
 
Nor can we use the IRIS--\planck\ FIR data, as the large-scale emission that fills the sky in that data is simply not present in the unfeathered \hersc\ data. For instance, the sky to the far south east of the SMC in the \hersc-PACS data has the same surface brightness as the sky to the far north west, due to the flattening imposed by the \hersc\ reduction -- but in the IRIS--\planck\ data, the surface brightness of these areas of sky differs by more than a factor of 2, due to variation diffuse Galactic emission. Subtracting this from the unfeathered \hersc\ maps would lead to major negative features in large portions of the map. However, the unfeathered \hersc\ maps do preserve {\it some} large scale foreground emission (especially for M\,31 and the SMC), so we cannot simply treat the background as being flat.
 
The tried-and-true method of placing an annulus around the target source to measure the sky level is also not ideal here. We are interested in diffuse emission at galaxy outskirts, and any emission in or beyond a sky annulus would therefore be subtracted. And because our targets are large (and sometimes irregular) galaxies, in relatively tight maps, any sky annulus we place would run into this problem. Instead, we mask out the central regions of each unfeathered \hersc\ map, leaving only a border along the inside edge of the \hersc\ footprint\footnote{Specifically within the same footprint region defined at the end of Section~\ref{Subsection:Feathering_IRIS-Planck_Herschel}, to exclude bright artefacts in the turnaround regions, etc.}, with width equal to 5\% the ${R_{25}}$ of the target galaxy. We then fit a 1\st-order 2-dimensional polynomial (ie, a tilted plane) to this border data, to model the foreground emission, which we then subtract. This maximizes our ability to recover diffuse emission in the outskirts of the target galaxies.

 \needspace{3\baselineskip} \subsection{The Change in Total Flux} \label{Subsection:Total_Flux_Change}

Our new maps change the photometry measured for the galaxies in our sample. The global fluxes we record are given in Table~\ref{Table:Fluxes}, which also lists the fluxes measured from the unfeathered maps, for comparison. 

For the LMC and SMC, these fluxes were measured by taking the total flux within the \hersc\ footprint of the foreground-subtracted feathered maps. We opted not to place a tighter aperture around the sources in order to not exclude any faint emission at large radii. Our foreground selection should be equally good across the entire \hersc\ footprint, so no systematic bias should be suffered from measuring the fluxes this way. Nonetheless, this will drive up the `aperture noise' associated with the resulting fluxes.

For M\,31 and M\,33, the fluxes were measured within elliptical apertures, centred on the coordinates given in Table~\ref{Table:Galaxy_Properties}. For M\,31, the aperture had $a = 126\arcmin$, $b = 46\arcmin$, $\theta = 38\degr$; for M\,33, the aperture had $a = 46\arcmin$, $b = 28\arcmin$, and $\theta = 20\degr$ (where $a$, $b$, and $\theta$ are the semi-major axis, semi-minor axis, and position angle, respectively). These apertures were designed to comfortably contain all the potential extended emission, especially as visible in the SPIRE maps\footnote{For the feathered maps, the photometric apertures extended beyond the feathered footprint in some bands (due to small PACS observing footprint, and the masked feathering edge effect areas). For these pixels, we use the low-resolution IRIS--\planck\ predictions of emission in the \hersc\ band.}. 

For the uncertainty on the recorded fluxes, we use the fractional uncertainty on the foreground subtraction, the cross-calibration correction, added in quadrature to the instrumental calibration uncertainty (7\% for PACS bands, 5.5\% for SPIRE).

For the SPIRE bands, the feathered fluxes are in close agreement with the unfeathered fluxes for the LMC and M\,33. For the SMC, the feathered fluxes are 9--21\% brighter; and for M\,31, the feathered SPIRE fluxes are 9--14\% brighter (differences decreasing towards longer wavelengths, as would be expected). 

In the PACS bands, there is relatively little change in the measured fluxes. Only the 160\,\micron\ flux for the LMC, and the 100\,\micron\ flux for M\,33, differ by larger than the uncertainty. In these cases, the feathered fluxes are actually less than the unfeathered measurements, by 18\% and 43\% respectively. We suspect this change is due to the feathered maps allowing better foreground subtraction; a filament of foreground cirrus passes east-west over M\,33, and a larger cloud of cirrus extends over the galaxy's southwest quadrant. For the other galaxies, where the change in PACS flux {\it is} within the uncertainty, the feathered fluxes are brighter in 5 out of 6 of them, for an average change of +7\%.

Note, however, that all of these comparisons are occurring after the application of the cross-calibration correction factors during the feathering process, which have therefore {\it not} been applied to the unfeathered maps. These correction factors impose a 14--31\% reduction in flux in the PACS bands, and a 0--16\% reduction in the SPIRE bands (Table~\ref{Table:Brightness_Corrections}). Therefore, had these corrections not been applied, the feathered data would have been much brighter, in almost all instances.

In short, whilst the surface brightness measured at any given point in our target galaxies will often have changed dramatically between the unfeathered and feathered maps (ie, Figures~\ref{Fig:Herschel+Planck+IRAS+COBE_Feathering_Output_250}, \ref{AppendixFig:Herschel+Planck+IRAS+COBE_Feathering_Output_100_160}, and \ref{AppendixFig:Herschel+Planck+IRAS+COBE_Feathering_Output_350_500}, the {\it total} measured flux tends to be quite similar -- thanks to the reduction in flux due to the cross-calibration correction being comparable to the increase in flux from restored emission. 

 \needspace{3\baselineskip} \subsection{Total Flux Compared to the Literature} \label{Subsection:Total_Flux_Literature}

For the LMC, our PACS fluxes agree within the uncertainties with those previously reported from IRAS \citep{Rice1988A}, COBE \citep{Israel2010B}, and \spitz\ \citep{Dale2009E}, but differ from those reported by \citet{Meixner2013A} with the HERITAGE data. At 100\,\micron, we find a flux 18\% larger than \citet{Meixner2013A}, and at 160\,\micron\ we find a flux 16\% less than \citet{Meixner2013A}. For SPIRE, however, our LMC fluxes agree closely with those of \citet{Meixner2013A} (which agrees well with COBE at 240--250\,\micron; \citealp{Israel2010B}).

For the SMC, our PACS fluxes once again agree with IRAS \citep{Rice1988A}, COBE \citep{Israel2010B}, and \spitz\ \citep{Leroy2007A,Dale2009E}, and once again differ from those reported by \citet{Meixner2013A} with the HERITAGE data; ours being 14\% brighter at 100\,\micron, and 49\% brighter at 160\,\micron. Our SPIRE fluxes for the the SMC are also larger than those reported by \citet{Meixner2013A}; by 29\%, 21\%, and 12\%, at 250, 350, and 500\,\micron\ respectively. Whereas our 250\,\micron\ flux agrees well with the published 240\,\micron\ COBE flux \citep{Israel2010B}. We note that the \citet{Meixner2013A} SMC SPIRE fluxes are, however, a close match to our unfeathered fluxes.

For the Magellanic Clouds, it appears that matching the calibration of our feathered data to well-calibrated all-sky surveys has fixed the discrepancy between photometry previously measured from \hersc, and photometry measured from other facilities (including all-sky surveys).

For M\,31, our fluxes agree with those reported by \citet{Viaene2014B} to within 10\% at 350\,\micron, and to with 5\% in all other bands; this is within our uncertainties in all cases. The agreement in the PACS bands indicates that the additional flux recovered in our maps closely matches the relative decrease in flux imposed by our cross-calibration, as suggested in Section~\ref{Subsection:Total_Flux_Change}.

For M\,33, our SPIRE fluxes agree to within 6.3\% with those reported by \citet{Hermelo2016A}; this is within the mutual uncertainties in each case. For the PACS bands, however, our fluxes are 12\% and 18\% fainter at 100 and 160\,\micron, respectively. This indicates that relatively less flux was restored by our feathering, as comapred to the cross-calibration decrease; this is not surprising, given that M\,33 is by far the most compact of our sources, and therefore will have suffered the least emission lost by filtering.
 
 \needspace{3\baselineskip} \section{Initial Results} \label{Section:Initial_Results}
 
Our new \hersc\ maps allow us to probe the FIR--submm emission from our target galaxies with a combination of accuracy and resolution that has not previously been possible. Here, we examine the surface brightness properties of the galaxies in the new maps, in comparison with the old maps, to see what differences there are -- and especially to inspect those properties out to low surface brightnesses that the old data could not probe.
 
Even with our feathered maps, the diffuse dust emission in the galaxies` peripheries is still faint, and typically has low SNR for individual pixels. Therefore, to get beneath the noise level, and study dust at densities otherwise too low to detect, we binned together the emission from pixels that have similar hydrogen column density.

\needspace{3\baselineskip} \subsection{Hydrogen Surface Density Maps} \label{Subsection:Hydrogen_Surface_Density_Maps}

We constructed $\Sigma_{\rm H}$ maps for each of our galaxies, using 21\,cm maps to trace the \HI, and CO maps to trace the ${\rm H_{2}}$. For the LMC, we used the \HI\ data of \citet{Kim2003A} and CO data of \citet{Wong2011C}. For the SMC, we used the \HI\ data of \citet{Stanimirovic1999A}, and the CO data of \citet{Mizuno2001C}. For M\,31 we used the \HI\ data of \citet{Braun2009A}, and the CO data of \citet{Nieten2006A}. For M\,33, we used the \HI\ data of \citet{Koch2018C}, CO map of \citet{Gratier2010C,Druard2014A}. We calculated molecular gas surface density, $\Sigma_{\rm H2}$ using the standard relation: 

\begin{equation}
\Sigma_{\rm H2}= \alpha_{\rm CO} I_{\rm CO(1-0)}  
\label{Equation:Sigma_H2}
\end{equation}

\noindent where $I_{\rm CO(1-0)}$ is the velocity-integrated main-beam brightness temperature of the CO(1-0) line (in ${\rm K\,km\,s^{-1}}$), $\alpha_{\rm CO}$ is the CO-to-${\rm H_{2}}$ conversion factor (in ${\rm K^{-1}\,km^{-1}\,s\,M_{\odot}\,pc^{-2}}$). 

All of the CO observations we used were of the CO(1-0) line, except for that of M\,33, which was of CO(2-1). For that data, we inferred $I_{\rm CO(1-0)}$ by applying a line ratio, $r_{2:1} = I_{\rm CO(2-1)} / I_{\rm CO(1-0)}$. In spiral galaxies, $r_{2:1}$ varies radially \citep{Casoli1991D,Leroy2009B}. We therefore use the procedure laid out in Section~3.4 of \citet{CJRClark2019B}, where $r_{2:1}$ is determined as a function of galactocentric radius as a fraction of the $R_{25}$, using measurements made by \citet{Leroy2009B} -- varying from 1.0 at the galaxy centre, to 0.55 at the $R_{25}$. We thereby computed $I_{\rm CO(1-0)}$ for each pixel in the M\,33 $I_{\rm CO(2-1)}$ map according to its position.

To compute $\Sigma_{\rm H2}$ for M\,31 and M\,33, we used the standard Milky Way value of $\alpha_{\rm CO} = 3.2\,{\rm K^{-1}\,km^{-1}\,s\,pc^{-2}}$, which tends to be applicable to high-metallicity spiral galaxies in general\footnote{For M\,33, some authors have $\alpha_{\rm CO}$ higher than the Milky Way value \citep{Bigiel2010D,Druard2014A}. However, \citet{Gratier2010C} find a value within 10\% of that of the Milky Way. Plus  \citet{Rosolowsky2003B} (who do not measure an an absolute $\alpha_{\rm CO}$, only its' apparent relative variation) find that $\alpha_{\rm CO}$ in the highest-metallicity regions of M\,33, at \textgreater\,1.4\,Z$_{\odot}$, is not systematically different from that in regions at \textless\,0.5\,Z$_{\odot}$; given that $\alpha_{\rm CO}$ at sigh high metallicities should be comparable to that of the Milky Way, it suggests this is not significantly different elsewhere in M\,33.} (see \citealp{Saintonge2011A}, \citealp{Bolatto2013B}, and references therein). For the  LMC and SMC, we used $\alpha_{\rm CO}$ values of 6.4 and 21 ${\rm K^{-1}\,km^{-1}\,s\,pc^{-2}}$ respectively \citep{Bolatto2013B}. From $\Sigma_{\rm H2}$ we then calculated ${\rm H_{2}}$ surface density

All of the \HI\ and ${\rm H_{2}}$ maps we use have resolutions in the region of 1\arcmin, being quite similar both to each other, and to our \hersc\ maps -- varying from 0.2\arcmin\ for the M\,33 ${\rm H_{2}}$ data, to 2.6\arcmin\ for the SMC ${\rm H_{2}}$ data. All of the \HI\ maps incorporate both interferometric and single-dish observations, ensuring that we won't miss any diffuse atomic structure due to lack of short-spacing data.

To make our maps of $\Sigma_{\rm H}$, we convolved the \HI\ and ${\rm H_{2}}$ maps for each galaxy to whichever resolution was worst out of the two maps, re-gridded them to the same projection, then added them together to produce the combined map of hydrogen surface density.
\begin{figure}
\centering
\includegraphics[width=0.23\textwidth]{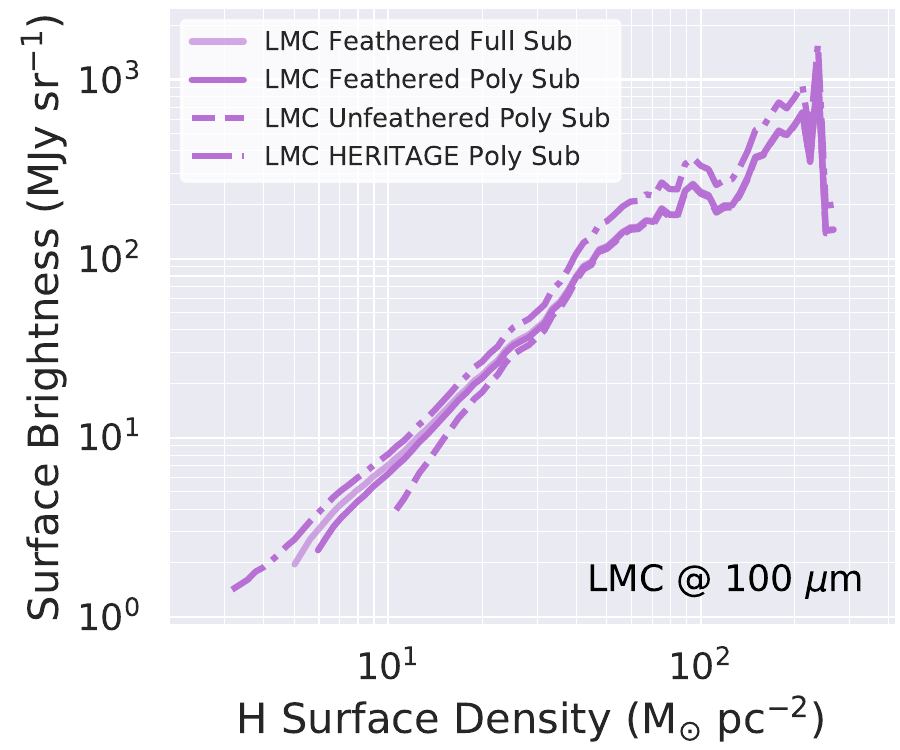}
\includegraphics[width=0.23\textwidth]{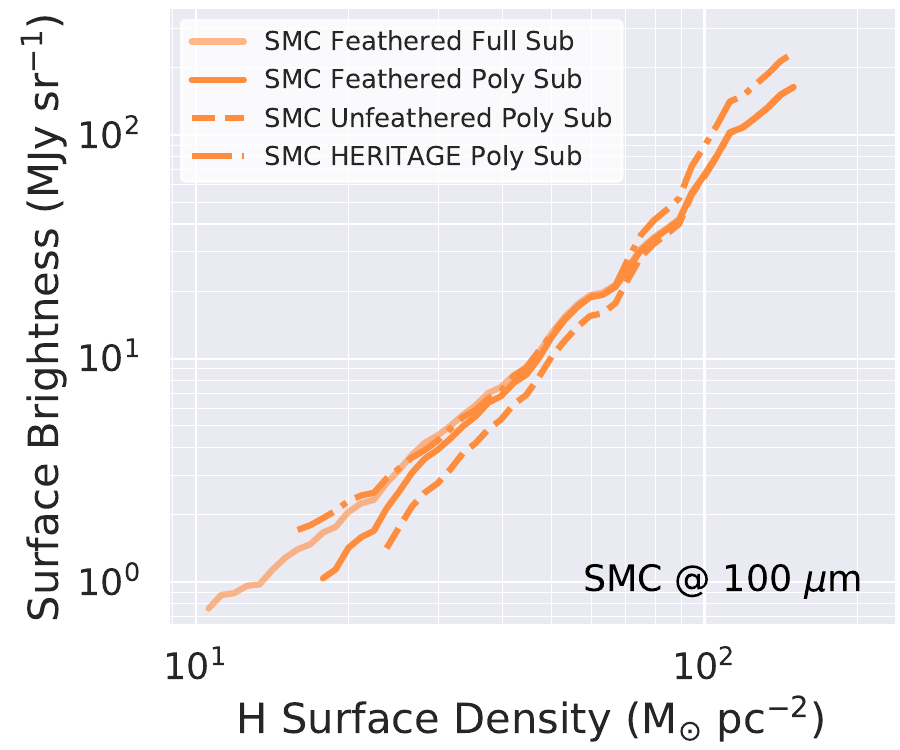}
\includegraphics[width=0.23\textwidth]{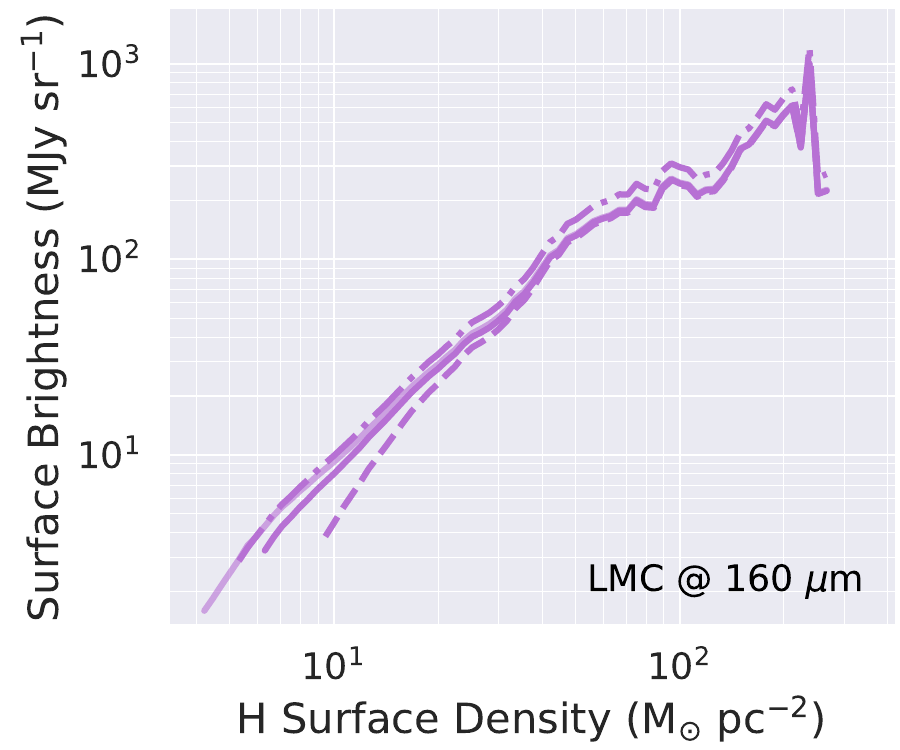}
\includegraphics[width=0.23\textwidth]{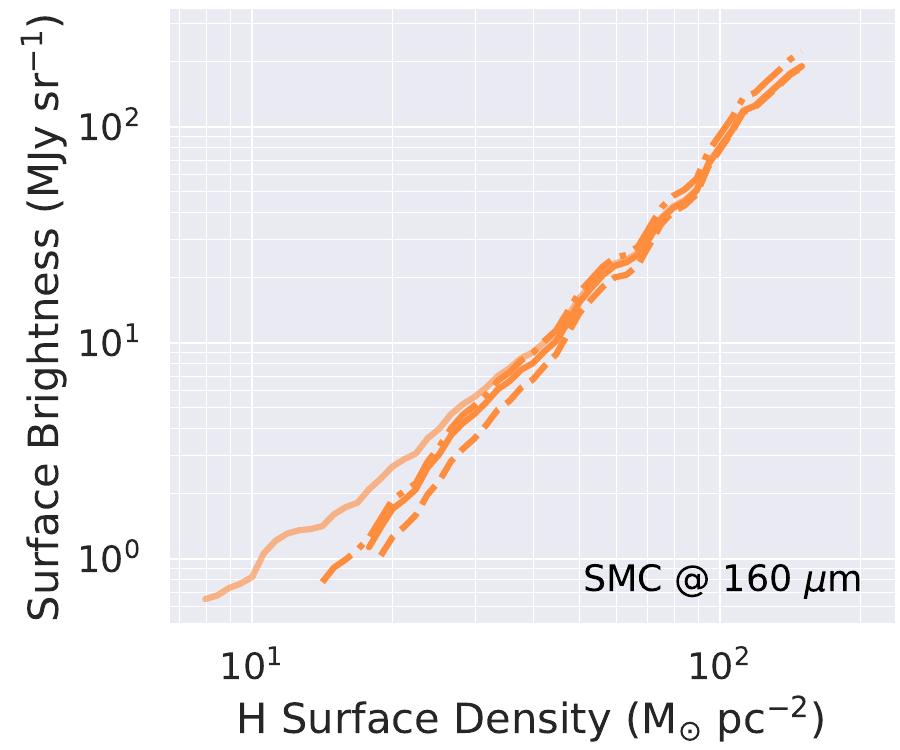}
\includegraphics[width=0.23\textwidth]{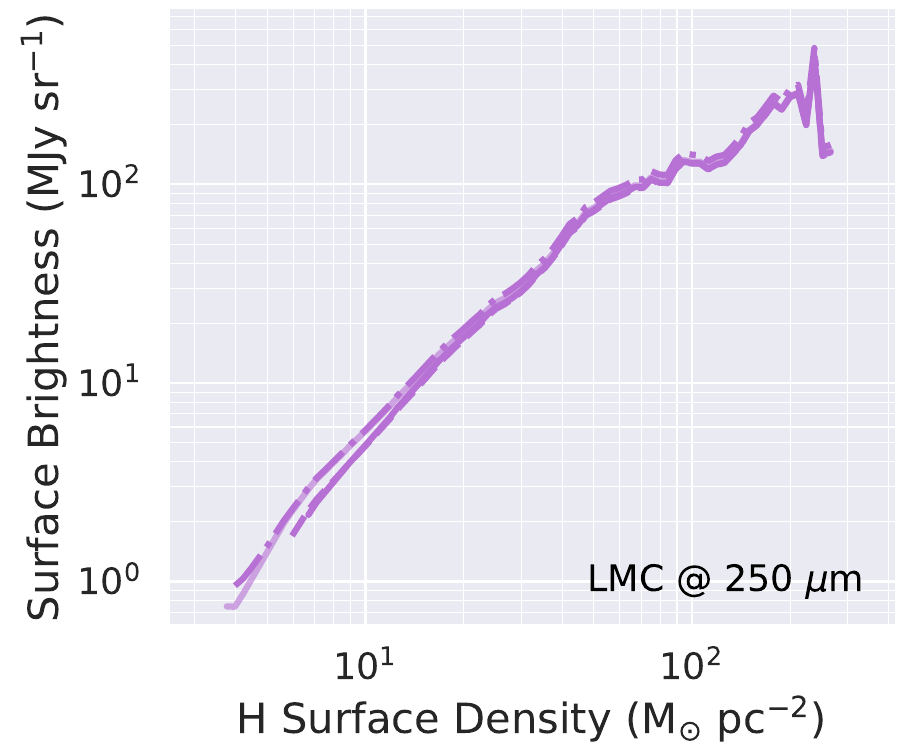}
\includegraphics[width=0.23\textwidth]{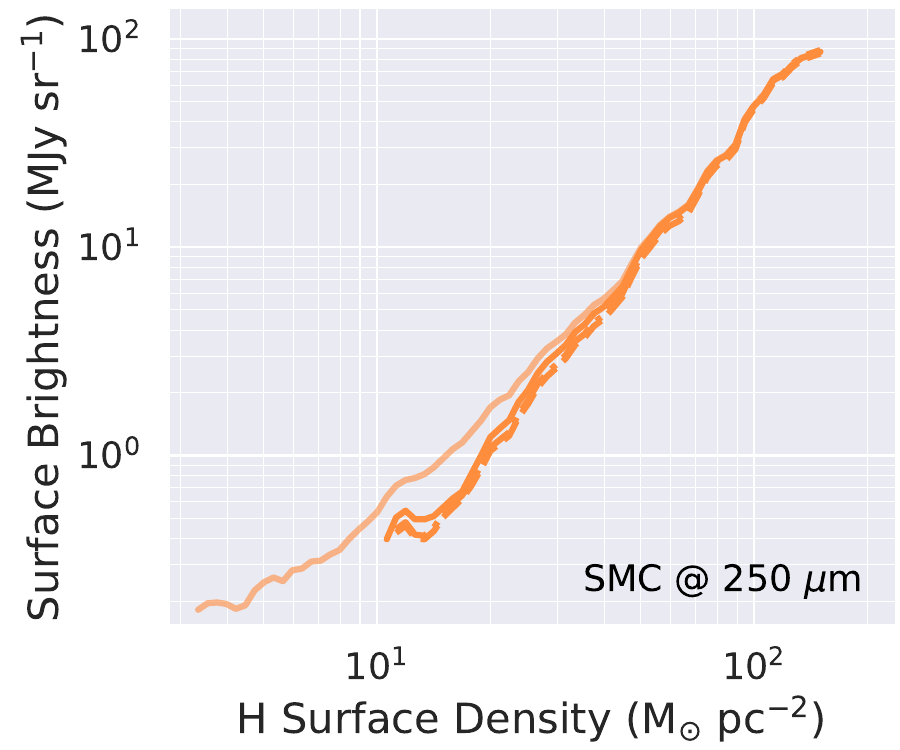}
\includegraphics[width=0.23\textwidth]{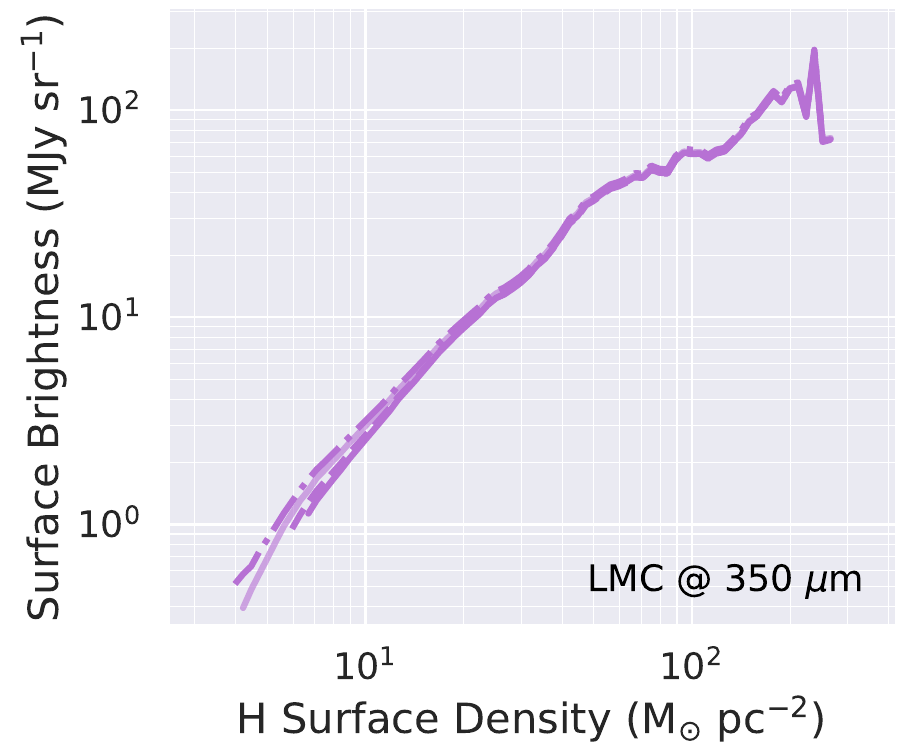}
\includegraphics[width=0.23\textwidth]{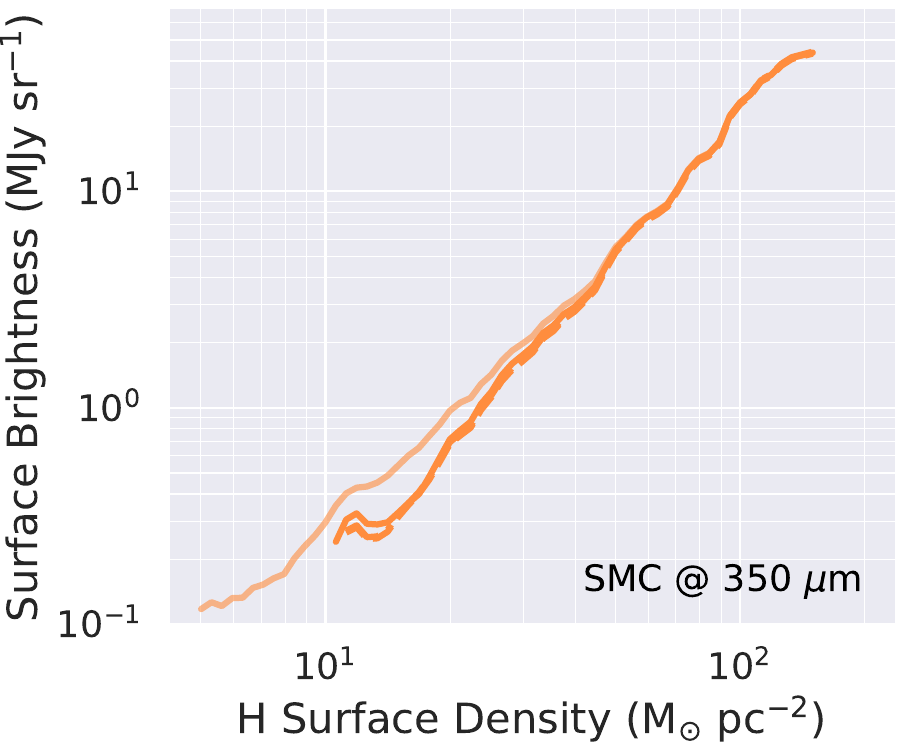}
\includegraphics[width=0.23\textwidth]{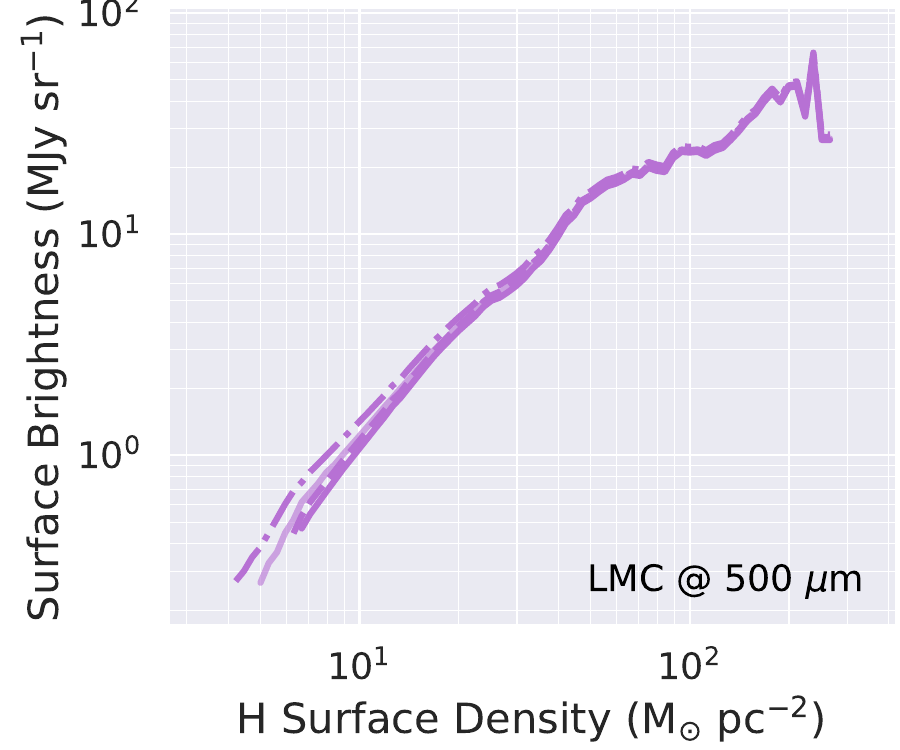}
\includegraphics[width=0.23\textwidth]{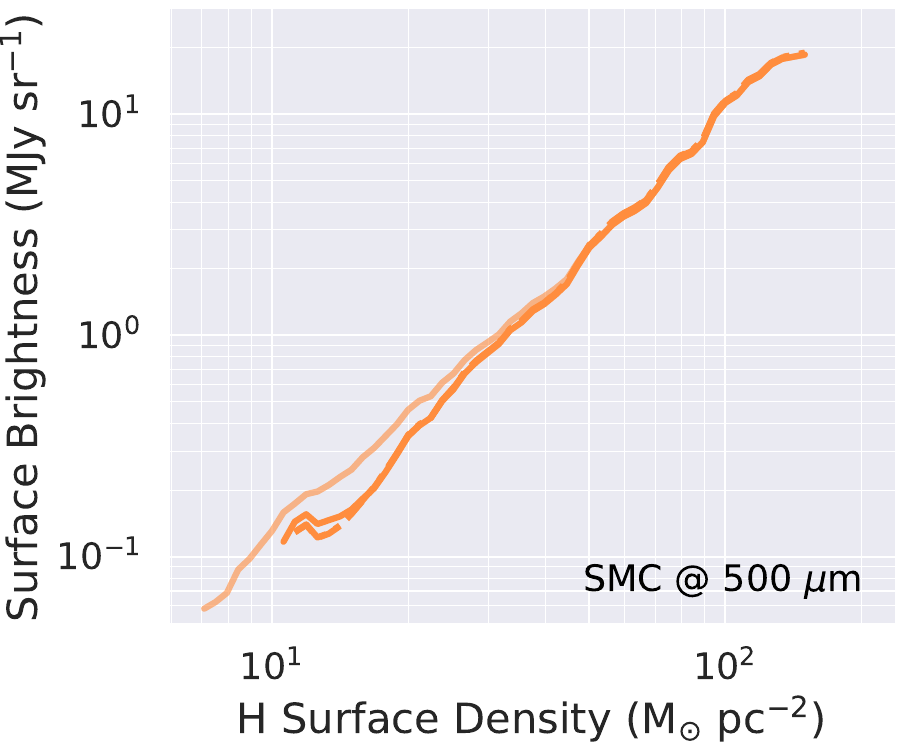}
\caption{Plots of surface brightness (averaged in bins of H density) against H surface density, for the LMC and SMC. In each plot, we show this relation for of full-foreground-subtracted feathered data, and for the polynomial-foreground-subtracted feathered, unfeathered, and HERITAGE data (to allow these three to be compared fairly, with the same foreground subtraction). We plot each relation down until the bin in which \textless\,67\% of the binned pixels have positive values; below this point, the line can `bounce' around due to the noise.}
\label{Fig:LMC_SMC_Apples_Brightness_versus_Gas}
\end{figure}

\needspace{3\baselineskip} \subsection{Tracing Dust Surface Brightness Down to the Lowest Hydrogen Surface Densities} \label{Subsection:Brightness_With_Density}

Dust and gas are typically well mixed in the ISM, across a wide range of surface densities \citep{Hildebrand1983C,Eales2012A,Williams2018A}. Given that ISM properties are strongly driven by density (see Section~\ref{Section:Introduction} and references therein), dust found at different locations, but sharing a given $\Sigma_{\rm H}$, should have similar properties.

We convolved our $\Sigma_{\rm H}$ and foreground-subtracted \hersc\ maps to the limiting resolution for each galaxy. For the LMC and SMC, the limiting data were the $\Sigma_{\rm H}$ maps, with resolutions of 1\arcmin\ for the LMC and 2.6\arcmin\ for the SMC. For M\,31 and M\,33, the limiting data was \hersc\ at 500\,\micron, with its 36\arcsec\ resolution. We then defined bins of $\Sigma_{\rm H}$, each 0.025\,dex wide. For each bin, all of the $\Sigma_{\rm H}$ pixels within that density range were identified, and all of the corresponding \hersc\ pixels had their mean surface brightnesses calculated, within each band.

For comparison, we also performed this process for our unfeathered maps -- plus, for the LMC and SMC, the original HERITAGE maps \citep{Meixner2013A}, to allow us to evaluate the difference from using the newer reduction calibrations, pipelines, and feathering. For the purposes of an `apples-to-apples' comparison, we applied the polynomial-based background subtraction used for our unfeathered maps in Section~\ref{Subsection:Naive_Foreground_Subtraction} to the feathered and HERITAGE maps also. By comparing the results for the feathered maps when using the polynomal versus full foreground subtractions, we are able to determine how much additional low-surface-brightness emission is retrievable thanks to the improved foreground subtraction that is only possible with the feathered maps.

\begin{figure}
\centering
\includegraphics[width=0.22\textwidth]{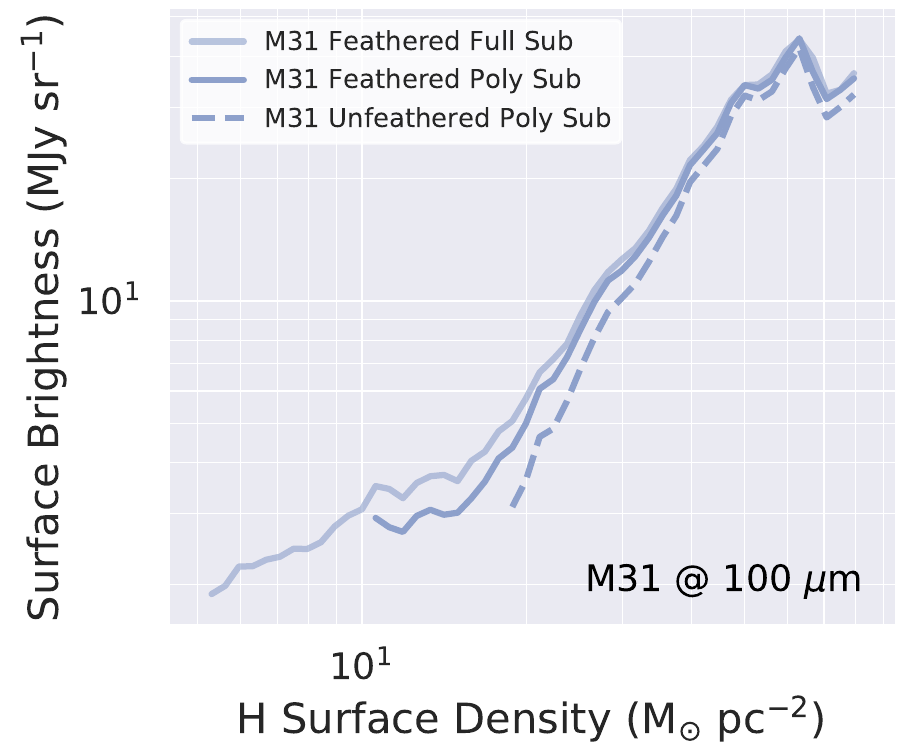}
\includegraphics[width=0.22\textwidth]{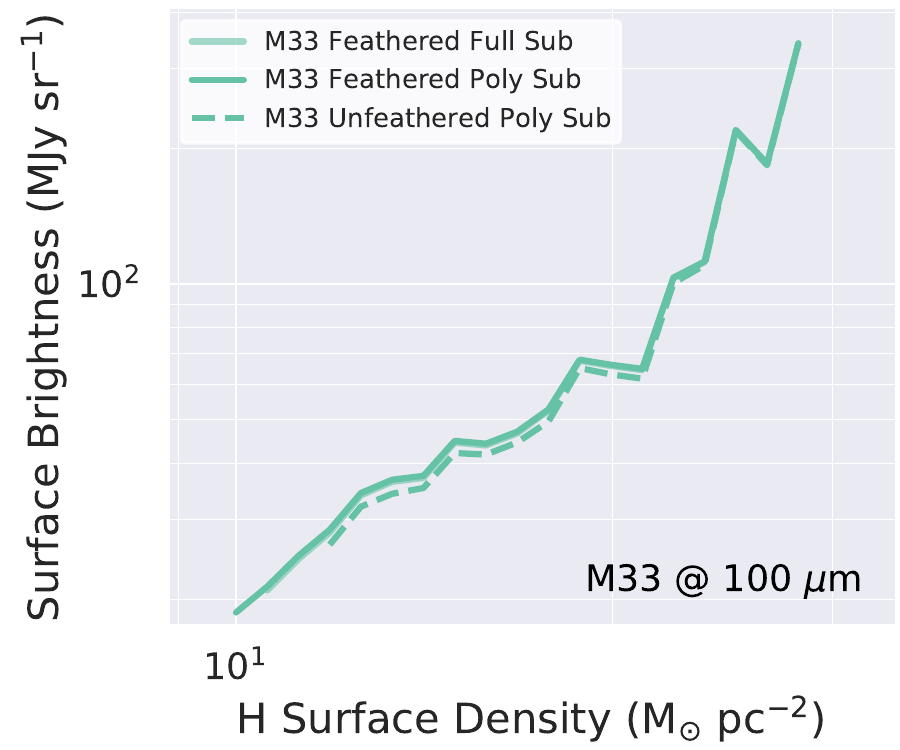}
\includegraphics[width=0.22\textwidth]{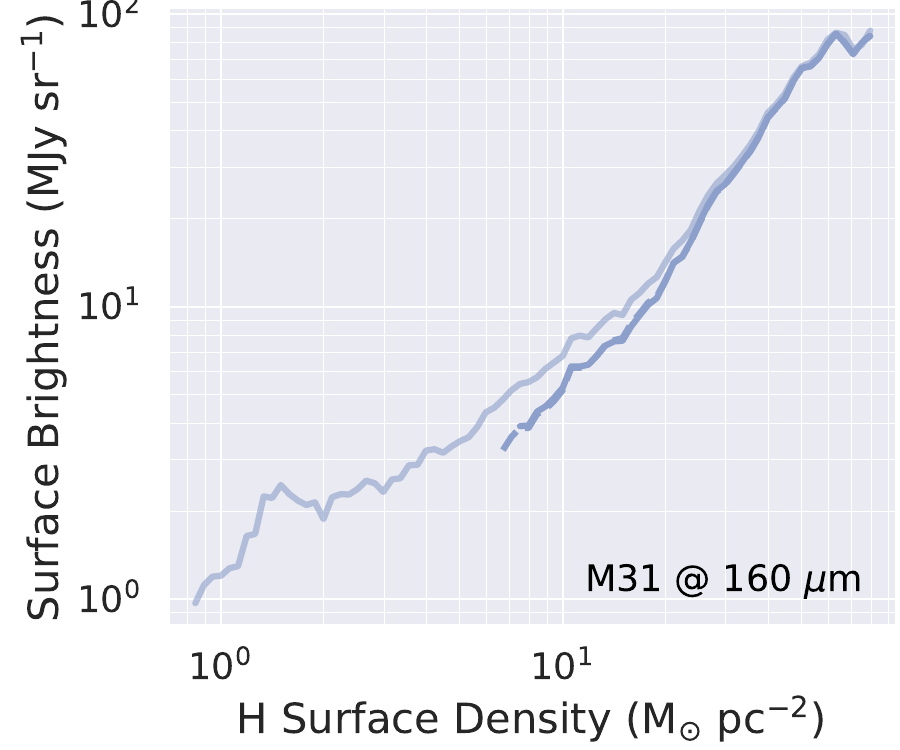}
\includegraphics[width=0.22\textwidth]{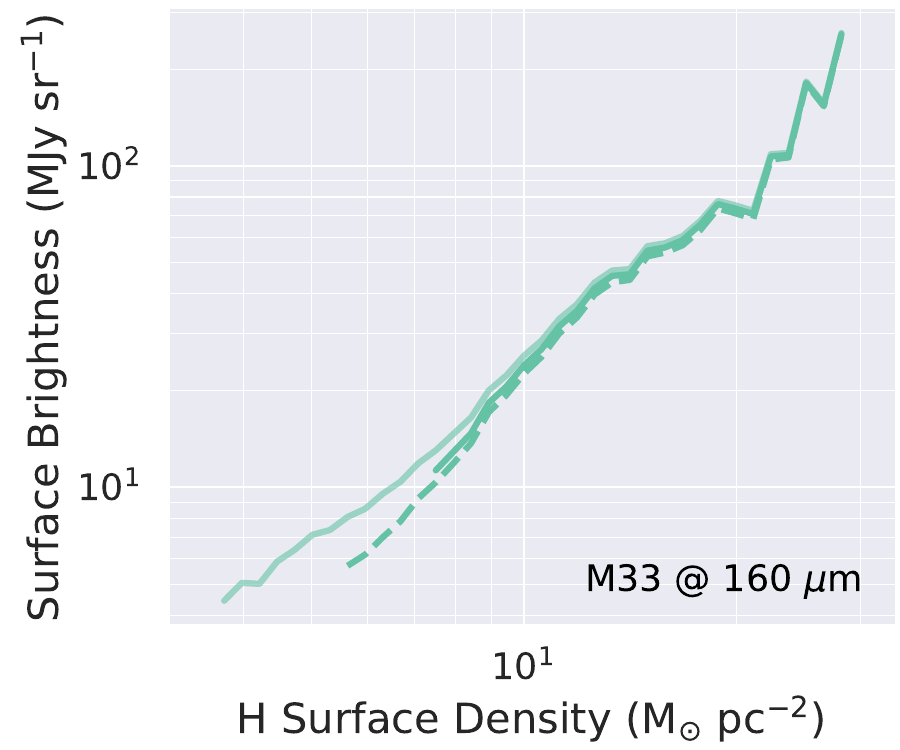}
\includegraphics[width=0.22\textwidth]{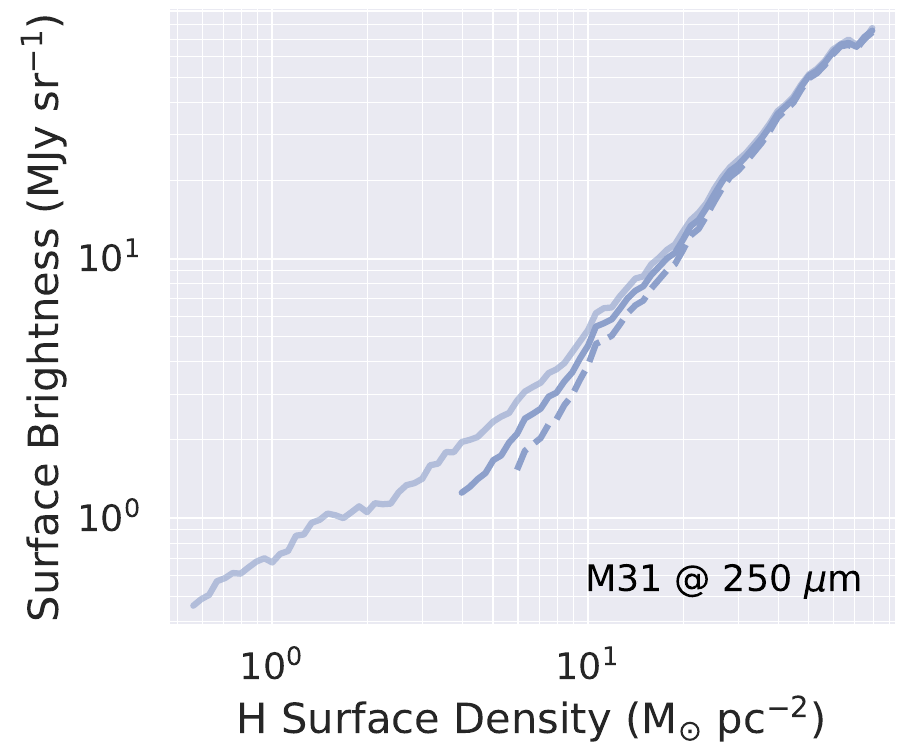}
\includegraphics[width=0.22\textwidth]{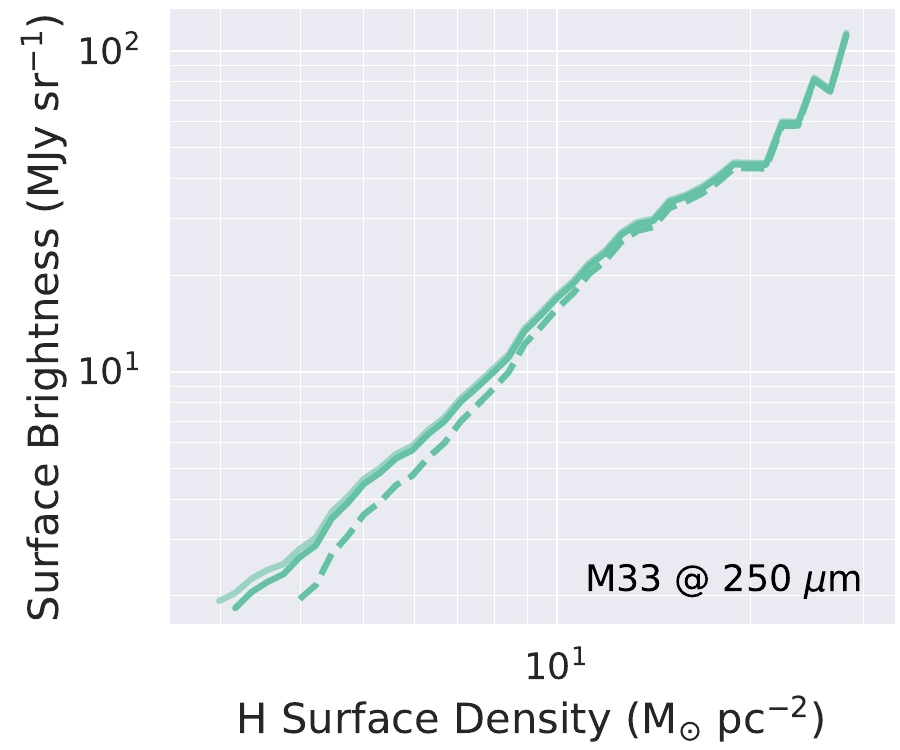}
\includegraphics[width=0.22\textwidth]{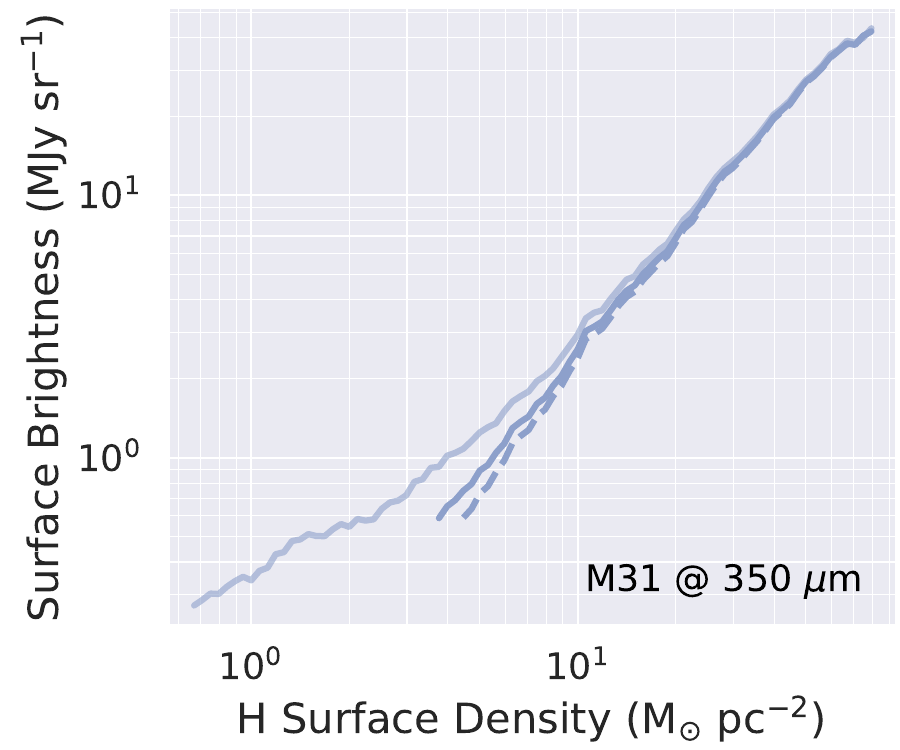}
\includegraphics[width=0.22\textwidth]{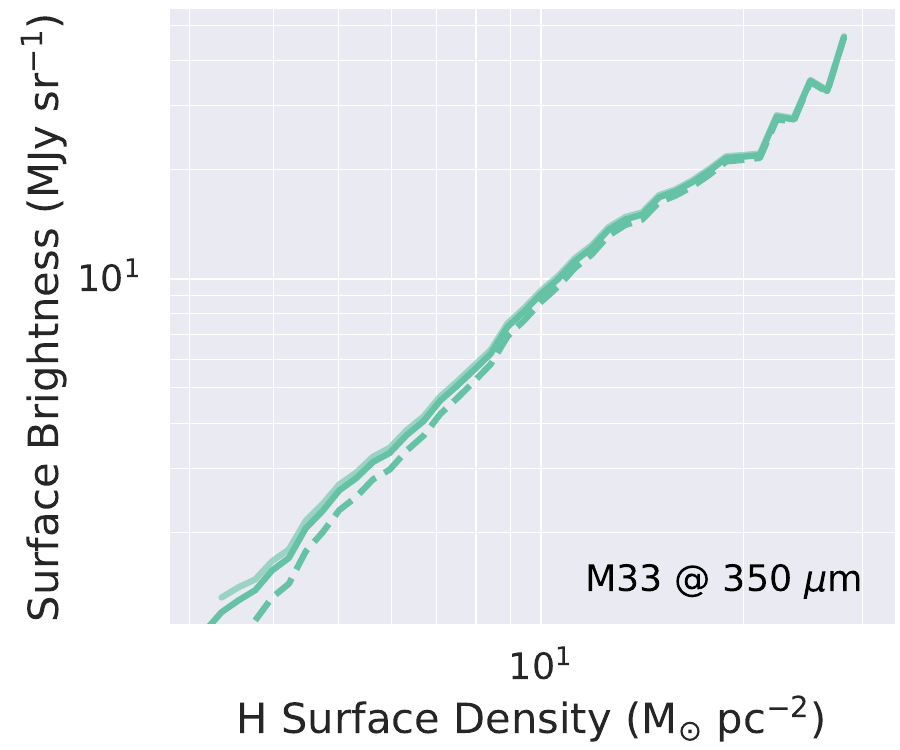}
\includegraphics[width=0.22\textwidth]{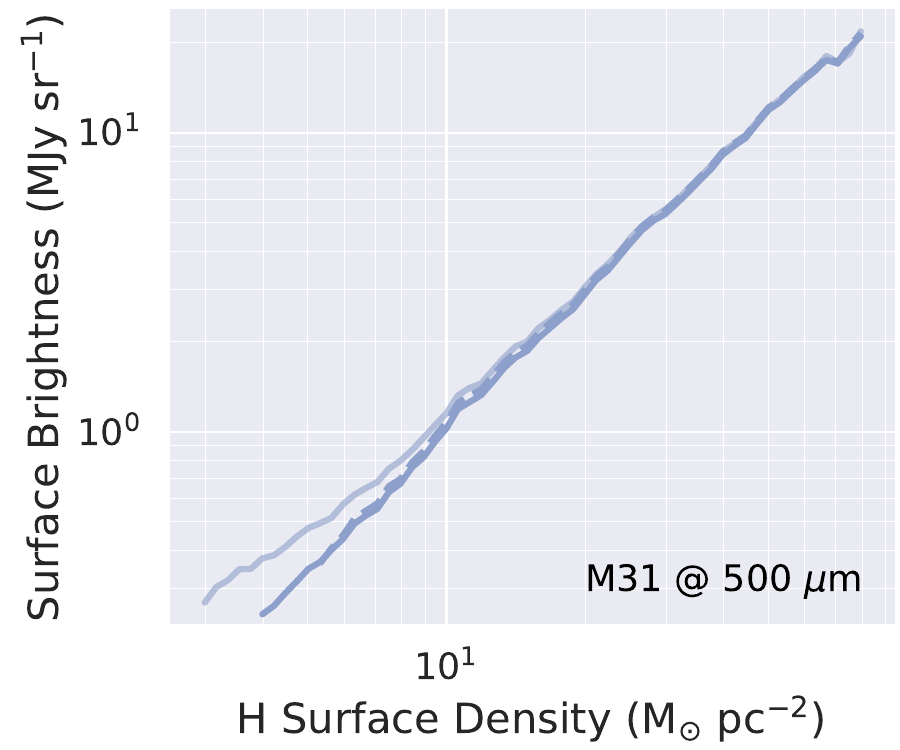}
\includegraphics[width=0.22\textwidth]{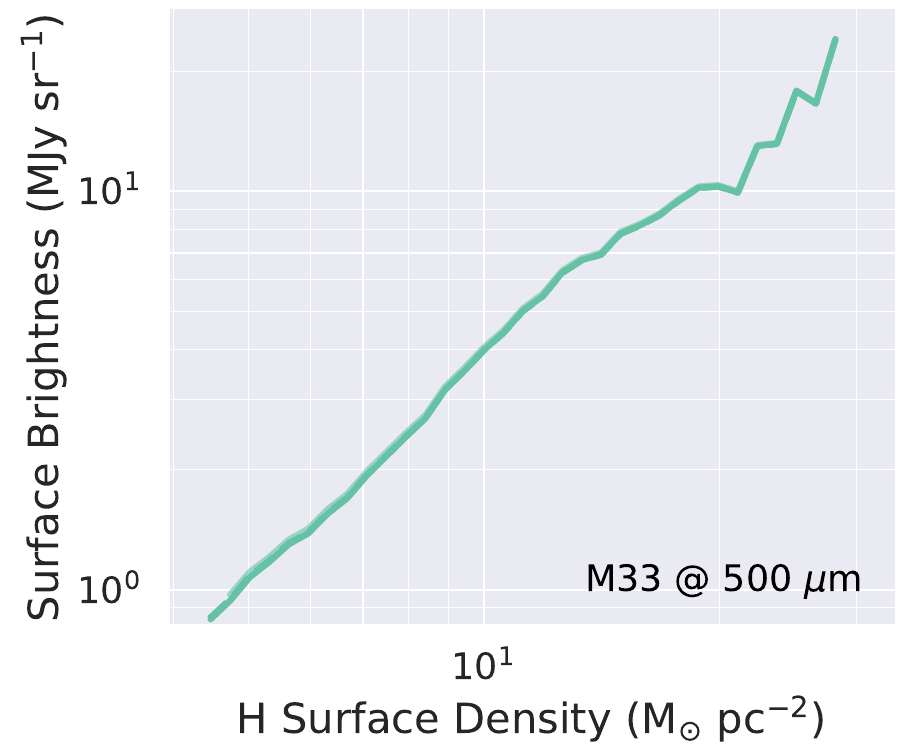}
\caption{Same as for Figure~\ref{Fig:LMC_SMC_Apples_Brightness_versus_Gas}, but for M\,31 and M\,33 (so therefore without HERITAGE data plotted).}
\label{Fig:M31_M33_Apples_Brightness_versus_Gas}
\end{figure}

In Figures~\ref{Fig:LMC_SMC_Apples_Brightness_versus_Gas} (LMC and SMC) and ~\ref{Fig:M31_M33_Apples_Brightness_versus_Gas} (M\,31, and M\,33), we compare the relationship between HI surface density and surface brightness for the polynomial-subtracted feathered, unfeathered, and HERITAGE maps, along with the feathered maps that had undergone our full \HI-based foreground subtraction. Note that it is hard to draw firm conclusions from differences between the HERITAGE reductions and our reductions, as HERITAGE used a much older version of HIPE (v7 for HERITAGE, compared to v12+ for our reductions), and there were some differences in calibration (eg, non-linearity corrections were added for PACS, bolometer relative gains were altered for SPIRE, and beam sizes were updated for both). However, it is still valuable to see what H column each dataset is able to trace down to. 

We only perform the binning for this comparison for pixels that are covered in all of the datasets (\hersc, \HI, and CO) for each given galaxy. As such, our ability to retrieve binned \hersc\ emission here is a lower limit (ie, we would be able to probe even deeper if able to bin over the entire map area).

For the SMC (right of Figure~\ref{Fig:LMC_SMC_Apples_Brightness_versus_Gas}) and M31 (left of Figure~\ref{Fig:M31_M33_Apples_Brightness_versus_Gas}), we see particularly impressive gains with the new data. For the SMC, averaged across all five bands, the full-foreground-subtracted feathered data can detect dust emission to a H surface density a factor of 2 lower than the HERITAGE data, and a factor of 2.3 lower than with the unfeathered data. Similarly, for M\,31, the feathered data allows us to detect dust emission to a mean factor of 5 lower in density (reaching a factor of 10 lower at 250\,\micron).

\begin{figure*}
\centering
\includegraphics[width=0.975\textwidth]{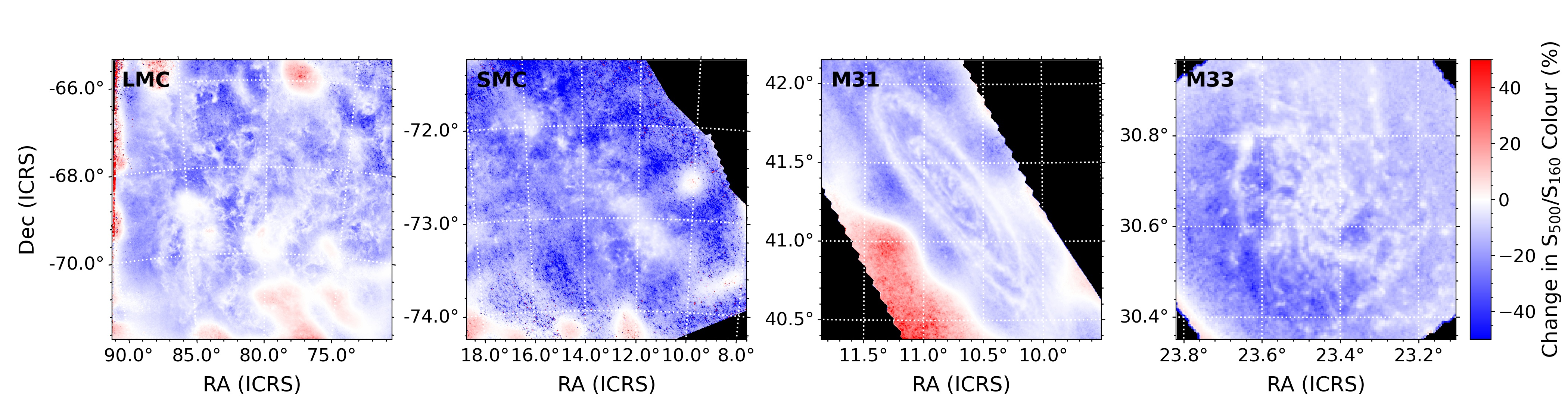}
\caption{Illustration of the change in the $S_{500}/S_{160}$ colour for each of our sample galaxies. Due to the greater amount of emission restored in the shorter-wavelength data, the bulk of the area is now bluer, especially in areas of more diffuse emission.}
\label{Fig:Herschel_Colour_Ratios}
\end{figure*}

For the LMC (left of Figure~\ref{Fig:LMC_SMC_Apples_Brightness_versus_Gas}), we see that there are rather minimal difference between the surface brightnesses, down to the lowest H density for which we are able to trace dust emission. In each band, both the polynomial-foreground-subtracted HERITAGE data or full-foreground-subtracted feathered data were each able to trace down to similar H densities. This may be due to the fact that the LMC has a relatively abrupt `edge' to its ISM disc, and the fact that the \hersc\ map is relatively tight compared to this disc, with little sky around the edges. There are also smaller gains for M\,33 in most bands, only probing to H densities 25\% lower, on average, than the unfeathered data (although the PACS bands specifically do go to 50\% lower densities). As with the LMC, the M\,33 maps are comparitively small, with M\,33 lying close to the edges.

We also note that the full-foreground-subtracted feathered data also traces dust emission out much lower H densities than the polynomial-foreground-subtracted feathered data. This well demonstrates the value of the fact that feathering the \hersc\ maps with all-sky survey maps allows the combined maps to properly trace the large-scale foreground cirrus -- therefore making it possible to accurately subtract it by comparison to Galactic \HI\ data.

For all galaxies and bands, our new data allows us to trace dust emission down to ISM surface densities of \textless\,10\,${\rm M_{\odot} pc^{-2}}$. It is interesting that the surface brightness versus H density relations do not drop off with decreasing H density; instead, the gradients remain the same, or even flatten out. This is suggestive of the dust abundance in these low-density environments -- which will be the central focus of the second paper in this series (Clark et al., {\it in prep.}). 

\needspace{3\baselineskip} \subsection{Much Bluer Galaxy Peripheries} \label{Subsection:Bluer_Peripheries}

The magnitude and distribution of the emission restored by the feathering process varies from band to band for each galaxy. In general, our feathering process restored more diffuse emission in the shorter-wavelength data than at longer wavelengths (where there was much less filtering). This restoration was most prominent at the galaxies` outskirts. As a result, the emission at the edges of these galaxies is much bluer than it was in the unfeathered data.

This is demonstrated in Figure~\ref{Fig:Herschel_Colour_Ratios}, which shows how much the $S_{500}/S_{160}$ colour (ie, the 500\,\micron\ to 160\,\micron\ surface brightness ratio) has changed, between the unfeathered maps and the feathered maps. Bright compact regions show no change in colour, as they suffered little filtering in the unfeathered data. However, diffuse regions, especially at the galaxies' peripheries, are much bluer in the feathered data, with $S_{500}/S_{160}$ often falling by 20-30\% over large areas. Note that this increase in blueness is {\it despite} the fact that the cross-calibration in Section~\ref{Subsection:Feathering_IRIS-Planck_Herschel} reduced brightnesses in the PACS bands by 25--35\% (see Table~\ref{Table:Brightness_Corrections}).

Dust temperatures derived from SED fitting FIR--submm data are strongly driven by colour \citep{Bendo2012A}. As such, the increased blueness of the new data can be expected to affect dust temperatures -- and also affect dust masses, and $\beta$ (due to the temperature--$\beta$ degeneracy; \citealp{Kelly2012B,Galliano2018B}). Exploring the dust SED properties will be a central focus of the second paper in this series (Clark et al., {\it in prep.}).

\needspace{3\baselineskip} \section{Conclusion} \label{Section:Conclusion}

We have produced new \hersc\ maps for the Local Group galaxies M\,31, M\,33, the LMC, and the SMC. These are some of the most heavily-studied galaxies in the sky, and represent key local laboratories on which we base our understanding of many systems across cosmic time. Because of these galaxies' large angular sizes, the standard \hersc\ reductions are vulnerable to severe loss of diffuse emission on large angular scales, filtered out as part of the reduction pipeline. To remedy this, we combined the latest \hersc\ reductions, in Fourier space, with \planck, IRAS, and DIRBE data, to restore any lost emission. From producing and analyzing these new `feathered' maps, our key findings are:

\begin{itemize}

\item Large amounts of diffuse dust emission was indeed missing from standard reductions, especially around the outskirts of the sample galaxies. Our new maps restore this missing dust. This restoration was particularly pronounced in the shorter-wavelength bands.

\item By restoring this diffuse emission, we find significantly different global fluxes measured for our galaxies, as compared to those measured from the unfeathered maps, by over 15\% in many cases.

\item However, from cross-calibrating the power spectra of \hersc\ data with that of absolutely-calibrated all-sky survey data, we find that \hersc-PACS maps seem to over-estimate the brightness of large-scale emission by 20--30\% (in line with similar findings by \citealp{Abreu-Vicente2017A}). Our new data corrects for this.

\item Previous \hersc\ photometry for the Magellanic Clouds, from HERITAGE \citep{Meixner2013A}, conflicted with published photometry from lower-resolution facilities such as COBE, IRAS, and \spitz, with differences of up to 50\%. Photometry from our new maps resolve this disagreement, with our new \hersc\ fluxes generally agreeing well with values from the other facilities 

\item When binning together pixels in the feathered maps according to hydrogen surface density, we find that we can detect dust emission down to ISM densities of $\Sigma_{\rm H} < 1\,{\rm M_{\odot} pc^{-2}}$ in some cases, and to at least $\Sigma_{\rm H} < 10\,{\rm M_{\odot} pc^{-2}}$ for all galaxies and bands. The quality of our feathered maps allows us to detect such emission to lower densities than with unfeathered data (down to densities a factor of 2.2 lower, on average), or with previous HERITAGE data for the Magellanic Clouds (a factor of (down to densities 50\% lower, on average), with particularly large improvements for the SMC and M\,31. We find no indication that the dust emission drops off more sharply at these lower ISM densities.

\item Because the restoration of large-scale emission was greater in the shorter-wavelength bans, the galaxies' far-infrared colours are now much bluer over large areas, especially in their peripheries. Specifically, we find the $S_{500}/S_{160}$ colour becoming \textgreater\,20\% bluer almost everywhere there is not bright compact emission present. 

\end{itemize}

With this new data in hand, the immediate future work will be to study the variation of dust abundance relative to gas, and how this relates to grain-growth, deplations, etc. This will be the focus of paper II in this series (Clark et al., {\it in prep.}). Additional investigations that will be enabled by this data include a study of how FIR dust emissivity compares to UV--optical extinction, as revealed by {\it Hubble} Space Telescope imaging programs such as Scylla \citep{Murray2019B} and the Panchromatic {\it Hubble} Andromeda Treasury \citep{Dalcanton2012C}, to trace variations in dust properties and composition; and constraining the dust mass opacity coefficient, $\kappa_{d}$, by comparison to gas and metallicity data, at spatial resolutions high enough to overcome significant temperature mixing \citep{Galliano2011B,Priestley2020B}.

The code used to carry out the feathering process presented here is available online at: \url{https://doi.org/10.5281/zenodo.4776266}. The maps themselves will be released alongside paper II in this series (Clark et al., {\it in prep.}).

{\small

\acknowledgments

The authors thank Oliver Lomax for sharing his adaptation of the old COBE-DIRBE reprojection algorithm from \texttt{FORTRAN} into \texttt{Python}, and thank Yumi Choi and Thomas Williams for valuable discussions. The authors also thank the anonymous referee for their thoughtful and helpful comments.

CJRC and JR-D acknowledge financial support from the National Aeronautics and Space Administration (NASA) Astrophysics Data Analysis Program (ADAP) grant 80NSSC18K0944.

This work used the Extreme Science and Engineering Discovery Environment (XSEDE; \citealp{Towns2014A}), which is supported by National Science Foundation grant number ACI-1548562. Specifically, this work used XSEDE Bridges-Large at the Pittsburgh Supercomputing Center, through allocation AST190060.

This research made use of \texttt{Astropy}\footnote{\url{https://www.astropy.org/}}, a community-developed core \texttt{Python} package for Astronomy \citep{astropy2013,astropy2019}. This research made use of \texttt{reproject}\footnote{\url{https://reproject.readthedocs.io}}, an \texttt{Astropy}-affiliated \texttt{Python} package for image reprojection. This research made use of \texttt{photutils}\footnote{\url{https://photutils.readthedocs.io}}, an \texttt{Astropy}-affiliated \texttt{Python} package for detection and photometry of astronomical sources \citep{Bradley2020C}. This research made use of \texttt{NumPy}\footnote{\url{https://numpy.org/}} \citep{VanDerWalt2011B,Harris2020A}, \texttt{SciPy}\footnote{\url{https://scipy.org/}} \citep{SciPy2001}, and \texttt{Matplotlib}\footnote{\url{https://matplotlib.org/}} \citep{Hunter2007A}. This research made use of the \texttt{pandas}\footnote{\url{https://pandas.pydata.org/}} data structures package for \texttt{Python} \citep{McKinney2010}. This research made use of \texttt{turbustat}\footnote{\url{https://turbustat.readthedocs.io/en/latest/}}, a \texttt{Python} package for turbulence statistics \citep{KochKoch2019C}. This research made use of \texttt{iPython}, an enhanced interactive \texttt{Python} \citep{Perez2007A}.

This research made use of sequential colour-vision-deficiency-friendly colourmaps from \texttt{cmocean}\footnote{\url{https://matplotlib.org/cmocean/}} \citep{Thyng2016A} and \texttt{CMasher}\footnote{\url{https://cmasher.readthedocs.io}} \citep{VanDerVelden2020A}

This research made use of \texttt{UNIMAP} \citep{Piazzo2012A,Piazzo2015A,Piazzo2015B,Piazzo2016B,Piazzo2016C,Piazzo2017A}, a development of the \texttt{ROMAMAP} pipeline \citep{Traficante2011A}.

This research made use of \texttt{TOPCAT}\footnote{\url{http://www.star.bris.ac.uk/~mbt/topcat/}} \citep{Taylor2005A}, an interactive graphical viewer and editor for tabular data.

This research made use of \texttt{Montage}\footnote{\url{https://montage.ipac.caltech.edu/}}, which is funded by the National Science Foundation under Grant Number ACI-1440620, and was previously funded by the NASA's Earth Science Technology Office, Computation Technologies Project, under Cooperative Agreement Number NCC5-626 between NASA and the California Institute of Technology.

This research made use of the SIMBAD database\footnote{\url{https://simbad.u-strasbg.fr/simbad/}}; \citealp{Wenger2000D}) and the VizieR catalogue access tool\footnote{\url{https://vizier.u-strasbg.fr/viz-bin/VizieR}} \citep{Ochsenbein2000B}, both operated at CDS, Strasbourg, France. This research has made use of the {\sc Nasa/ipac} Extragalactic Database\footnote{\url{https://ned.ipac.caltech.edu/}} (NED), operated by the Jet Propulsion Laboratory, California Institute of Technology, under contract with NASA. This research made use of the HyperLEDA database\footnote{\url{http://leda.univ-lyon1.fr/}} \citep{Makarov2014A}.

This research makes use of data from \planck, a project of the European Space Agency, which received support from: ESA; CNES and CNRS/INSU- IN2P3-INP (France); ASI, CNR, and INAF (Italy); NASA and DoE (USA); STFC and UKSA (UK); CSIC, MINECO, JA, and RES (Spain); Tekes, AoF, and CSC (Finland); DLR and MPG (Germany); CSA (Canada); DTU Space (Denmark); SER/SSO (Switzerland); RCN (Norway); SFI (Ireland); FCT/MCTES (Portugal); ERC and PRACE (EU).

\hersc\ is an ESA space observatory with science instruments provided by European-led Principal Investigator consortia and with important participation from NASA. The \hersc\ spacecraft was designed, built, tested, and launched under a contract to ESA managed by the \hersc/\planck\ Project team by an industrial consortium under the overall responsibility of the prime contractor Thales Alenia Space (Cannes), and including Astrium (Friedrichshafen) responsible for the payload module and for system testing at spacecraft level, Thales Alenia Space (Turin) responsible for the service module, and Astrium (Toulouse) responsible for the telescope, with in excess of a hundred subcontractors.

}

\newpage

\bibliography{ChrisBib}{}
\bibliographystyle{aasjournal}

\appendix
\restartappendixnumbering

\needspace{3\baselineskip} \section{Herschel Observations Used In Reductions} \label{AppendixSection:Herschel_Observations}

In Sections~\ref{Subsection:Herschel-PACS} and \ref{Subsection:Herschel-SPIRE}, we discuss the \hersc\ PACS and SPIRE data used in this work. Here we present the specifics \hersc\ observation IDs that we used to reduce our data; they are listed in Table~\ref{AppendixTable:Herschel_Observations}. These observations incorporate data from: \citet{Kramer2010A,Boquien2011C,Fritz2012A,Meixner2013A}.

\begin{table}[b]
\centering
\caption{\hersc\ observation IDs we use for data reduction for each of our sample galaxies. Data from all bands observed by a given observation were used, unless otherwise noted.}
\label{AppendixTable:Herschel_Observations}
\begin{tabular}{llll}
\toprule \toprule
\multicolumn{4}{c}{Observation IDs} \\
\cmidrule(lr){1-4}
\multicolumn{1}{c}{LMC} &
\multicolumn{3}{c}{} \\
\cmidrule(lr){1-1}
1342195668 & 1342195669 & 1342195683 & 1342195707 \\
1342195708 & 1342195684 & 1342195712 & 1342195713 \\
1342195728 & 1342202086 & 1342202087 & 1342202202 \\
1342202203 & 1342202216 & 1342202217 & 1342202224 \\
1342202225 & 1342202243 & 1342202244 & \\
\multicolumn{1}{c}{SMC} &
\multicolumn{3}{c}{} \\
\cmidrule(lr){1-1}
1342192680 & 1342192681 & 1342192697 & 1342192698 \\
1342192699 & 1342198565 & 1342198566 & 1342198590 \\
1342198591 & 1342198863 & 1342205049 & 1342205050 \\
1342205054 & 1342205055 & 1342205092 & \\
\multicolumn{1}{c}{M\,31} &
\multicolumn{3}{c}{} \\
\cmidrule(lr){1-1}
1342211294 & 1342211309 & 1342211319 & 1342213207 \\
\multicolumn{1}{c}{M\,33} &
\multicolumn{3}{c}{} \\
\cmidrule(lr){1-1}
1342189079$^{\rm a }$ & 1342247408$^{\rm a }$ & 1342189079 & 1342247409 \\
\bottomrule
\end{tabular}

\justifying
$^{\rm a }$ Used for PACS 100\,\micron\ only.\\
\end{table}

\needspace{3\baselineskip} \section{Regions Masked During SPIRE Destriping} \label{AppendixSection:SPIRE_Masked_Regions}

\begin{table}
\centering
\caption{Positions and radii of circular regions masked when running the destriper during reduction of \hersc-SPIRE data for the LMC and SMC.}
\label{AppendixTable:SPIRE_Masked_Regions}
\begin{tabular}{rrr}
\toprule \toprule
\multicolumn{1}{c}{$\alpha$ (J2000)} &
\multicolumn{1}{c}{$\delta$ (J2000)} &
\multicolumn{1}{c}{Radius (arcmin)} \\
\cmidrule(lr){1-3}
\multicolumn{1}{c}{LMC} &
\multicolumn{1}{c}{} &
\multicolumn{1}{c}{} \\
\cmidrule(lr){1-1}
85.425 & -69.551 & 53.8 \\
82.776 & -71.136 & 13.2 \\
82.882 & -68.509 & 15.5 \\
83.592 & -67.623 & 21.4 \\
80.581 & -67.942 & 14.0 \\
81.422 & -66.173 & 12.9 \\
74.219 & -66.379 & 21.4 \\
72.905 & -69.266 & 18.7 \\
\multicolumn{1}{c}{SMC} &
\multicolumn{1}{c}{} &
\multicolumn{1}{c}{} \\
\cmidrule(lr){1-1}
11.849 & -73.198 & 37.5 \\
13.412 & -72.715 & 38.6 \\
15.468 & -71.963 & 33.5 \\
17.104 & -72.803 & 33.6 \\
19.715 & -73.213 & 25.8 \\
21.929 & -73.454 & 25.8 \\
\bottomrule
\end{tabular}
\end{table} 

In Section~\ref{Subsubsection:SPIRE_Re-Reduction}, we describe how, for our \hersc-SPIRE reduction process, we mask a number of bright regions in the LMC and SMC when running the destriper. This prevents artefacts being added, instead of removed, around these regions, and increases the speed with with the destriping algorithm can converge on a solution. These regions were identified manually, based on both their brightness, and their prominence relative to their surroundings. The positions and radii of the regions (all of which are circles) are listed in Table~\ref{AppendixTable:SPIRE_Masked_Regions}.  

\needspace{3\baselineskip} \section{Apodisation} \label{AppendixSection:Apodisation}

\begin{figure*}
\centering
\includegraphics[width=0.975\textwidth]{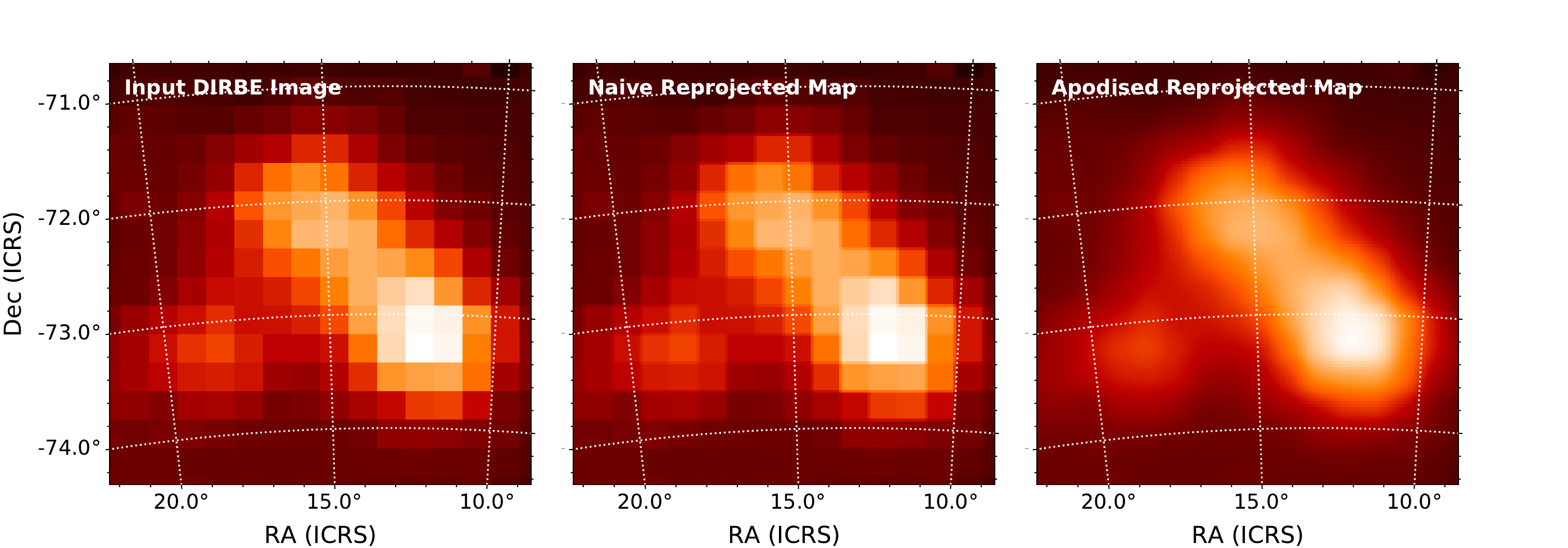}
\caption{{\it Left:} DIRBE 100\,\micron\ data for the SMC. {\it Centre:} DIRBE data reprojected to the IRIS pixel grid; the DIRBE pixels have diameters 10 times greater than the IRIS pixels, so each `big' pixel in this image is in fact made up of 10$\times$10 IRIS-sized pixels, with the transitions between each `big'' pixel still being quite sharp. {\it Right:} The reprojected DIRBE data, apodised to remove the pixel edge artefacts.}
\label{AppendixFig:SMC_Apodise_Example}
\end{figure*}

Unfeathered reprojection of low-resolution data to the pixel grid of high-resolution data can be problematic. An example of this, in the case of DIRBE and IRIS data, is shown in Figure~\ref{AppendixFig:SMC_Apodise_Example}. The DIRBE pixels are so large compared to the IRIS pixels (150\arcmin\ versus 1.5\arcmin) that the pixel edges remain extremely prominent in the reprojection. If this data is used for feathering, these pixel edge artefacts would still be partially visible in the final feathered data. A similar problem is encountered when reprojecting IRIS and \planck\ data to a \hersc\ pixel grid.

We prevent the problem by apodising the reprojected data -- smoothing it with a kernel smaller than the low-resolution pixel size. This has the effect of suppressing the sharp pixel edges, whilst having only a very minor impact on the effective resolution of the low-resolution data. Specifically, we smooth the reprojected data with a Gaussian kernel with a standard deviation of $p_{\it low} / (p_{\it high} 2 \sqrt{2})$ pixels, where $p_{\it low}$ and $p_{\it high}$ are the pixel widths of the low- and high-resolutions maps respectively; this corresponds to 0.35 times the low-resolution pixel width. To make sure the feathering process can properly incorporate this change in the low-resolution effective PSF, we also convolved the low-resolution PSF with the apodisation kernel. There are other reprojection methods that result in a similarly `smooth' output map (spline interpolation, etc) without requiring this after-the-fact apodisation; however, this method allows us to explicitly account for the resulting change in the PSF; this is important given that successful feathering depends upon the PSF being well-constrained.

\needspace{3\baselineskip} \section{Feathering In-Out Tests} \label{AppendixSection:Feathering_In-Out_Tests}

\begin{figure*}
\centering
\includegraphics[width=0.935\textwidth]{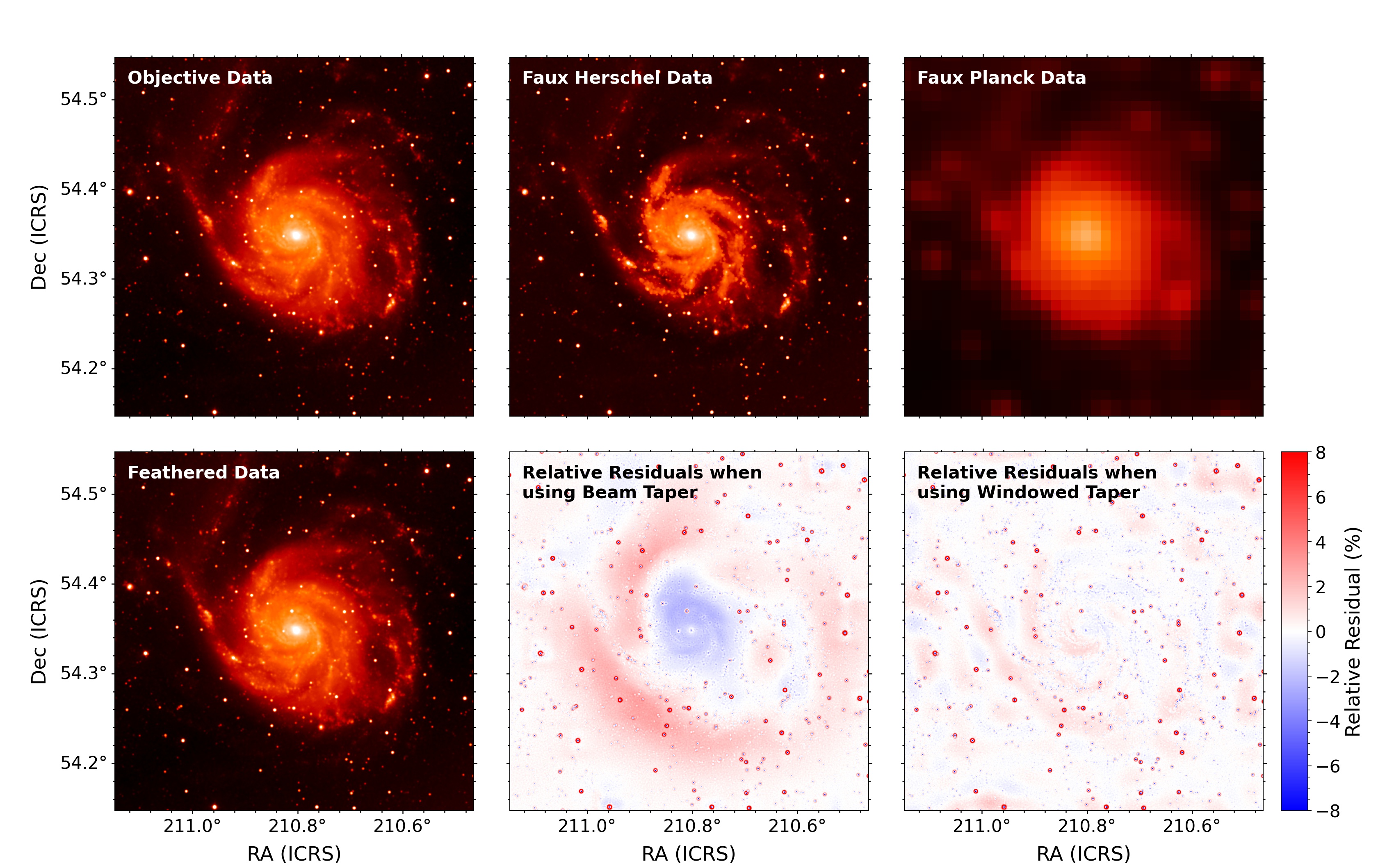}
\caption{Illustration of our in-out test of our feathering process, as tested on SDSS {\it r}-band data of M\,101. {\it Upper left:} The objective data, at 6\arcsec\ resolution, with no emission filtered out. {\it Upper centre:} The `faux' \hersc\ data, at 6\arcsec\ resolution, with emission on scales \textgreater\,200\arcsec\ filtered out. {\it Upper right:} The `faux' \planck\ data, at 100\arcsec\ resolution. {\it Lower left:} Feathered output, combining the two `faux' observations, when using a windowed taper (the result when using a beam taper is visually extremely similar, so we show the one example). {\it Lower middle:} Relative residuals between the feathered map and the objective map, when using a beam taper. {\it Lower right:} Relative residuals between the feathered map and the objective map, when using a windowed taper.}
\label{AppendixFig:M101_Feathering_Sim}
\end{figure*}

In order to test our feathering process, we performed a test whereby we generated three sets of test data: an objective map with \hersc-like resolution but with all large-scale emission preserved; a `faux' \hersc\ map, with \hersc-like resolution and with large-scale emission removed; and a `faux` \planck\ map, with no emission removed but with a low resolution. By feathering together the `faux' \hersc\ and \planck\ maps, we should be able to re-create the objective map. How closely the feathered map matches the objective map allows us to evaluate how well the feathering has performed.

To create these data, we started with Sloan Digital Sky Survey (SDSS; \citealp{York2000B,Eisenstein2011B}) {\it r}-band observations of M\,101. Obviously this is not FIR data, but it does contains both highly compact (individual stars, H\textsc{ii}-regions) and highly extended (disc, stellar streams) emission. Plus, the SDSS resolution of $\approx$1.3\arcsec\ is good enough, compared the \hersc- and \planck-like resolutions to which we will be smoothing it, that it can be treated as an effectively-infinite resolution `ground truth'.

We convolved the SDSS {\it r}-band data to a resolution of 6\arcsec\ to create our objective map. To create our `faux' \planck\ map, we convolved the data to a resolution of 100\arcsec. Whilst 100\arcsec\ is considerably better resolution than the $\approx$300\arcsec\ resolution of \planck, M\,101 is so compact that smoothing to that resolution would render it too close to being a point source to be useful for feathering. Plus, the factor of 16.7 difference in resolution between 6\arcsec\ and 100\arcsec\ well-matches the factor 16.3 difference in resolution between the \hersc\ 250\,\micron\ resolution of 18\arcsec, and the \planck\ 850\,\micron\ resolution of 293\arcsec\ (to which we smoothed all of our IRIS--\planck\ data in Section~\ref{Subsection:Feathering_DIRBE_IRIS}).

To create out `faux' \hersc\ map, we filtered large scale emission from the objective map; we did this by masking compact sources from the objective map, smoothing that map to a resolution of 200\arcsec, multiplying that map's surface brightnesses by 0.5, then subtracting this map from a copy of the original objective map. This produces a map with the same 6\arcsec\ resolution as the objective map, but with emission on scales larger than 200\arcsec\ filtered out (but with little negative bowling, much like real \hersc\ data). These three maps are shown in the top row of Figure~\ref{AppendixFig:M101_Feathering_Sim}.

We then feathered together the `faux' \hersc\ and \planck\ maps. We did this two different ways. The first time we used the low-resolution beam to taper the transition from the high- to-low resolution data in Fourier space (as we did for feathering DIRBE with IRIS in Section~\ref{Subsection:Feathering_DIRBE_IRIS}). The second time we tapered within a window of angular scales to which both data were sensitive (as we did for feathering IRIS--\planck\ with \hersc\ in Section~\ref{Subsection:Feathering_DIRBE_IRIS}). Specifically, we used a tapering window of 120\arcsec\ to 180\arcsec; this is above the 100\arcsec\ resolution of the `faux' \planck\ data, beneath and 200\arcsec\ filtering scale of the `faux' \hersc\ data. Plus the upper and lower bounds of this window differ only by a factor of 1.5, compared to the factor of 2 difference for the window we actually use in Section~\ref{Subsection:Feathering_IRIS-Planck_Herschel}, therefore the results in this test should be worse than those obtained in reality (specifically, a narrower window should cause more ringing), and so provide a conservative assessment.

To evaluate how accurately the feathering performed, we found the relative residual between the two output feathered maps, and the objective map. An example feathered output, and the two residual maps, are shown in the lower row of Figure~\ref{AppendixFig:M101_Feathering_Sim}. 

When using the beam taper, there are some large-scale residuals; the surface-brightness in the centre of M\,101 was underestimated, and the surface brightness in the outskirts of overestimated. However these effects are reasonably small, ranging from -2.5\% to +3.0\%. When using a tapering window, the residuals are much smaller in angular scale, tracing M\,101's spiral arms, and also very small in magnitude, ranging from -0.9\% to +1.1\%. 

This in-out test indicates that, for both methods, the total error introduced as a result of feathering is small – smaller than the instrumental calibration uncertainty for all instruments. Whilst the beam tapering seems to be somewhat less accurate, we are not able to window taper in the case of our DIRBE--IRIS feathering, as described in Section~\ref{Subsection:Feathering_DIRBE_IRIS}. As noted above, this in-out test has been designed to be conservative, and will likely be introducing slightly larger errors than our Local Group galaxy feathering will suffer in practice.

In the case of both methods, compact ringing residuals appear around bright point sources (in this case, stars), with a negative residual in the centre, surrounded by a ring of positive residual. However, total flux is conserved correctly overall for each source, and the entire artefact is always smaller than the instrumental PSF. Moreover, for our actual feathering with \hersc\ data, there are very few bright point sources of this kind, so we are not concerned about this kind of ringing affecting our final data products.

\needspace{3\baselineskip} \section{IRAS-IRIS Response Artefacts Around 30 Doradus} \label{AppendixSection:IRIS_Around_30-Dor}

\begin{figure*}
\centering
\includegraphics[width=0.305\textwidth]{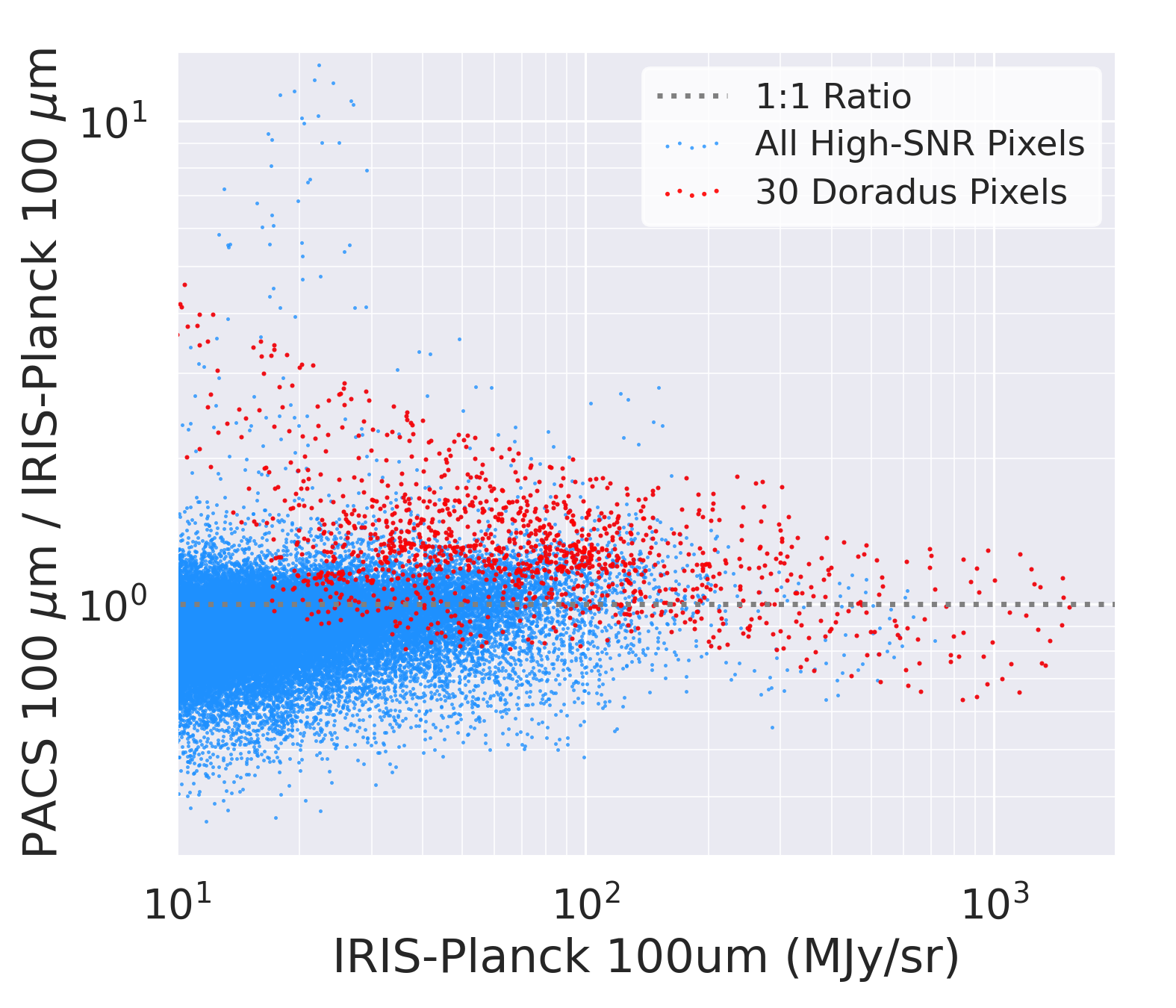}
\includegraphics[width=0.305\textwidth]{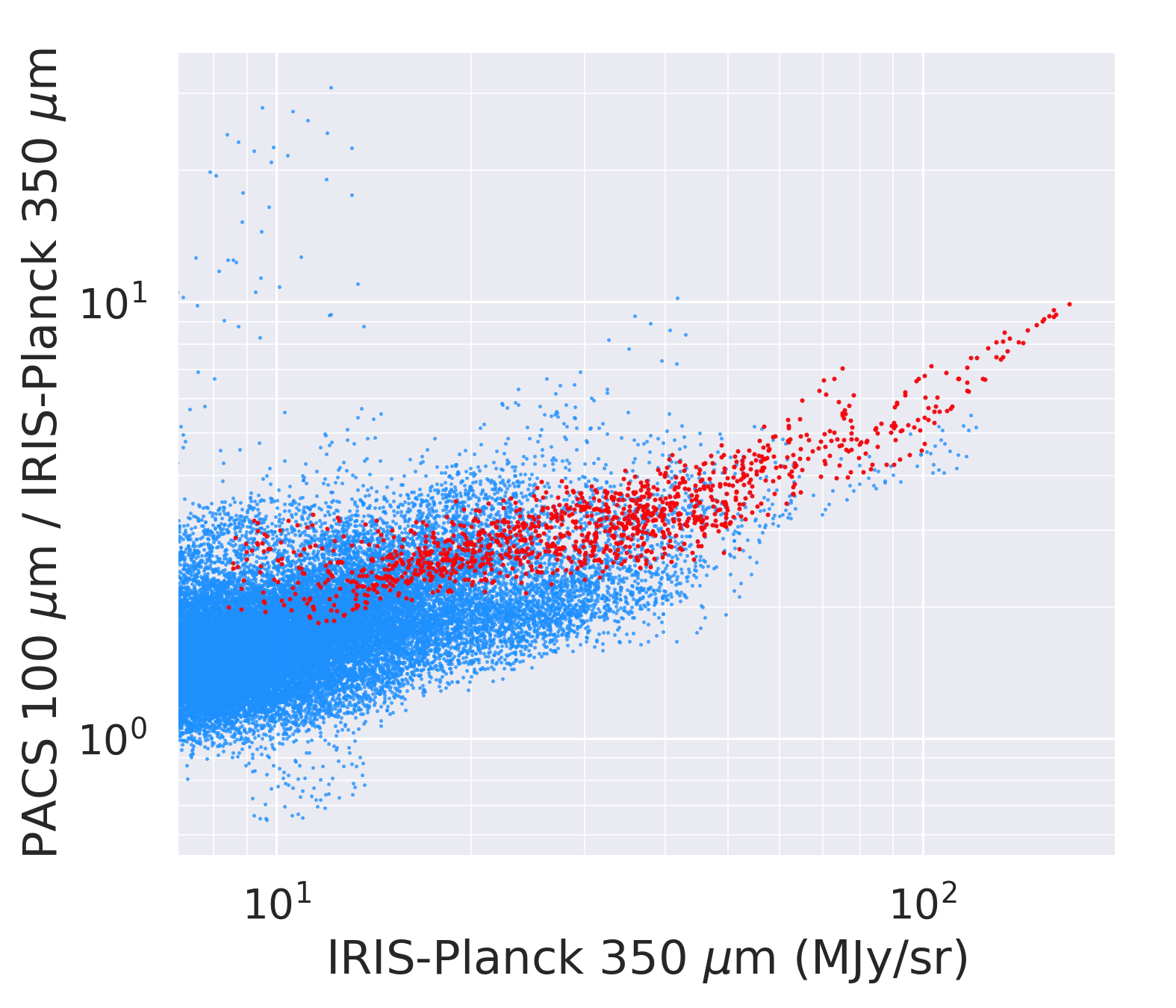}
\includegraphics[width=0.305\textwidth]{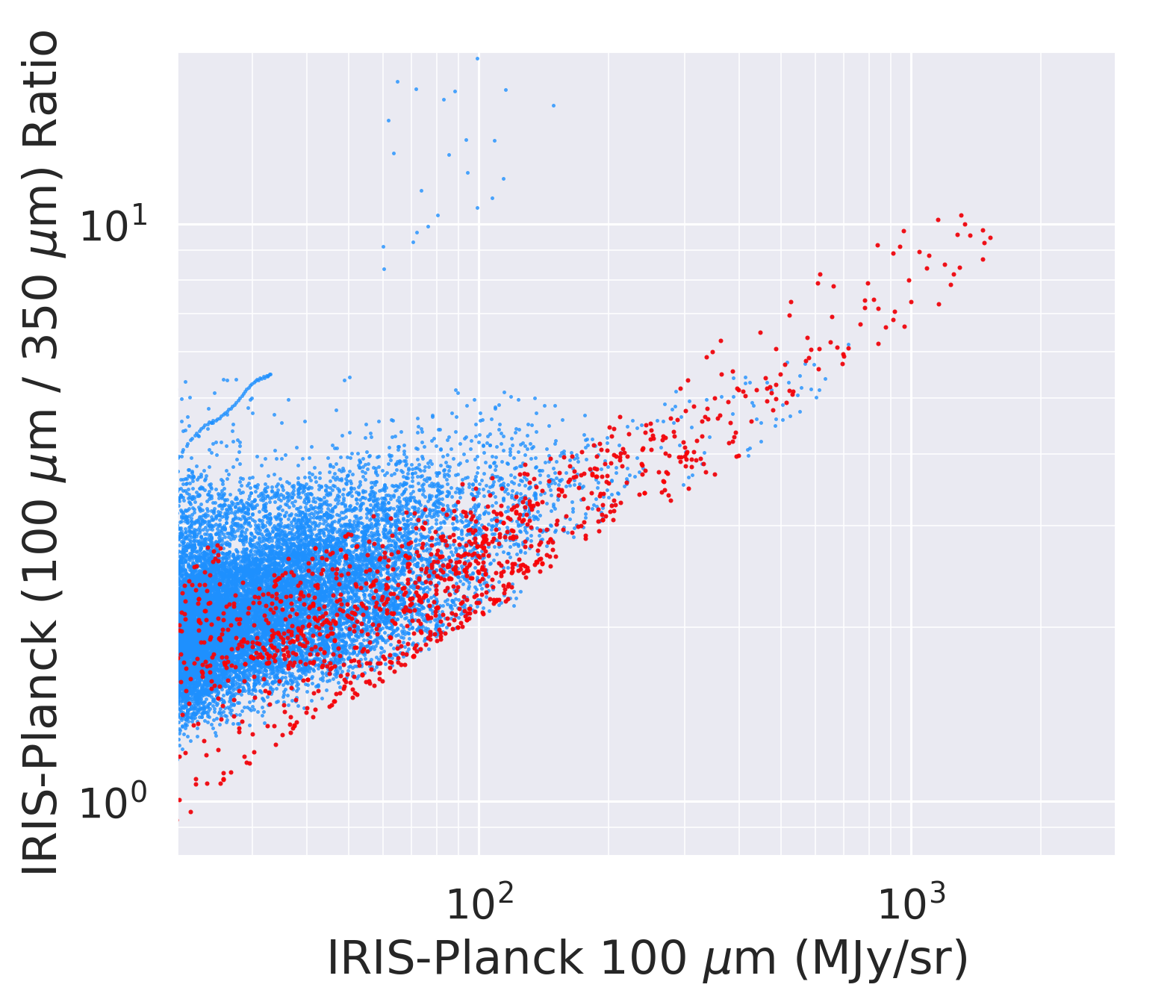}
\caption{Plots used to diagnose the origin of artefacts around the 30 Doradus star forming region in the LMC. In each case, all data was convolved to the same resolution and projected to the same pixel grid before comparison. {\it Left:} IRIS--\planck\ 100\,\micron\ surface brightness plotted against the PACS 100\,\micron\ / IRIS--\planck\ 100\,\micron\ ratio. {\it Centre:} IRIS--\planck\ 350\,\micron\ surface brightness plotted against the PACS 100\,\micron\ / IRIS--\planck\ 350\,\micron\ ratio. {\it Right:} IRIS--\planck\ 100\,\micron\ surface brightness plotted against the IRIS--\planck\ 100\,\micron\ / IRIS--\planck\ 350\,\micron\ ratio. Pixels located with 0.5\degr\ of the centre of 30 Dor are highlighted in each plot.}
\label{AppendixFig:30-Dor_Non-Linearity}
\end{figure*}

Upon first producing feathered maps for the LMC in the 100 and 160\,\micron\ \hersc-PACS bands, we noticed large negative bowls around the 30 Doradus star-forming complex, especially at 100\,\micron. There was also considerable ringing around 30 Dor (and to a lesser extent around the star-forming region LHA 120-N 55A) in the maps of the residuals between the feathered data and the IRIS--\planck\ data. This ringing manifested as a large flux deficit around the perimeter of 30 Dor (hence the negative bowls), with some flux enhancement in the centre.

By comparing the differences in surface brightness between several of our datasets over this region, we realised that these artefacts were actually originating in the IRIS--\planck\ data. (Here, we again use the term `IRIS--\planck\ data' to refer to the data that infers emission in the \hersc\ bands from SED-fitting to the IRIS and \planck\ maps.)

In the left panel of Figure~\ref{AppendixFig:30-Dor_Non-Linearity}, we plot the IRIS--\planck\ 100\,\micron\ surface brightness against the \hersc-PACS 100\,\micron\ / IRIS--\planck\ 100\,\micron\ surface brightness ratio; the datasets were convolved to the same resolution and projected to the same pixel grid before comparison, and limited to high-SNR pixels (\textgreater\,10\,MJy\,sr$^{-1}$ in IRIS--\planck\ 100\,\micron). For the most part the two maps agree quite closely, as would be expected, with the ratio remaining roughly constant (NB, there is no true zero level for the \hersc-PACS data, so the surface-brightness ratio values on the y-axis are unavoidably somewhat arbitrary). However for pixels within 0.5\degr\ of 30 Dor, highlighted in red, there is strong disagreement, with \hersc-PACS being much brighter than the IRIS--\planck\ for fainter pixels, and somewhat fainter for brighter pixels.

In the central panel of Figure~\ref{AppendixFig:30-Dor_Non-Linearity}, we plot IRIS--\planck\ 350\,\micron\ surface brightness against the \hersc-PACS 100\,\micron\ / IRIS--\planck\ 350\,\micron\ surface brightness ratio. The IRIS--\planck\ 350\,\micron\ data will naturally be dominated by the contribution of the \planck\ data to the IRIS--\planck\ SED fitting, and therefore should be well insulated against any IRIS-specific effects. This plot shows that the \hersc-PACS 100\,\micron\ data around 30 Dor is actually much better correlated with the IRIS--\planck\ 350\,\micron\ data than it was with the IRIS--\planck\ 100\,\micron\ data. We see that brighter pixels tend to have higher 100\,\micron/350\,\micron\ ratios (as expected due to colour effects), but that this trend is much the same for 30 Dor as for the rest of the LMC. This strongly indicates that the problem is with the IRIS--\planck\ 100\,\micron\ data, not the \hersc-PACS 100\,\micron\ data.

The right panel of figure Figure~\ref{AppendixFig:30-Dor_Non-Linearity} plots the IRIS--\planck\ 100\,\micron\ surface brightness against the IRIS-\planck\ 100\,\micron/350\,\micron\ ratio. Once again, pixels in 30 Dor show markedly different behaviour than pixels in the rest of the LMC. Fainter pixels around 30 Dor can actually be {\it fainter} in IRIS--\planck\ 100\,\micron\ than in 350\,\micron; the inverse is true for the brighter pixels. This reflects what was seen in the first panel of the figure. 

In short, the IRIS--\planck\ 100\,\micron\ data appears to be aberrant from both the \hersc-PACS 100\,\micron\ data {\it and} the IRIS--\planck\ 350\,\micron\ data.

The feathering processes was correcting the artefacts in the IRIS--\planck\ 100\,\micron\ data at smaller scales, where the erroneous emission in the IRIS--\planck\ map was replaced by the artefact-free emission in the \hersc-PACS map. But at larger scales, where the feathered map did not incorporate the \hersc\ emission, the artefacts persisted -- hence the negative bowls. Our working assumption is that some aspect of our DIRBE+IRIS feathering gave rise to these artefacts -- possibly because we had to use beam-mediated tapering, as opposed to a tapering window (see Section~\ref{Subsection:Feathering_DIRBE_IRIS} and Appendix~\ref{AppendixSection:Feathering_In-Out_Tests}). That is why the negatve bowls were strong in the 100\,\micron\ feathered map, and somewhat present in the 160\,\micron\ feathered map (where the impact of the 100\,\micron\ DIRBE+IRIS data on the inferred 160\,\micron\ surface brightness was lessened)

By extending the tapering window out to larger angular scales, we were able to increase the range of angular scales over which the artefacts were replaced by the artefact-free \hersc-PACS data. Fortunately, for the LMC, there is not an immediate, precipitous loss of power in the \hersc-PACS data as soon as one progresses to angular scales sampled by the IRIS--\planck\ (see Figure~\ref{Fig:IRIS_Planck_Herschel_Power_Spec}, and Section~\ref{Subsubsection:Validating_Feathering_DIRBE_IRIS}); rather, it seems that the \hersc-PACS data is mainly missing data on only the very largest scales. Therefore, we were able to employ a slightly larger tapering window to fix the artefacts, without excluding emission missed by \hersc-PACS. We experimented, to find a window that would remove the negative bowls from the final data whilst not causing residuals on the larger scales, and found that 45\arcmin--90\arcmin\ appears to achieve this.

\needspace{3\baselineskip} \section{Feathering Outputs} \label{AppendixSection:Feathering_Outputs}

\begin{figure*}[p]
\centering
\includegraphics[width=0.8\textwidth]{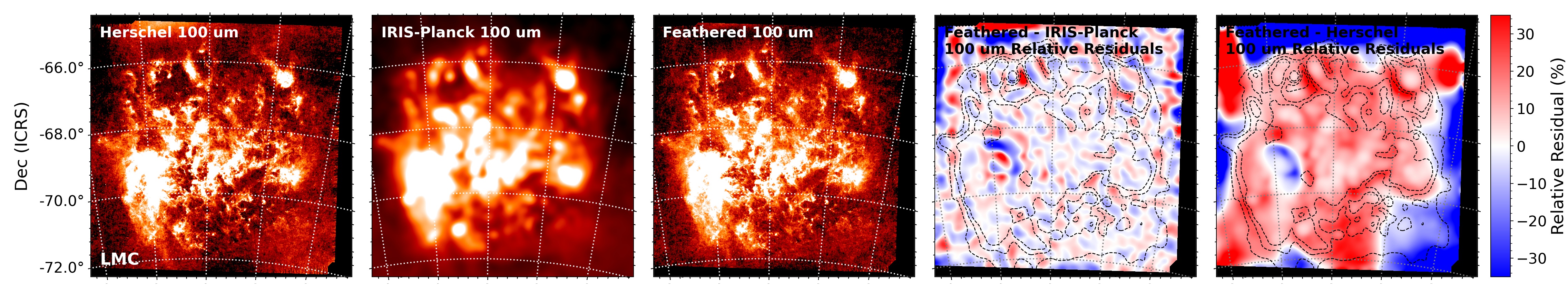}
\includegraphics[width=0.8\textwidth]{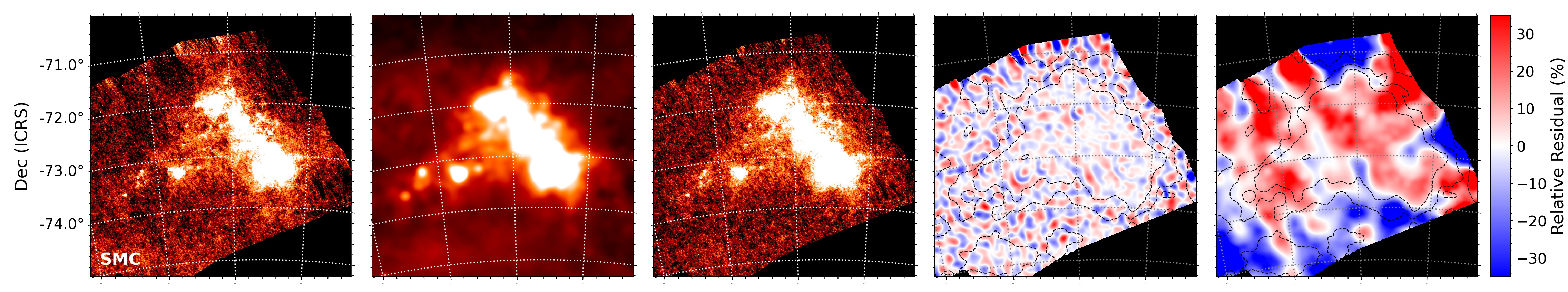}
\includegraphics[width=0.8\textwidth]{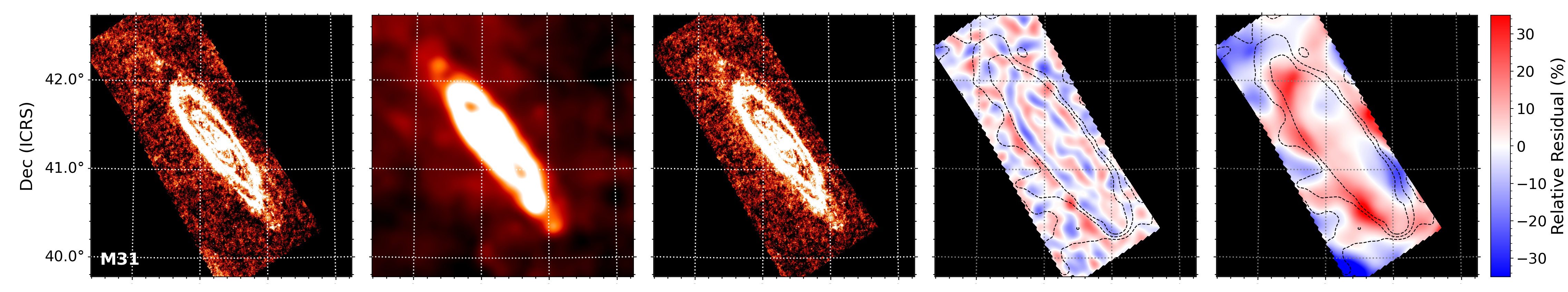}
\includegraphics[width=0.8\textwidth]{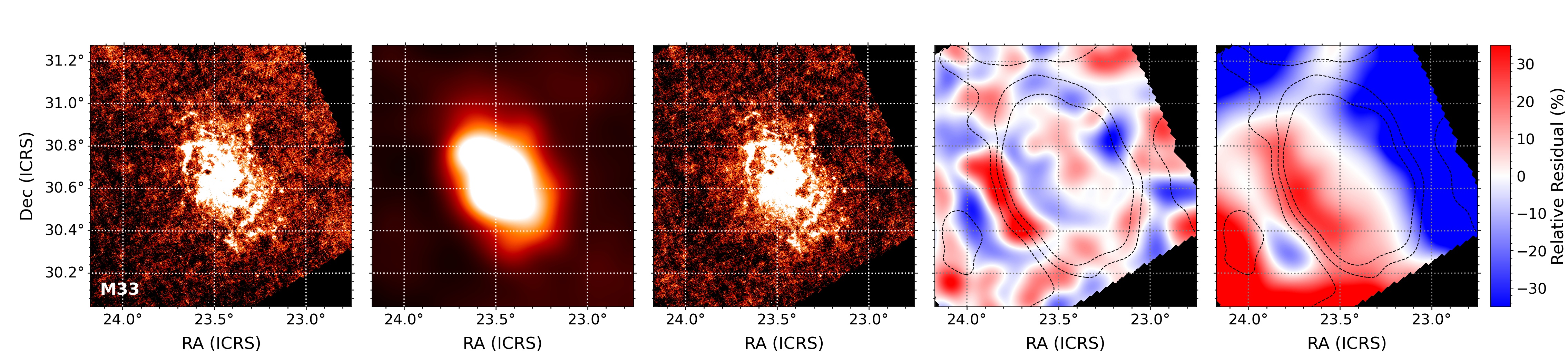}
\\
\includegraphics[width=0.01\textwidth]{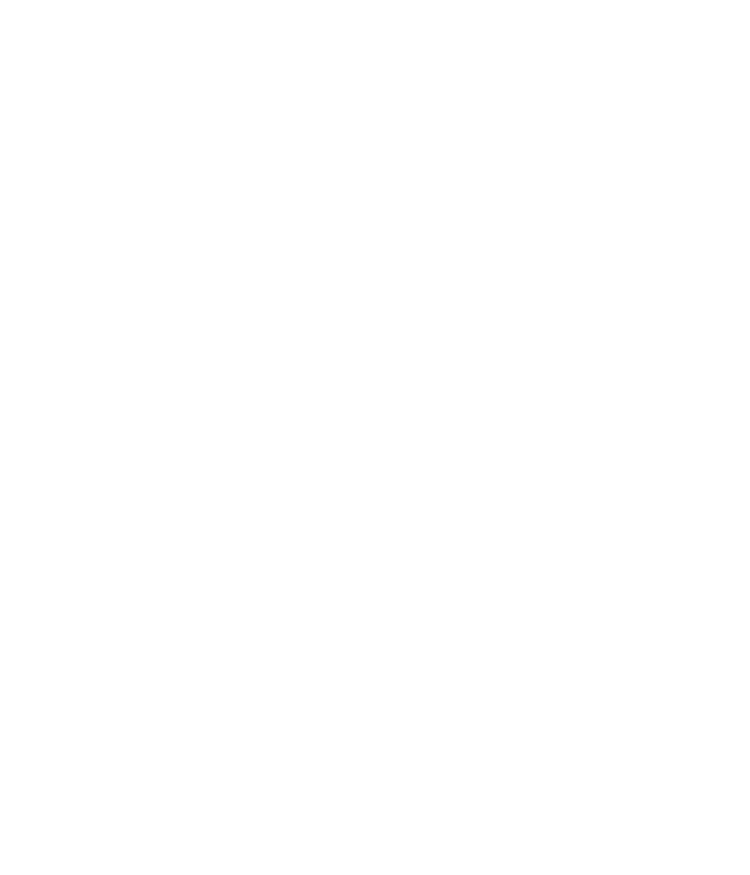}
\\
\includegraphics[width=0.8\textwidth]{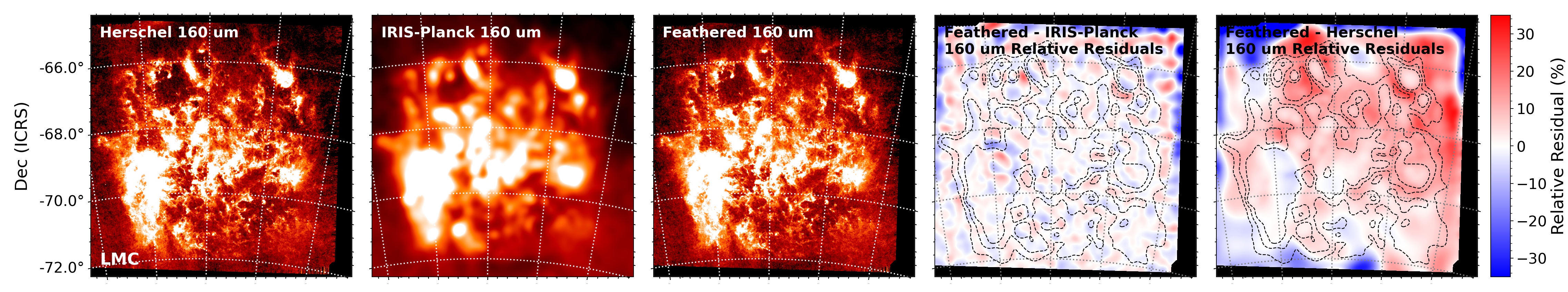}
\includegraphics[width=0.8\textwidth]{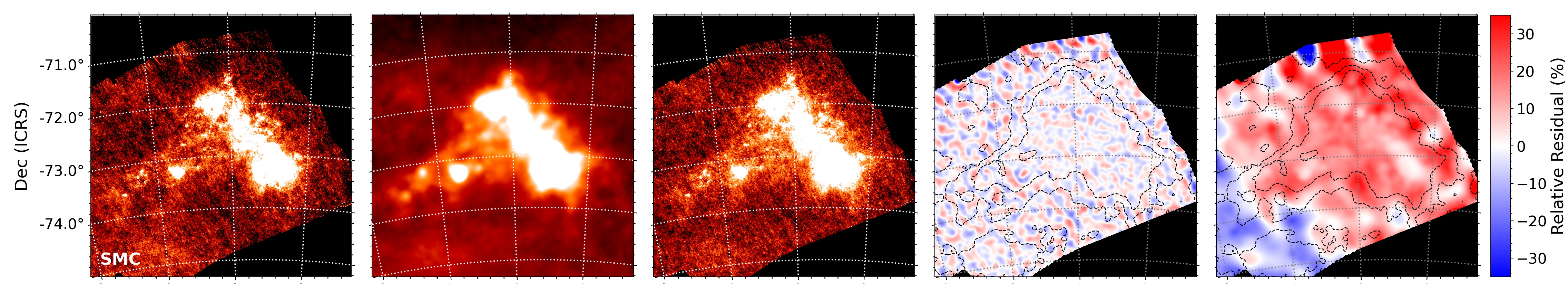}
\includegraphics[width=0.8\textwidth]{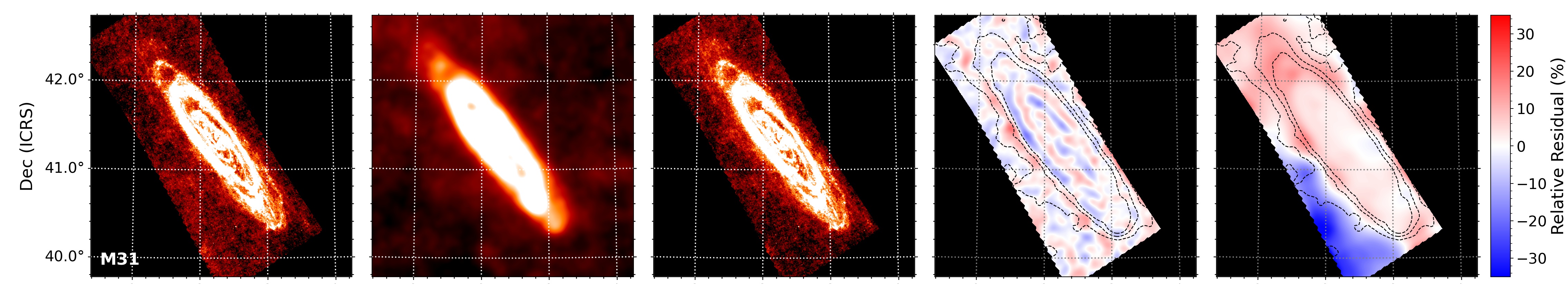}
\includegraphics[width=0.8\textwidth]{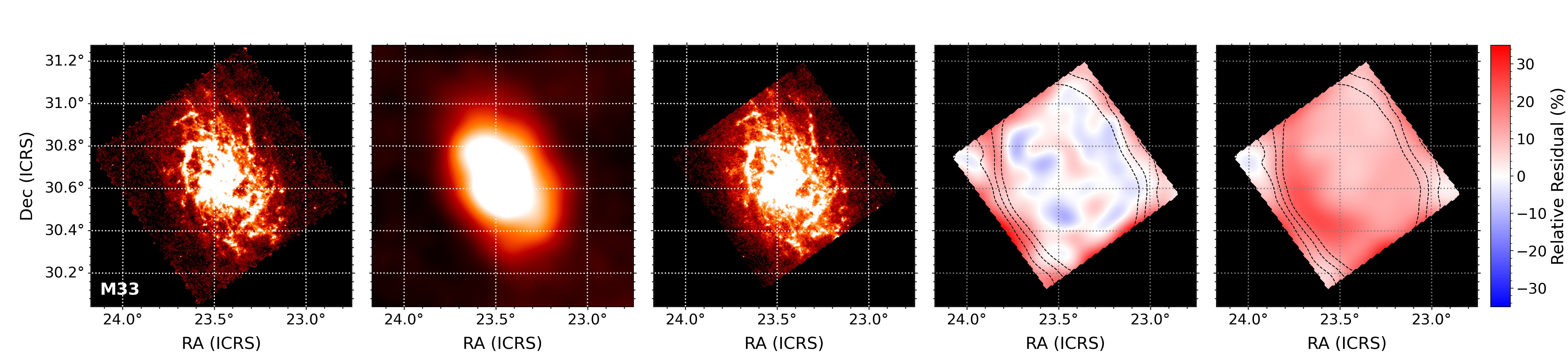}
\caption{The results of feathering together \hersc\ with IRIS--\planck\ data at 100\,\micron\ (top four rows) and 160\,\micron\ (bottom four rows) for each of our galaxies, along with diagnostic plots. Description of panels same as per Figure~\ref{Fig:Herschel+Planck+IRAS+COBE_Feathering_Output_250}.}
\label{AppendixFig:Herschel+Planck+IRAS+COBE_Feathering_Output_100_160}
\end{figure*}

\begin{figure*}[p]
\centering
\includegraphics[width=0.8\textwidth]{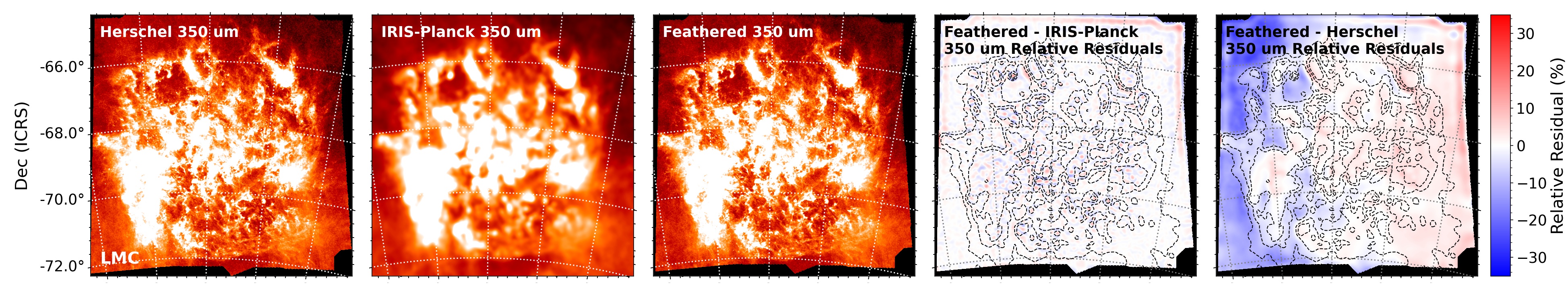}
\includegraphics[width=0.8\textwidth]{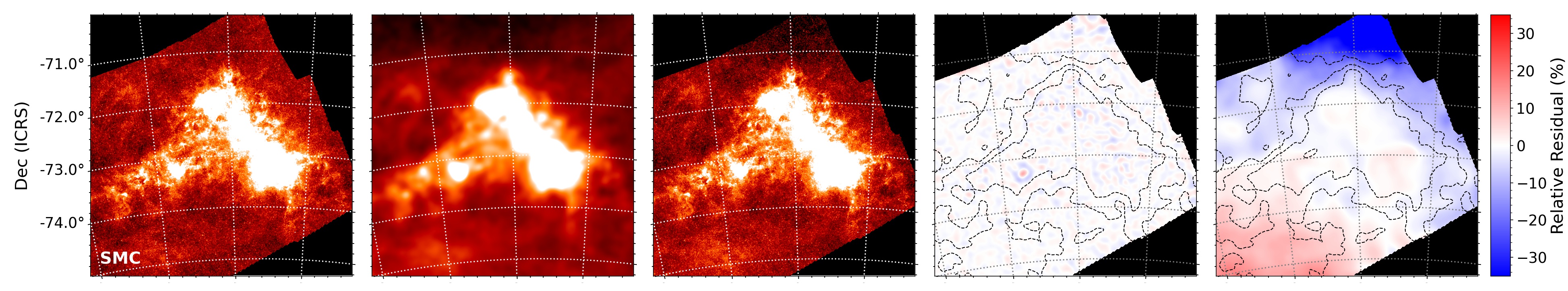}
\includegraphics[width=0.8\textwidth]{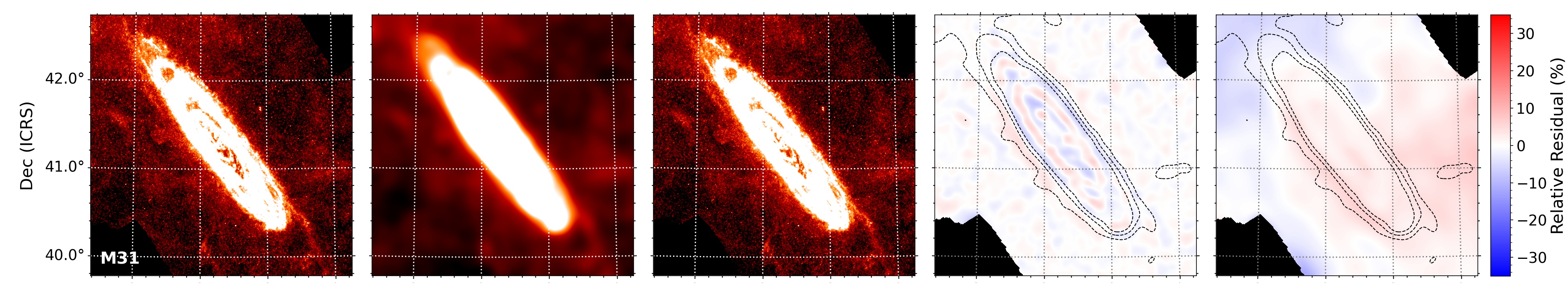}
\includegraphics[width=0.8\textwidth]{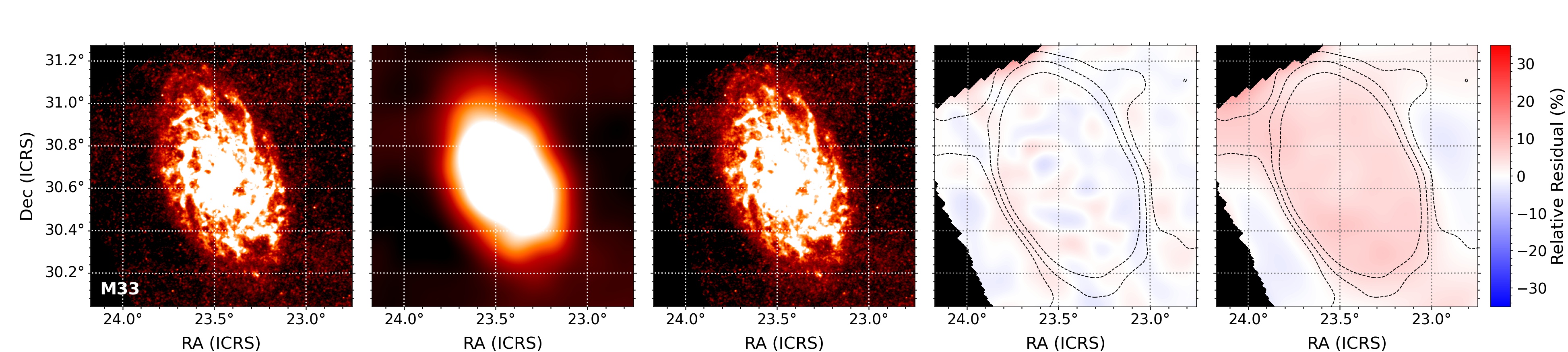}
\\
\includegraphics[width=0.01\textwidth]{Blank.png}
\\
\includegraphics[width=0.8\textwidth]{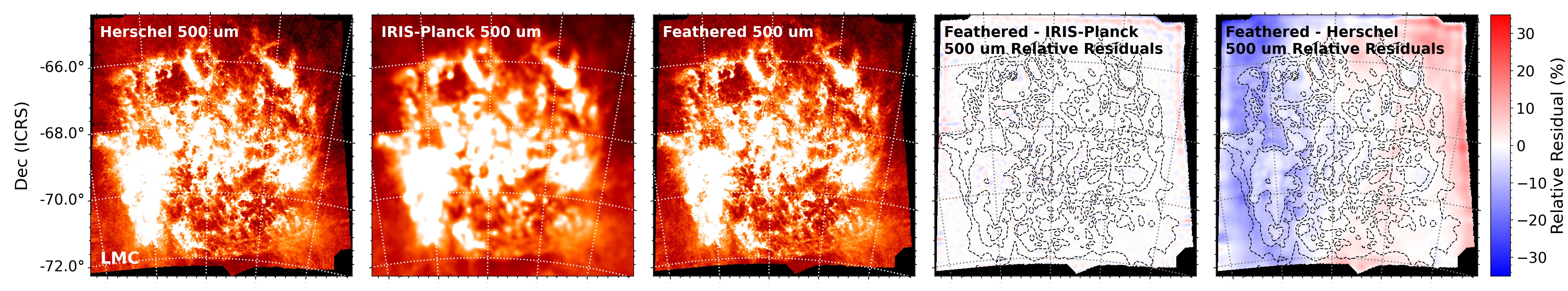}
\includegraphics[width=0.8\textwidth]{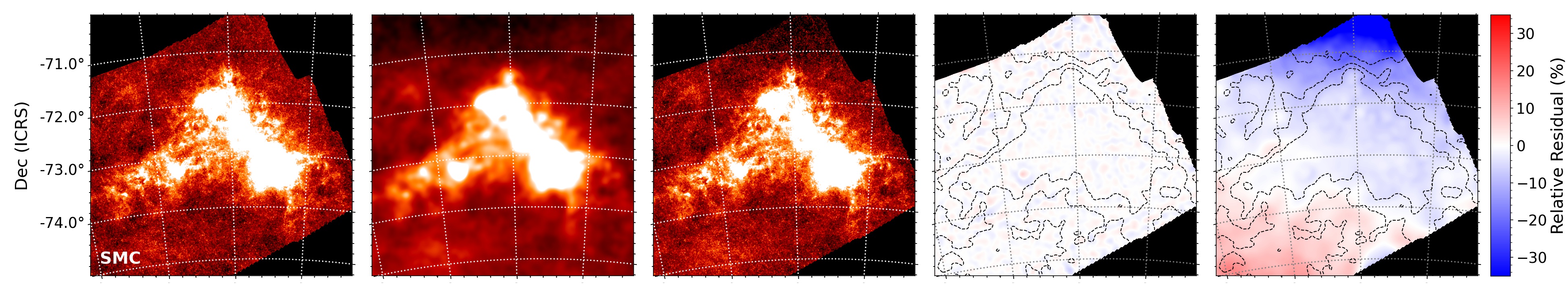}
\includegraphics[width=0.8\textwidth]{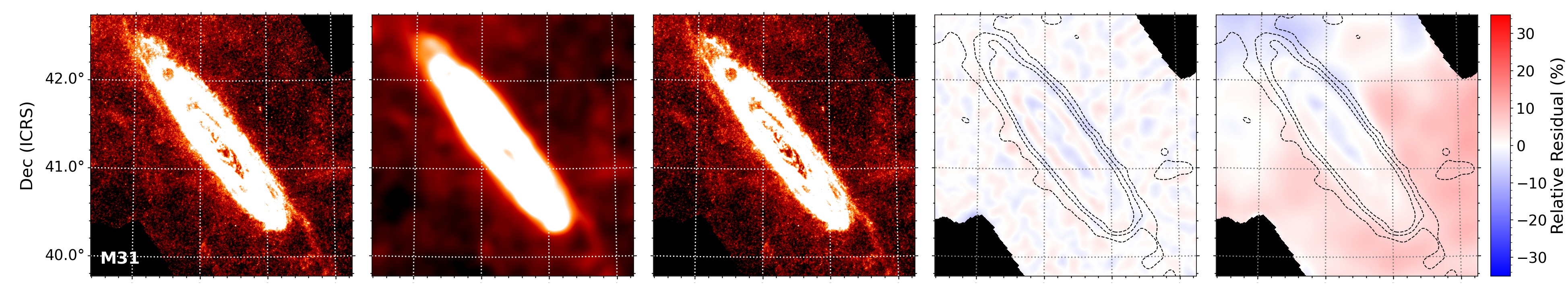}
\includegraphics[width=0.8\textwidth]{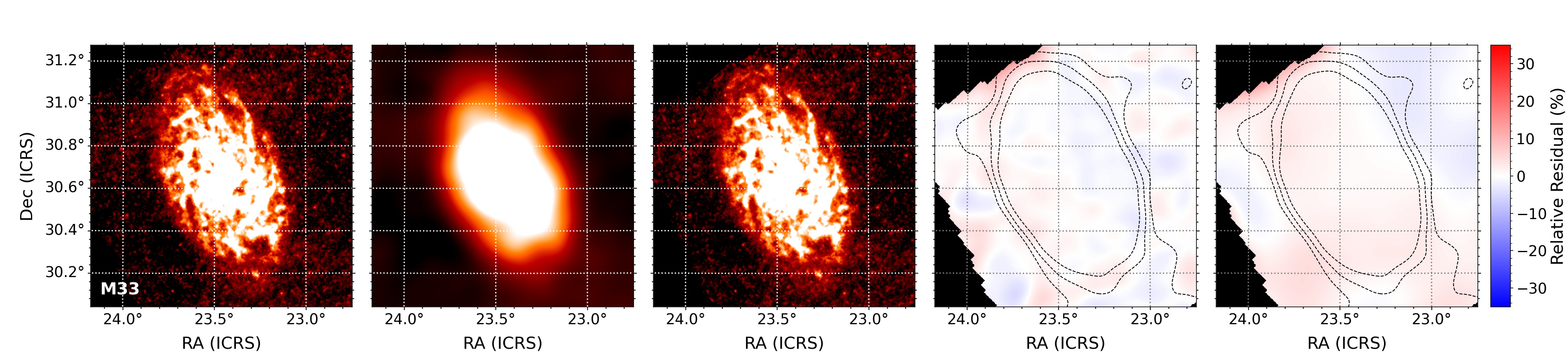}
\caption{The results of feathering together \hersc\ with IRIS--\planck\ data at 350\,\micron\ (top four rows) and 500\,\micron\ (bottom four rows) for each of our galaxies, along with diagnostic plots. Description of panels same as per Figure~\ref{Fig:Herschel+Planck+IRAS+COBE_Feathering_Output_250}.}
\label{AppendixFig:Herschel+Planck+IRAS+COBE_Feathering_Output_350_500}
\end{figure*}

Here we provide illustrations of the outputs of our feathering \hersc\ data with IRIS--\planck\ data, along the associated diagnostic plots, at 100, 160, 350, and 500\,\micron\ (Figures~\ref{AppendixFig:Herschel+Planck+IRAS+COBE_Feathering_Output_100_160} and \ref{AppendixFig:Herschel+Planck+IRAS+COBE_Feathering_Output_350_500}). The 250\,\micron\ outputs are shown in Figure~\ref{Fig:Herschel+Planck+IRAS+COBE_Feathering_Output_250}.

\end{document}